\documentclass[12pt,a4paper]{article}
\usepackage{graphicx}
\usepackage{times}
\usepackage{subfigure}
\usepackage{amsmath, amsthm, amssymb}
\usepackage{natbib}
\usepackage{setspace}
\usepackage[a4paper]{geometry}
\usepackage{float}
\usepackage{bm}
\usepackage[bookmarksnumbered]{hyperref}
\usepackage{appendix}

\newtheorem*{Theorem}{\bf Theorem}

\begin{document}

\title{
    \bf Maximum leave-one-out likelihood estimation for
    location parameter of unbounded densities
}
\author{Thanakorn Nitithumbundit\footnote{Corresponding author. Email: T.Nitithumbundit@maths.usyd.edu.au} \ and
    Jennifer S.K. Chan \\
    \small{School of Mathematics and Statistics, University of Sydney, NSW 2006, Australia}}
\date{\today}

\maketitle

{\bf Abstract:}
Maximum likelihood estimation of a location parameter fails when the density have unbounded mode.
An alternative approach is considered by
leaving out a data point to avoid the unbounded density in the full likelihood. This modification give rise to the
leave-one-out likelihood.
We propose an ECM algorithm which maximises the leave-one-out likelihood.
It was shown that the estimator which maximises the leave-one-out likelihood is consistent and super-efficient.
However, other asymptotic properties such as the optimal rate of convergence and asymptotic distribution is still under question.
We use simulations to investigate these asymptotic properties of the location estimator using our proposed algorithm.


\section{Introduction}

Asymptotic properties of maximum likelihood estimators for location parameters are well known for the case when the likelihood is bounded and even non-differentiable \citep{Rao1968},
but the methodology breaks down when the likelihood is unbounded at certain points.
Alternate approaches for the unbounded case have been considered in
\citet{Ibragimov1981unbounded,Ibragimov1981StatisticalEstimation}
and in
\cite{Rao1966asymptotic}
where they proved consistency results
using the Bayesian approach.

Under the likelihood approach however, modifications to the full likelihood is necessary.
A possible solution is to leave out a data point closest to the location parameter in the full likelihood which might cause the density to become unbounded.
This modification leads to a concept known as the leave-one-out (LOO) likelihood proposed by \citet{Podgorski2015}.
They proved consistency and super-efficiency of the location estimator that maximises the LOO likelihood.
More precisely, they have found a lower bound for the rate of convergence of the location estimator.
However, other asymptotic properties such as optimal rate of convergence and the asymptotic distribution are yet to be proven.

Our main objective of the paper is to
propose an expectation/conditional maximisation (ECM) algorithm \citep{Meng1993} to obtain the maximum LOO estimator of parameters from variance gamma (VG) distribution \citep{MadanSeneta1990}.
This proposed algorithm is an extension to the EM algorithm
for estimating the location parameter of symmetric generalised Laplace distribution in \citet{Podgorski2015}.
Additionally, they have not yet supplied simulations results using their algorithm.
The convergence properties of the ECM algorithm for the LOO likelihood is similar to the ECM algorithm for the full likelihood.
Our other objective is to
analyse the asymptotic behaviour of the maximum LOO likelihood estimator for the location parameter,
by applying our proposed algorithm to simulated data from a VG distribution with different samples sizes and shape parameters.

There are two important reasons why we consider parameter estimation from VG distribution.
Firstly, it is part of a more general class of distributions called generalised hyperbolic (GH) distribution where it has a normal mean-variance mixture representation \citep{Barndorff-Nielsen1982}.
Not only that, it is an important special case that corresponds to the unbounded case of the GH distribution.
In order for the GH distribution to approach the VG distribution, it needs to have one of its shape parameters approach the boundary of the parameter space.
So the regular EM algorithm that estimates parameters from GH distribution proposed by \citet{Protassov2004} does not truly capture the unbounded density.
Secondly, it has applications in many areas such as financial data, signal processing and quality control.
See \citet{Kotz2001} for other applications and further details on generalised Laplace distribution which are fundamentally equivalent to VG distribution.

It is worth emphasising that not only can this methodology deal with estimation of location parameter of unbounded densities, but can deal with other extreme cases where the parameter estimate approaches the boundary of the parameter space, potentially causing the density to become unbounded.
One particular example is based on the singularity problem in finite mixture of normals model \citep{Seo2012}.

In summary,
Section \ref{Section: Variance Gamma Distribution} summarises some important properties of the multivariate skewed VG distribution.
Section \ref{Section: Maximum Leave-one-out Likelihood} formulates the maximum LOO likelihood framework for location parameter estimation of distributions with unbounded densities.
Section \ref{Section: ECM algorithm for LOO likelihood} introduces the ECM algorithm using the LOO likelihood to estimate parameters from the multivariate skewed VG distribution.
Section \ref{Section: Simulation study of asymptotic distribution} presents the simulation study to analyse the asymptotic behaviour of the maximum LOO likelihood estimator for location parameter of the VG distribution.
We conclude the paper with further remarks in Section \ref{Section: Conclusion}.

\section{Variance gamma distribution}
\label{Section: Variance Gamma Distribution}

We will first discuss some important properties of the multivariate skewed VG (MSVG) distribution.
The probability density function (pdf) of a $d$-dimensional MSVG distribution is given by
\small
\begin{align} \label{VGpdf}
    f_{VG} (\bm y) =&
        \frac{
            2^{1-\frac{d}{2}}  \nu ^{\nu }
        }{
            \left|\bm\Sigma\right|^{\frac{1}{2}} \pi^{\frac{d}{2}} \Gamma (\nu )
        }
        \frac{
            K_{\nu-\frac{d}{2}}\left(\sqrt{[2\nu + \bm \gamma'\bm\Sigma^{-1}\bm \gamma]
                (\bm y - \bm\mu)'\bm\Sigma^{-1}(\bm y - \bm\mu) }\right)
            \exp\left((\bm y - \bm\mu)'\bm\Sigma^{-1}\bm \gamma\right)
        }{
            (2 \nu + \bm \gamma'\bm\Sigma^{-1}\bm \gamma)^{\frac{1}{4} (2 \nu -d)}
            [(\bm y - \bm\mu)'\bm\Sigma^{-1}(\bm y - \bm\mu)] ^{\frac{1}{4}(d-2\nu)}
        }
\end{align}
\normalsize
where $\bm\mu\in\mathbb R^d$ is the location parameter,
$\bm\Sigma$ is a $d\times d$ positive definite symmetric scale matrix,
$\bm\gamma\in\mathbb R^d$ is the skewness parameter, $\nu>0$ is the shape parameter, $\Gamma(\cdot)$ is the gamma function and
$K_\eta(\cdot)$ is the modified Bessel function of the second kind with index $\eta$
\citep[\S9.6]{Abramowitz2007}.

The MSVG distribution has a normal mean-variance mixtures representation given by
\begin{equation}
    \bm y_i | \lambda_i \sim \mathcal N_d(\bm\mu+\bm\gamma\lambda_i, \lambda_i \bm\Sigma), \quad \lambda_i \sim \mathcal G(\nu,\nu) \label{scale mixture rep of VG}
\end{equation}
where $\mathcal G(\alpha,\beta)$ is a Gamma distribution with shape parameters $\alpha>0$, rate parameter $\beta>0$ and pdf
\begin{equation*}
    f_{G}(\lambda) = \frac{\beta^\alpha}{\Gamma(\alpha)} \lambda^{\alpha-1} \exp(-\beta \lambda) , \text{ for } \lambda>0.
\end{equation*}
The mean and covariance matrix of a MSVG random vector $\bm Y_i$ are given by
\begin{align*}
    \mathbb E (\bm Y_i) = \bm\mu + \bm\gamma \quad \text{and} \quad
    \mathbb C \mbox{ov}(\bm Y_i) = \bm\Sigma + \tfrac{1}{\nu} \bm\gamma \bm\gamma',
\end{align*}
respectively.
The pdf in (\ref{VGpdf}) as $\bm y_i \rightarrow \bm\mu$ is given by
\begin{align}
    f_{VG}(\bm y_i) \sim
        \begin{cases}
            \displaystyle2^{\nu-d} \pi^{-\frac{d}{2}} \left|\bm\Sigma\right|^{-\frac{1}{2}}
                \frac{\Gamma \left( \nu - \tfrac{d}{2} \right)}{\Gamma\left(\nu\right)}
                \frac{ \nu^\nu  }
                    {(2\nu + \bm \gamma'\bm\Sigma^{-1}\bm \gamma)^\frac{2\nu-d}{2}}
                & \text{ if } \nu>\tfrac{d}{2}, \\ \vspace{2mm}
            \displaystyle-2^{1-d} \pi^{-\frac{d}{2}} \left|\bm\Sigma\right|^{-\frac{1}{2}}
                \frac{d^\frac{d}{2}}{\Gamma\left(\tfrac{d}{2}\right)}
                \log \left(z_i \right)
                & \text{ if } \nu=\tfrac{d}{2}, \\ \vspace{2mm}
            \displaystyle2^{-\nu} \pi^{-\frac{d}{2}} \left|\bm\Sigma\right|^{-\frac{1}{2}}
                \frac{\Gamma \left(\tfrac{d}{2} - \nu \right)}{\Gamma\left(\nu\right)}
                \nu^\nu z_i^{2\nu-d}
                & \text{ if } \nu<\tfrac{d}{2}, \label{density_asym}
        \end{cases}
\end{align}
where
\begin{equation}    \label{z2, Mahalanobis distance}
    z^2_i = (\bm y_i - \bm\mu)'\bm\Sigma^{-1}(\bm y_i - \bm\mu).
\end{equation}

So the density becomes unbounded for the case when $\nu\leq\frac{d}{2}$.
This poses some technical difficulty when working with the MSVG distribution as it is unclear whether the shape parameter will fall into the unbounded range.

\begin{figure}[htbp]
    \begin{center}
         \subfigure[log-likelihood comparison]
            {\label{}
            \includegraphics[width=0.48\textwidth]
            {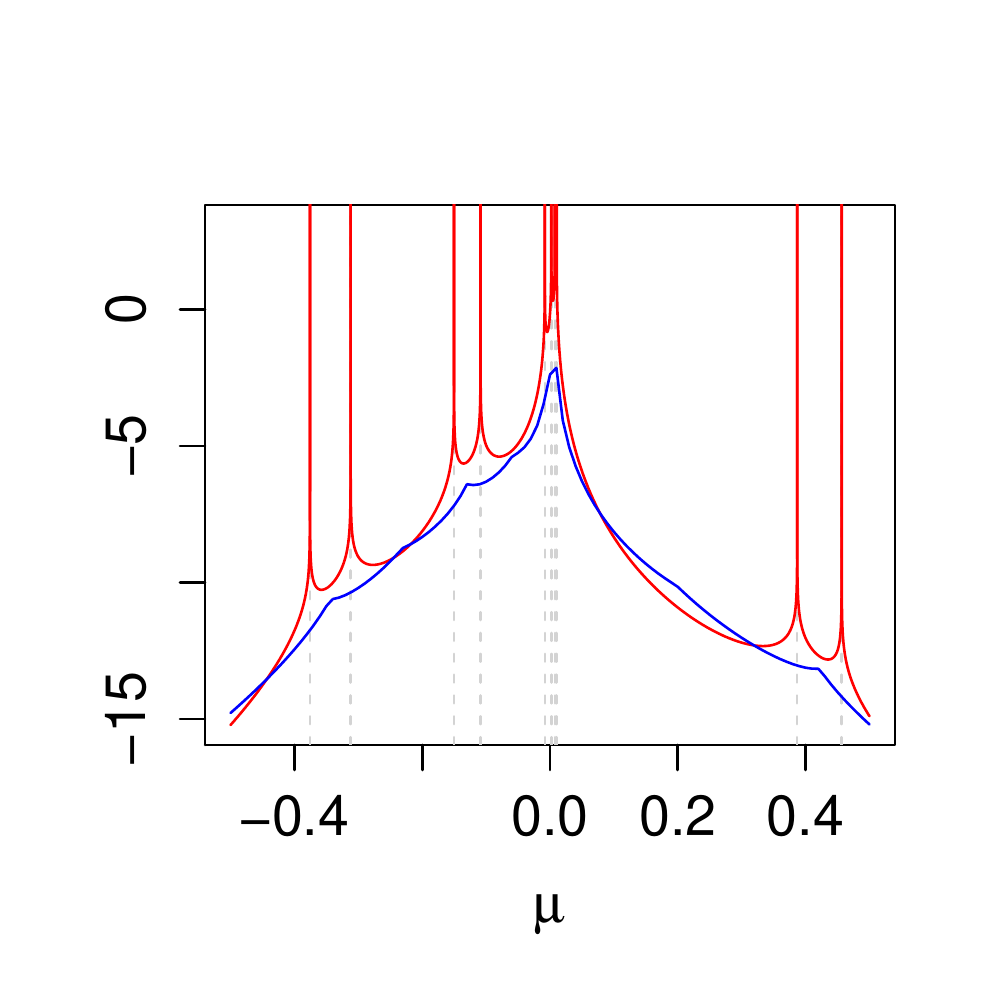}}
         \subfigure[close-up on the LOO log-likelihood]
            {\label{Plot_LOO_likelihood_closer_4x4}
            \includegraphics[width=0.48\textwidth]
            {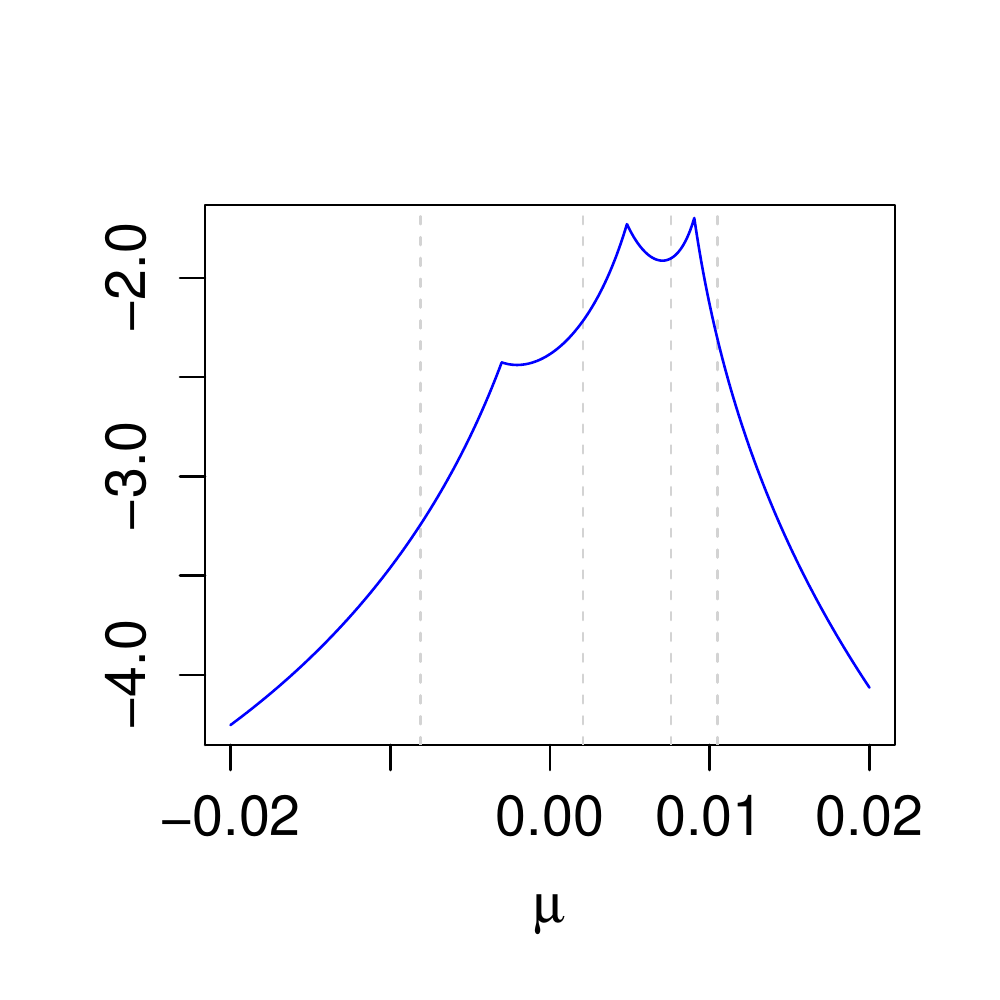}}
    \end{center}
    \caption{
        Left: Comparing full log-likelihood (red) vs.
            LOO log-likelihood (blue) of simulated data from standardised VG distribution with
            $\nu=0.2$ and sample size of ten (vertical grey dashed lines).
        Right: Close-up of the left figure at around $\mu=0$ focusing on the LOO log-likelihood.
    }
    \label{LOO likelihood plots}
\end{figure}

This problem is illustrated in Figure \ref{LOO likelihood plots}, we first generate ten standardised VG samples with $\nu=0.2$ using the normal mean-variance mixture representation.
Then we plot both the full
log-likelihood along with the leave-one-out (LOO) log-likelihood with respect to the location parameter.
We see that leaving the data point out essentially smooths out the unbounded points of the log-likelihood so the maximum can be well defined.
Additionally, if we zoom in at around $\mu=0$, observe that cusps tend to occur between data points.

\section{Maximum leave-one-out likelihood}
\label{Section: Maximum Leave-one-out Likelihood}

Let us suppose
$\bm y=(\bm y_1, \cdots, \bm y_n)$ be observed data from MSVG distribution with corresponding missing parameters
$\bm\lambda=(\lambda_1, \cdots, \lambda_n)$, and
$\bm\theta=(\bm\mu, \bm\Sigma, \bm\gamma, \nu)$ be parameters from MSVG distribution in parameter space $\bm\Theta$.
The density of the MSVG distribution is unbounded at $\bm\mu$ when $\nu\leq\frac{d}{2}$.
So the maximum likelihood estimate is not well defined since there are multiple unbounded points in the likelihood function.
Thus the Fisher information matrix with respect to $\bm\mu$ is also not well defined.
Instead we consider the incomplete Fisher information matrix defined by
\begin{align}
    \mathcal{I}_{\epsilon}(\bm\theta)=\mathbb{E}\left[\left.\left(\frac{\partial }{\partial \theta }\log f (X|\theta )\right)^2\right|\left| X\right| >\epsilon \right]
\end{align}
for $\epsilon>0$.

We aim to provide a methodology to estimate parameters from MSVG distribution even with the presence of unboundedness. Although in general this methodology can also apply to distributions which satisfies the following assumptions \citep{Podgorski2015}:

(A1)    $f(x)=p(x)|x|^\alpha$, $\alpha\in(-1,0)$, $p$ has bounded derivative on $\mathbb R\backslash\{0\}$ and,
for some $\epsilon>0$, $f$ is non-zero and continuous either on $[-\epsilon,0]$ or on $[0,\epsilon]$.

(A2)    There exist $b>0$ such that $f(x)=O(|x|^{-b-1})$ when $|x|\rightarrow\infty$.

(A3)    For all $\epsilon>0$, the incomplete Fisher information is finite.

\subsection{Leave-one-out likelihood}
Let the observed LOO likelihood is defined as
\begin{align}
    L^{\rm LOO}(\bm\theta; \bm y)
        = \prod_{i\ne k(\bm\mu)} f(\bm y_i)
\end{align}
where we define the LOO index
\begin{align}
    k(\bm\mu) = \underset{k\in \{1,\text{...},n\}}{\text{argmin}} (\bm y_k - \bm\mu)^T \bm\Sigma^{-1} (\bm y_k-\bm\mu).
\end{align}
For the case where there are more than one indices, we choose the smallest index.
Let the observed LOO log-likelihood be defined as
$\ell^{\rm LOO}(\bm\theta; \bm y) = \log L^{\rm LOO}(\bm\theta; \bm y)$.

Let us define the maximum leave-one-out likelihood estimator denoted as
$\hat{\bm\theta}^{\rm MLLE}_n$ (or simply $\hat{\bm\theta}_n$) to be the estimator that maximises the LOO likelihood with respect to $\bm\theta$.
Some main properties of the location estimator $\hat{\mu}_n$ is consistency and super-efficient rate of convergence.
These properties follow from the main theorem established by \citet{Podgorski2015}.
\begin{Theorem}
    Let $f$ satisfies the assumptions
    (A1) to (A3) and
    let $\hat{\mu}_n$ be the maximiser of
    $L^{\rm LOO}(\mu;\bm{y})$.
    Then $\hat{\mu}_n$ is consistent estimator of $\mu$ and
    for any $\beta<1/(1+\alpha)$,
    \begin{align}
        \underset{n\rightarrow\infty}{\lim} n^\beta(\hat{\mu}_n-\mu)\overset{p}{=}0.
    \end{align}
\end{Theorem}

By the main theorem, the lower bound of the rate of convergence for the maximum LOO likelihood location estimator is attained,
but doesn't state the optimal rate of convergence.
By setting $\beta=1/(1+\alpha)$, this possibly gives us the optimal rate of convergence.
For comparison purposes, we will call this the {\it proposed optimal rate}.
Additionally,
$n^{\beta_0}(\hat{\mu}_n-\mu)$
will converge to some asymptotic distribution for some suitable choice of $\beta_0$.
We will investigate the asymptotic properties later in Section \ref{Section: Simulation study of asymptotic distribution} using simulations from
VG distribution.

\section{ECM algorithm for LOO likelihood}
\label{Section: ECM algorithm for LOO likelihood}

Finding the maximum LOO likelihood estimator $\hat{\bm\theta}_n$ can be difficult as
the LOO likelihood has many cusps when $\nu\leq d$, and the LOO index $k(\bm\mu)$ makes derivatives tedious to work with
since the summation and the differential can't simply be interchanged.
Alternatively, we can maximise the complete-data LOO likelihood which allows the implementation of the ECM algorithm.

Using the normal mean-variance mixture representation in Section \ref{Section: Variance Gamma Distribution}, we can represent the complete-data LOO log-likelihood as
\begin{align}
    \ell^{\rm LOO}(\bm\theta; \bm y,\bm\lambda)
        =& \ell^{\rm LOO}_{N}(\bm\mu,\bm\Sigma,\bm\gamma;\bm y,\bm\lambda)
            + \ell^{\rm LOO}_{G}(\nu;\bm\lambda)
\end{align}
where the LOO log-likelihood of the conditional normal distribution is given by
\begin{align}
    &\ell^{\rm LOO}_{N}(\bm\mu, \bm\Sigma, \bm\gamma;\bm y , \bm\lambda) \nonumber \\
        &= -\frac{n-1}{2}\log|\bm\Sigma| - \frac{1}{2}\sum_{i\ne k(\bm\mu)}
                \frac{1}{\lambda_i}(\bm y_i - \bm\mu - \lambda_i\bm\gamma)'
                    \bm\Sigma^{-1} (\bm y_i - \bm\mu - \lambda_i\bm\gamma)
                    -\frac{(n-1)d}{2}\log\pi
\end{align}
and the LOO log-likelihood of the conditional gamma distribution is given by
\begin{align}
    \ell^{\rm LOO}_{G}(\nu;\bm\lambda)  \label{Gamma log-likelihood}
        = (n-1)(\nu\log\nu - \log\Gamma(\nu)) + (\nu-1)\sum_{i\ne k(\bm\mu)} \log\lambda_i - \nu\sum_{i\ne k(\bm\mu)} \lambda_i.
\end{align}

The outline of the ECM algorithm of MSVG distribution using the full likelihood is given in \citet{Nitithumbundit2015}.
However, modifications to the algorithm is necessary when using the LOO likelihood.
We will discuss the necessary modifications needed in order to
attain local and global convergence of the algorithm.

\subsection{E-step}

By analysing the conditional posterior distribution of $\lambda_i$ given $\bm y_i$ which has density
\begin{align} \label{pdfGIG}
    f(\lambda_i| \bm y_i, \bm\theta)
        \propto& \ \lambda_i^{\nu-\frac{d}{2}-1}
            \exp\left[-\frac{1}{2\lambda_i}(\bm y_i - \bm\mu)'\bm\Sigma^{-1}(\bm y_i - \bm\mu)
                - \frac{\lambda_i}{2}\left(2\nu + \bm\gamma'\bm\Sigma^{-1}\bm\gamma \right)\right]
\end{align}
which corresponds to the pdf of a generalised inverse Gaussian distribution \citep{Embrechts1983}, we can calculate the following conditional expectations:
\begin{align}
    \widehat{\lambda_i} \label{E-step lambda1}
        =& \ \mathbb{E} \left(\lambda_i | \bm y, \bm\theta \right)
        =  \frac{z_i K_{\nu-\frac{d}{2}+1}\left(\sqrt{2\nu + \bm\gamma'\bm\Sigma^{-1}\bm\gamma} z_i\right)}
                {\sqrt{2\nu+\bm\gamma'\bm\Sigma^{-1}\bm\gamma} K_{\nu-\frac{d}{2}}\left(\sqrt{2\nu+\bm\gamma'\bm\Sigma^{-1}\bm\gamma}z_i\right)} , \\
    \widehat{1/\lambda_i} \label{E-step lambda2}
        =&\ \mathbb{E} \left(\frac{1}{\lambda_i} \Bigg| \bm y, \bm\theta \right)
        =  \frac{\sqrt{2\nu+\bm\gamma'\bm\Sigma^{-1}\bm\gamma} K_{\nu-\frac{d}{2}-1}\left(\sqrt{2\nu+\bm\gamma'\bm\Sigma^{-1}\bm\gamma}z_i\right)}
                {z_i K_{\nu-\frac{d}{2}}\left(\sqrt{2\nu+\bm\gamma'\bm\Sigma^{-1}\bm\gamma}z_i\right)} , \\
    \widehat{\log\lambda_i} \label{E-step lambda3}
        =& \ \mathbb{E} \left(\log\lambda_i | \bm y, \bm\theta \right)
        = \log\left(\frac{z_i}{\sqrt{2\nu+\bm\gamma'\bm\Sigma^{-1}\bm\gamma}}\right)
            + \frac{K_{\nu-\frac{d}{2}}^{(1,0)}(\sqrt{2\nu+\bm\gamma'\bm\Sigma^{-1}\bm\gamma}z_i)}
                {K_{\nu-\frac{d}{2}}(\sqrt{2\nu+\bm\gamma'\bm\Sigma^{-1}\bm\gamma}z_i)}
\end{align}
where $\displaystyle K_{\nu}^{(1,0)}(z) = \tfrac{\partial}{\partial\alpha} K_\alpha (z)\big|_{\alpha=\nu}$
which is approximated using the second-order central difference approximation
\begin{equation}    \label{central difference formula for BesselK}
    K_{\nu}^{(1,0)}(z) \approx \frac{K_{\nu+h}(z) - K_{\nu-h}(z)}{2h}
\end{equation}
where we let $h=10^{-5}$.

\subsection{Derivative of LOO log-likelihood}
\label{Section: Derivative of LOO log-likelihood}

Derivatives of $\ell^{\rm LOO}_{N}$ with respect to $(\bm\Sigma, \bm\gamma)$ are straight forward to calculate using matrix differentiation.
Here we will show some difficulties with the derivative with respect to $\bm\mu$.
The first-order derivative of the complete-data LOO log-likelihood with respect to $\bm\mu$ is
\begin{align}
    \frac{\partial}{\partial\bm\mu}\ell ^{\rm LOO}_{N}
    = - \frac{1}{2} \left(\frac{\partial}{\partial\bm\mu}\sum_{i\ne k(\bm\mu)}
                \frac{1}{\lambda_i}(\bm y_i - \bm\mu - \lambda_i\bm\gamma)'
                    \bm\Sigma^{-1} (\bm y_i - \bm\mu - \lambda_i\bm\gamma)\right).
\end{align}
The problem is that the summation index depends on $\bm\mu$,
so the differential and the summation cannot simply be interchanged.
Thus the CM-step for $\bm\mu$ does not have a closed form solution.

Alternatively, we can approximate the derivative by simply considering the summation index to be fixed.
At the $t$-th iteration, suppose we have
$\bm\mu^{(t)}$ as our current estimate for $\bm\mu$.
We can fix the summation index so that we leave out the data point closest to $\bm\mu^{(t)}$ instead of $\bm\mu$. This gives us an approximation to the derivative
\begin{align}
    \frac{\partial}{\partial\bm\mu}\ell ^{\rm LOO}_{N}
    &\approx - \frac{1}{2} \left(\frac{\partial}{\partial\bm\mu}\sum_{i\ne k(\bm\mu^{(t)})}
                \frac{1}{\lambda_i}(\bm y_i - \bm\mu - \lambda_i\bm\gamma)'
                    \bm\Sigma^{-1} (\bm y_i - \bm\mu - \lambda_i\bm\gamma)\right)\\
    &= \bm\Sigma^{-1} \sum_{i\ne k(\bm\mu^{(t)})}\frac{1}{\lambda_i}\left(\bm y_i - \bm\mu - \lambda_i \bm\gamma\right).
\end{align}

Similarly, applying the approximate derivative to
$\ell^{\rm LOO}_{N}$ and
$\ell^{\rm LOO}_{G}$
with respect to other parameters and solving the approximate derivatives at zero gives us the following CM-steps.

\subsection{CM-step}
\label{SubSection: CM-step}

\noindent {\bf CM-step for $\bm\mu, \bm\Sigma, \bm\gamma$:}

Suppose that the current iterate is $\bm\theta^{(t)}$ and
    $\bm\lambda$ is given.
After equating each component of the approximate partial derivatives of $\ell^{\rm LOO}_{N}(\bm\mu, \bm\Sigma, \bm\gamma|\nu, \bm y , \bm\lambda)$ to zero, we obtain the following estimates:
\begin{align}
    \bm\mu^{(t+1)} \label{CM-step mu}
        =& \ \frac{S_{\bm y/\lambda} S_{\lambda} - (n-1) S_{\bm y}}
                {S_{1/\lambda} S_{\lambda} - (n-1)^2} , \\
    \bm\gamma^{(t+1)} \label{CM-step gamma}
        =& \ \frac{S_{\bm y} - (n-1)\bm\mu^{(t+1)}}{S_{\lambda}} , \\
    \bm\Sigma^{(t+1)} \label{CM-step Sigma}
        =& \ \frac{1}{n-1} \sum_{i\ne k(\bm\mu^{(t)})} \frac{1}{\lambda_i} (\bm y_i - \bm\mu^{(t+1)})(\bm y_i - \bm\mu^{(t+1)})'
            - \frac{1}{n-1} \bm\gamma^{(t+1)}\left(\bm\gamma^{(t+1)}\right)' S_{\lambda},
\end{align}
where the complete data sufficient statistics are:
\begin{align} \label{suffstats}
    S_{\bm y} = \hspace{-0.2cm}
        \sum_{i\ne k(\bm\mu^{(t)})} \bm y_i , \quad
    S_{\bm y/\lambda} = \hspace{-0.2cm}
        \sum_{i\ne k(\bm\mu^{(t)})} \frac{1}{\lambda_i} \bm y_i , \quad
    S_{\lambda} = \hspace{-0.2cm}
        \sum_{i\ne k(\bm\mu^{(t)})} \lambda_i , \quad
    S_{1/\lambda}  = \hspace{-0.2cm}
        \sum_{i\ne k(\bm\mu^{(t)})} \frac{1}{\lambda_i}.
\end{align}

But these estimates won't guarantee the monotonic convergence of the LOO log-likelihood,
since we used the approximate derivatives.
However, we can apply a line search to guarantee the monotonic convergence of the ECM algorithm.
See Section \ref{SubSection: Line Search} for more details about the lines search.

\noindent {\bf CM-step for $\nu$:} \\
Given the mixing parameters $\bm\lambda$,
the estimate $\nu^{(t+1)}$ can be obtained by numerically maximising $\ell^{\rm LOO}_{G}(\nu|\bm\lambda)$ in \eqref{Gamma log-likelihood} with respect to $\nu$
using Newton-Raphson (NR) algorithm where the approximate derivatives is given by:
\begin{align}
    \frac{\partial}{\partial\nu} \ell^{\rm LOO}_{G} &=
        \ (n-1)\left(1 + \log\nu - \psi(\nu) \right)
            + S_{\log \lambda} - S_\lambda,\\
    \frac{\partial^2}{\partial\nu^2} \ell^{\rm LOO}_{G} &=
        \ (n-1)\left(\frac{1}{\nu} - \psi'(\nu)\right)
\end{align}
where $\psi(x)=\frac{d}{dx} \log \Gamma(x)$ is the digamma function and
\begin{equation}    \label{Sloglam; suffstats}
    S_{\log\lambda}  = \hspace{-0.2cm}
        \sum_{i\ne k(\bm\mu^{(t)})} \log\lambda_i.
\end{equation}

\subsection{Local point search}
\label{SubSection: Local Point Search}

Even when the LOO likelihood smooths out the unbounded points from the full likelihood, there still exist cusps in the LOO likelihood.
So we cannot completely rely on derivative based methods to find the global maximum of LOO likelihood with respect to the location parameter.
Nevertheless, these cusp in the LOO likelihood typically occur between data points as seen in Figure \ref{Plot_LOO_likelihood_closer_4x4}.
So for simplicity, we search for data points around the current iterate $\hat{\bm\mu}^{(t)}$ and
choose the one that increases the LOO likelihood.

\noindent {\bf Local point search algorithm:}
Let $(\bm\mu^{(t)}, \bm\Sigma^{(t)}, \bm\gamma^{(t)}, \nu^{(t)})$ be our current location estimates:

(i)\quad Calculate the Mahalanobis distance between $\bm y_i$ and $\hat{\bm\mu}^{(t)}$
\begin{align}
    (\bm y_i - \hat{\bm\mu}^{(t)})^T (\bm\Sigma^{(t)})^{-1} (\bm y_i - \hat{\bm\mu}^{(t)})
\end{align}
and choose the least $m$ with corresponding data points $\bm y_{i_1} , ..., \bm y_{i_{m}}$.
We choose $m=20$ for our simulation study.
Additionally, let $\bm y_{i_0}=\hat{\bm\mu}^{(t)}$ for notational convenience.

(ii)\quad
Update the location estimate by choosing $\bm\mu$ out of
$\{\bm y_{i_0},\text{...},\bm y_{i_m}\}$
such that it maximises the LOO log-likelihood
\begin{align}
    \underset{\bm\mu\in \{\bm y_{i_0},\text{...},\bm y_{i_m}\}}{\text{argmax}}
        \ell^{\rm LOO}\left(\bm\mu, \bm\Sigma^{(t)}, \bm\gamma^{(t)}, \nu^{(t)};\bm y\right).
\end{align}

\subsection{Line search}
\label{SubSection: Line Search}

Using the approximate derivatives for the CM-steps does not necessarily increase the LOO log-likelihood. So we need to implement a line search to guarantee the monotonic convergence of the ECM algorithm after each CM-step.
Here we abuse the notation by representing
$\bm\theta^{(t)}$ as the current estimate and
$\bm\theta^{(t+1)}$ as the updated estimate after the CM-step in Section \ref{SubSection: CM-step}.

Let us construct the line search by defining
\begin{align}
    \bm\theta^* = \bm\theta^{(t)} +
        \alpha \left(\bm\theta^{(t+1)}-\bm\theta^{(t)}\right)
\end{align}
where
$\alpha\in I\subset \mathbb R$ and
the interval $I$ is chosen so that
    $\bm\theta^{*}\in\bm\Theta$.
    For simplicity, we consider the interval $I=[0,1]$.

Using the \texttt{optimise} function in \texttt{R}, find $\alpha$ such that it maximises the LOO log-likelihood
\begin{align}
    \alpha^*
        = \underset{\alpha\in I}{\text{argmax}}
            \ell^{\rm LOO}(\bm\theta^{*}).
\end{align}
Although finding the maximum of a non-smooth likelihood function is difficult, so alternatively we can choose $\alpha^*$ such that
\begin{align}
    \ell^{\rm LOO}(\bm\theta^{*};\bm y)
        \geq \ell^{\rm LOO}(\bm\theta^{(t)};\bm y).
\end{align}

\subsection{ECM algorithm}
\label{Subsection: ECM Algorithm}

Combining the steps we introduced earlier gives us the ECM algorithm for MSVG distribution using the LOO likelihood:

\noindent{\bf Initialisation step:}
Choose suitable starting values $(\bm\mu_0, \bm\Sigma_0, \bm\gamma_0, \nu_0)$ .
It is recommended to choose starting values
$(\bar {\bm y}, \text{cov}(\bm y),\bm 0, 4d)$
where $\bar {\bm y}$ and $\text{cov}(\bm y)$ denote the sample mean and sample variance-covariance matrix of $\bm y$ respectively.
For more leptokurtic data, it is recommended to use more robust measure of location and scale.

\noindent{\bf ECM algorithm for MSVG:}
At the $t$-th
iteration with current estimates
$(\bm\mu^{(t)}, \bm\Sigma^{(t)}, \bm\gamma^{(t)}, \nu^{(t)})$: \par
{\bf Local Point Search:} Update the estimate to $\bm\mu^{(t+1/2)}$
using local point search in Section \ref{SubSection: Local Point Search}.

{\bf E-step 1:}
    Calculate $\widehat{\lambda}_i^{(t+1/3)}$ and $\widehat{1/\lambda_i}^{(t+1/3)}$ for $i=1,...,n$
    in \eqref{E-step lambda1} and \eqref{E-step lambda2} respectively using
            $(\bm\mu^{(t+1/2)}, \bm\Sigma^{(t)}, \bm\gamma^{(t)}, \nu^{(t)})$.
    Calculate also the sufficient statistics
        $S_{\bm y/\lambda}^{(t+1/3)}$,
        $S_{\lambda}^{(t+1/3)}$ and
        $S_{1/\lambda}^{(t+1/3)}$ in \eqref{suffstats}.

{\bf CM-step 1:} Update the estimates to
        $(\bm\mu^{(t+1)}, \bm\gamma^{(t+1)})$
    in \eqref{CM-step mu} and \eqref{CM-step gamma} respectively using the sufficient statistics in E-step 1.

{\bf E-step 2:}
    Same as E-step 1, calculate
    $\widehat{\lambda}_i^{(t+2/3)}$ and $\widehat{1/\lambda_i}^{(t+2/3)}$ for $i=1,...,n$ , and sufficient statistics
        $S_{\bm y/\lambda}^{(t+2/3)}$,
        $S_{\lambda}^{(t+2/3)}$ and
        $S_{1/\lambda}^{(t+2/3)}$ in \eqref{suffstats}.

{\bf CM-step 2:} Update the estimate to
        $\bm\Sigma^{(t+1)}$
    in \eqref{CM-step Sigma} using the sufficient statistics in E-step 2.

{\bf E-step 3:}
    Calculate $\widehat{\lambda_i}^{(t+1)}$ and $\widehat{\log\lambda_i}^{(t+1)}$ for $i=1,...,n$
    in \eqref{E-step lambda1} and \eqref{E-step lambda3} respectively using the updated estimates
        $(\bm\mu^{(t+1)}, \bm\Sigma^{(t+1)}, \bm\gamma^{(t+1)}, \nu^{(t)})$.
    Calculate also the sufficient statistics
        $S_{\lambda}^{(t+1)}$ and $S_{\log\lambda}^{(t+1)}$
    in \eqref{suffstats} and \eqref{Sloglam; suffstats}.

{\bf CM-step 3:} Update the estimate to $\nu^{(t+1)}$
    using the NR algorithm in Section \ref{SubSection: CM-step}.

\noindent{\bf Stopping rule:} Repeat the procedures until the relative increment of LOO log-likelihood function is smaller than tolerance level $10^{-8}$.

After each CM-step,
we apply the line search in Section \ref{SubSection: Line Search} to ensure
the local convergence of the ECM algorithm. The local point search ensures the global convergence of the ECM algorithm.

We will use this algorithm for studying
the optimal rate of convergence and
the asymptotic distributions of $\hat{\mu}_n$
in the Section \ref{Section: Simulation study of asymptotic distribution}.

\subsection{Convergence of ECM algorithm}
Just like with EM algorithm for the full likelihood in \citet{Dempster1977},
we also have monotonic convergence for the ECM algorithm using LOO likelihood. To see this,
consider the two fundamental facts for EM algorithm for LOO likelihood
\begin{align}
    \ell^{\rm LOO}(\bm\theta;\bm y)
        &= Q^{\rm LOO}(\bm\theta;\bm\theta^{(t)})
            - H^{\rm LOO}(\bm\theta;\bm\theta^{(t)})
\end{align}
and
\begin{align}
    H^{\rm LOO}(\bm\theta;\bm\theta^{(t)})
        &\leq H^{\rm LOO}(\bm\theta^{(t)};\bm\theta^{(t)})
\end{align}
where we let
\begin{align}
    Q^{\rm LOO}(\bm\theta;\bm\theta^{(t)})
        =& \int\ell^{\rm LOO}(\bm\theta;\bm y,\bm\lambda)
            f(\bm\lambda|\bm y;\bm\theta^{(t)}) \, {\rm d}\bm\lambda
\end{align}
with
$f(\bm\lambda|\bm y;\bm\theta^{(t)})
    =\prod_{i=1}^{n}f(\lambda_i|\bm y_i;\bm\theta^{(t)})$ , and
\begin{align}
    H^{\rm LOO}(\bm\theta;\bm\theta^{(t)})
        =& \int \ell^{\rm LOO}(\bm\theta;\bm\lambda|\bm y)
            f(\bm\lambda|\bm y;\bm\theta^{(t)}) \, {\rm d}\bm\lambda
\end{align}
with
$\ell^{\rm LOO}(\bm\theta;\bm\lambda|\bm y)
    =\sum_{i\ne k(\bm\mu)} \log f(\lambda_i|\bm y_i;\bm\theta)$.

The idea of the proof for the two fundamental facts are exactly the same as in \citet{Wu1983}.
Just simply interchange the full likelihood with the LOO likelihood.

Using these fundamental facts will guarantee the monotonic convergence of the EM algorithm for LOO likelihood.
In fact monotonic convergence still holds for generalised EM (GEM) algorithm where instead we define $\bm\theta^{(t+1)}$ to be the parameter update such that
\begin{align}
    \ell^{\rm LOO}(\bm\theta^{(t+1)};\bm y)
        \geq \ell^{\rm LOO}(\bm\theta^{(t)};\bm y).
\end{align}
Moreover, similar to the ECM algorithm in \citet{Meng1993},
we can deduce by induction that ECM is a GEM for LOO likelihood. So all the convergence properties in GEM is retained in the ECM algorithm.

\section{Simulation study of asymptotic distribution}
\label{Section: Simulation study of asymptotic distribution}

\citet{Podgorski2015} have proved the consistency and super-efficiency of the location estimator using the maximum LOO likelihood. The aim of this section is to determine whether the optimal rates in the main theorem is consistent with simulations, and analyse the asymptotic distribution of the location estimator.

We present the set-up of the simulation below:
\begin{enumerate}
    \item   Set the true shape parameters $\nu$ to be one of the 50 shape parameters \\
        $\{0.02, 0.04 \, ... , 0.98, 1\}$.
    \item  For each shape parameter, set the sample size $n$ to be one of the 20 sample sizes \\
        $\{500, 1000, ..., 9500, 10000 \}$.
    \item   For each pair of $(\nu,n)$, generate 20000 different sets of samples, each set from standardised univariate symmetric VG distribution with
    shape parameter $\nu$ and
    sample size $n$.
    \item  For each set of samples, estimate $\hat{\mu}_n$ using
    steps in the ECM algorithm in Section \ref{Subsection: ECM Algorithm}
    which only involve the location parameter.
    That is, we use the following steps in the ECM algorithm: local point search, E-step 1, and CM-step 1 with the line search
    where the other parameters $(\sigma^2,\gamma,\nu)$ are fixed.
\end{enumerate}
This gives us 20000 $\hat{\mu}_n$'s for each pair of $(\nu,n)$.

\subsection{Optimal rate}
Since the scale of asymptotic distribution of $\hat{\mu}_n$ increases under a power law with respect to $n$,
we fit a power curve to estimate the optimal rate $\beta$.
We choose the interquartile range (IQR) as a robust measure of spread.

Each pair of $(\nu,n)$ have 20000 $\hat{\mu}_n$'s.
So first fix $\nu$, then take the IQR of the 20000 $\hat{\mu}_n$'s for each $n$.
We want to fit a power curve to $n$ vs ${\rm IQR}$,
or in other words, find parameters $a$ and $b$ such that
${\rm IQR}=a n^b$.
This is equivalent to fitting a simple linear regression model to $\log n$ vs. $\log {\rm IQR}$.
That is, we want to find parameters $(\widehat{\log a}, \hat{b})$ to fit the linear model
\begin{align}
    \log {\rm IQR}=\log a +b \log n.
\end{align}

After obtaining estimates $(\widehat{\log a}, \hat{b})$, letting
$\hat{\beta} = -\hat{b}$
gives us our estimate for the optimal rate for a given $\nu$.
We repeat this process for other $\nu$'s.

\begin{figure}[htbp]
    \begin{center}
         \subfigure[log of optimal rate vs. log of $\nu$]
         {\label{}\includegraphics[width=0.48\textwidth]
            {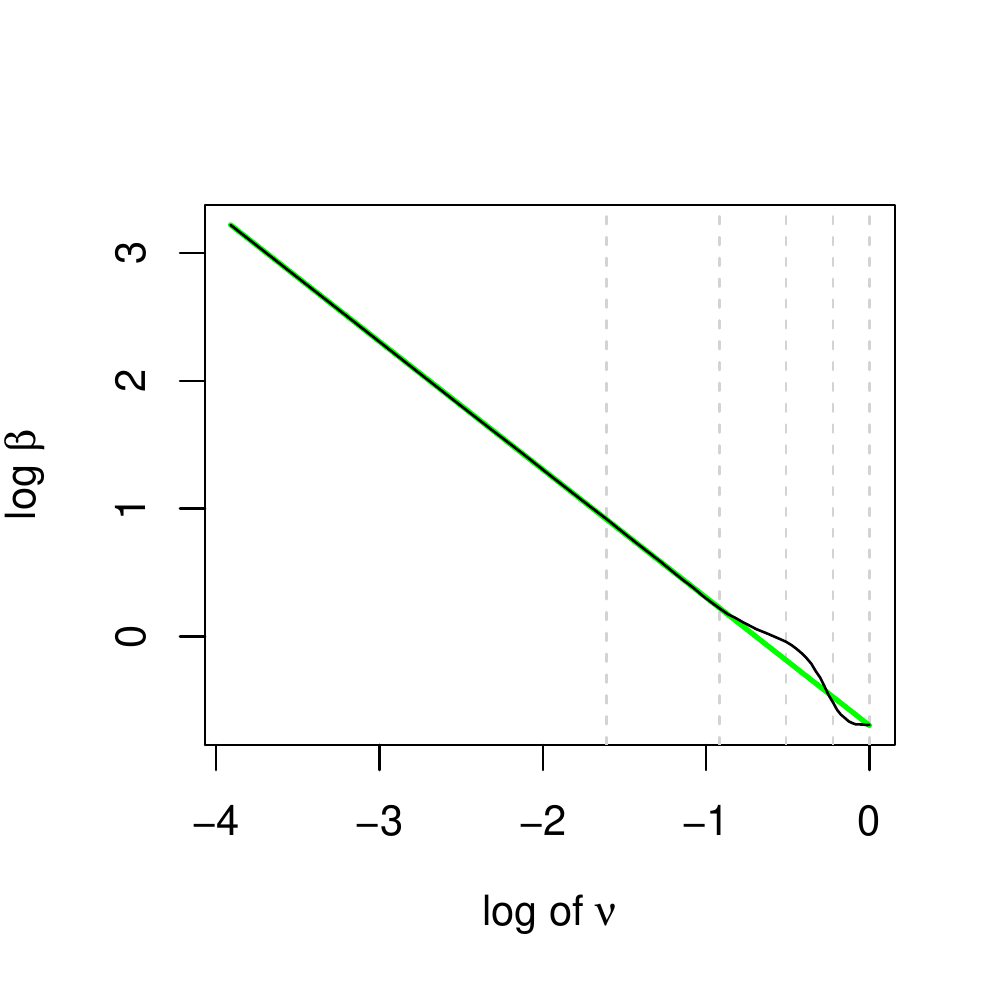}}
         \subfigure[relative error of optimal rate vs. $\nu$]
         {\label{}\includegraphics[width=0.48\textwidth]
            {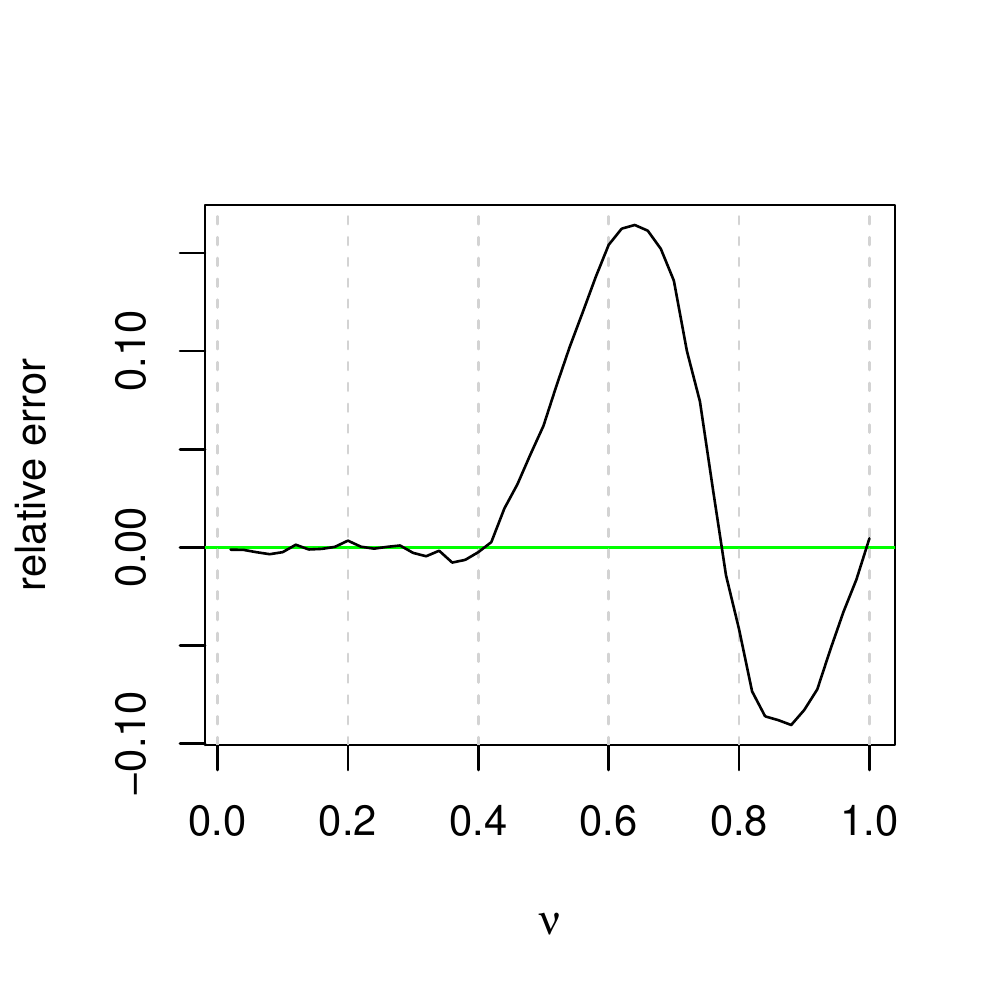}}
    \end{center}
    \caption{Plotting the estimated optimal rate against $\nu$.
        The proposed optimal rate $\beta=\frac{1}{2\nu}$ (green) is also included for comparison.
        The grey broken lines represents grid lines for
        $\nu$=\{0, 0.2, 0.4, 0.6, 0.8, 1\}.
        The relative error is also calculated and plotted using the formula
        $\frac{\hat\beta-\beta}{\beta}$.
        }
    \label{Plot_optimal_rate}
\end{figure}

In Figure \ref{Plot_optimal_rate}, the optimal rate estimate in the simulation appears to follow the proposed optimal rate when $0<\nu\leq0.4$.
However when $0.4<\nu<1$, the optimal rate estimate appears slightly different with a sinusoidal pattern.
In fact for $0.4<\nu\leq0.76$, optimal rate estimate appears to be greater than the proposed optimal rate.
As $\nu$ approaches to 1, the optimal rate estimate approaches the convergence rate for asymptotic normality.
Although for $0.76\leq\nu<1$, optimal rate estimate appears to be less than the proposed optimal rate which contradicts the main theorem.
The reason for this is yet to be known.
So to investigate this unusual behaviour further, we need to analyse the asymptotic distribution from the simulation study.

\subsection{Asymptotic distribution}

\begin{figure}[htbp]
    \begin{center}
            \includegraphics[width=0.48\textwidth]{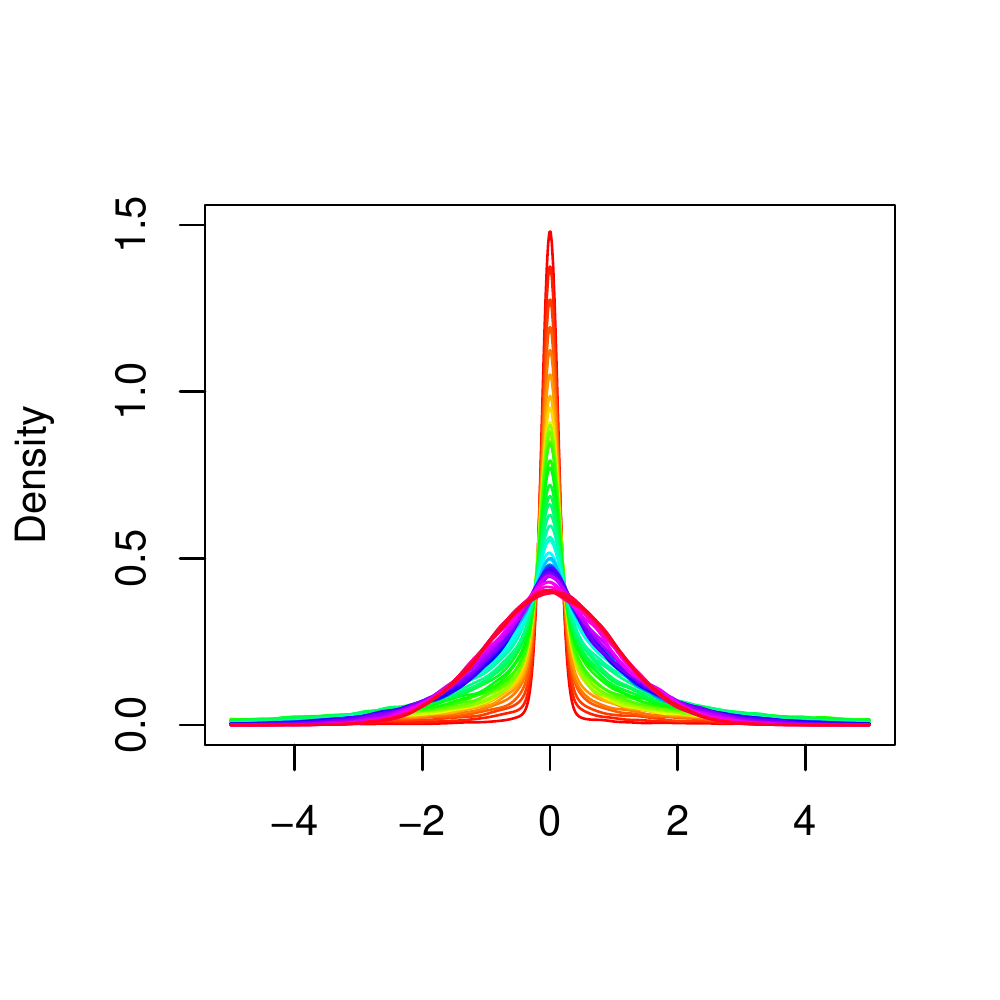}
    \end{center}
    \caption{Density plots of the simulated $n^{\hat\beta}\hat{\mu}_{n}$ with its scale standardised using IQR for each $\nu$ where $n=10000$.
        We use a rainbow colour scheme ranging from red ($\nu$=0.02) to magenta ($\nu$=1).
    }
    \label{Plot_DensityPlots}
\end{figure}

In Figure \ref{Plot_DensityPlots},
we plot the kernel density estimation of the simulated $n^{\hat\beta}\hat{\mu}_{n}$'s with
its scale standardised using IQR and $n=10000$ for each $\nu$ using a Gaussian kernel.
Notice that it exhibits heavier tails and sharper peaks at the expense of intermediate tails as $\nu$ decreases,
which has similar behaviour to the VG distribution.
We will test this claim by applying the ECM algorithm to fit the simulated $n^{\hat\beta}\hat{\mu}_n$ to the VG distribution for each pair $(\nu,n)$, then observe the Q-Q plots.

\begin{figure}[htbp]
    \begin{center}
         \subfigure[$\nu$ vs. estimated log$(\sigma_{\hat\mu})$ of $n^{\hat\beta}{\hat\mu}_n$]{
            \label{Plot_df_vs_logsigmuMLE_othernlist}
            \includegraphics[width=0.48\textwidth]
            {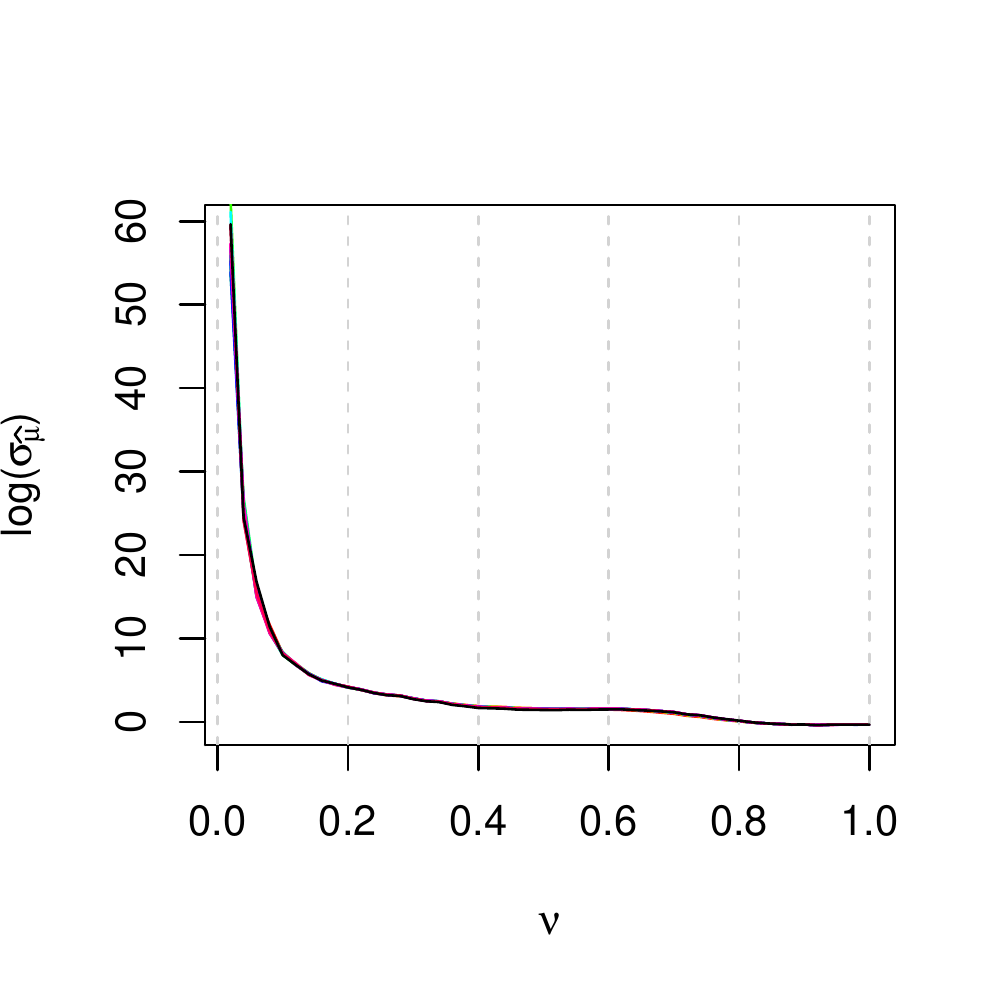}}
         \subfigure[$\nu$ vs. estimated log$(\nu_{\hat\mu})$ of $n^{\hat\beta}{\hat\mu}_n$]{
            \label{Plot_df_vs_logdfMLE_othernlist}
            \includegraphics[width=0.48\textwidth]
            {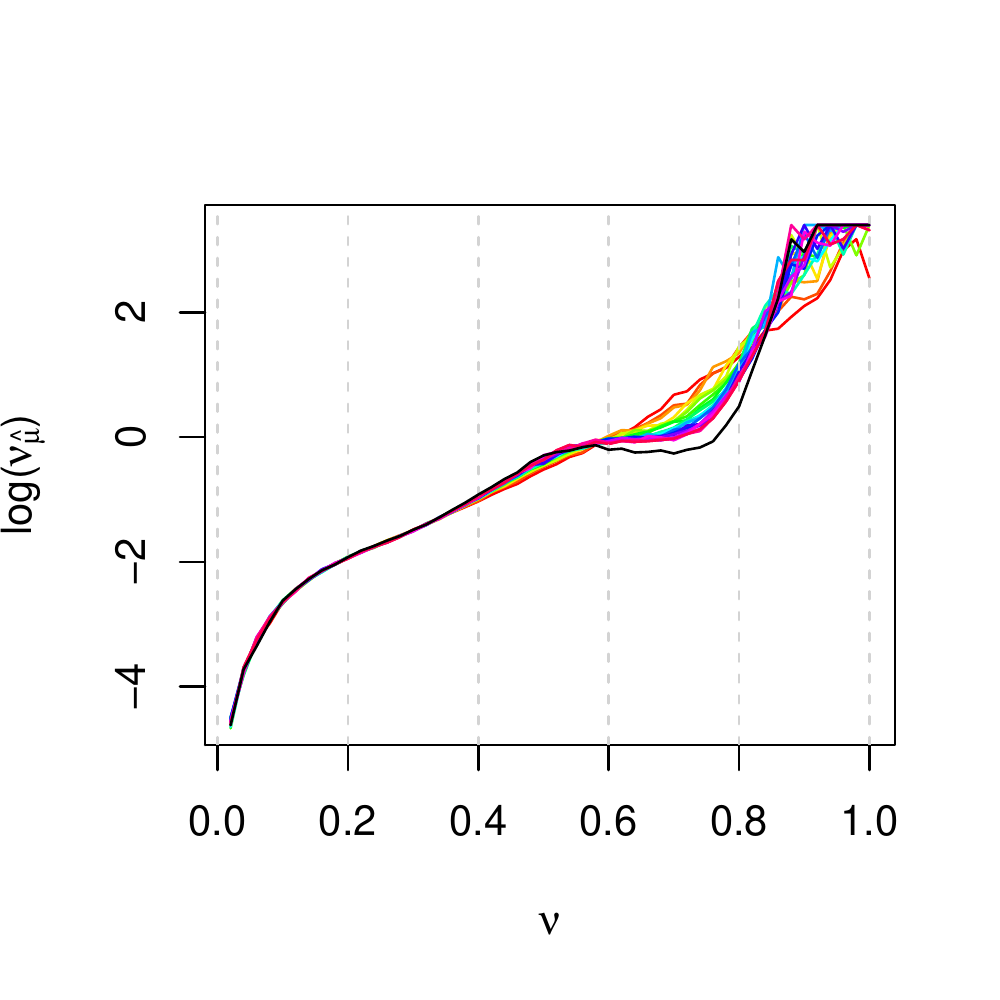}}
    \end{center}
    \caption{Plots of $\nu$ against estimates of VG distribution fitted to $n^{\hat\beta}{\hat\mu}_n$.
        We use a rainbow colour scheme ranging from red (n=500) to magenta (n=10000).
        Additionally, we have a black line to represent n=20000.
    }
    \label{Plot_df_vs_estVGmuMLE_othernlist}
\end{figure}

Let us denote the VG scale and shape parameter estimates of
$n^{\hat\beta}\hat{\mu}_n$ to be $(\sigma_{\hat\mu}, \nu_{\hat\mu})$.
For simplicity, we will set the location and skewness parameter to be 0 when applying the ECM algorithm to
reduce the number of parameters.
The Q-Q plots is generated empirically by plotting
the ordered monte carlo samples of size 20000 from the estimated VG distribution with scale and shape parameters $(\sigma_{\hat\mu}, \nu_{\hat\mu})$
against
the ordered simulated $n^{\hat\beta}{\hat\mu}_n$ for $n=10000$.
Note that we only plot for $n=10000$ as the other sample sizes exhibits similar distributional behaviour.
The plots and tables of the VG estimates of $n^{\hat\beta}\hat{\mu}_n$ is given in Figure \ref{Plot_df_vs_estVGmuMLE_othernlist}
and Table \ref{Table 1} and \ref{Table 2} respectively,
and the Q-Q plots is given in Figure \ref{QQ plots 1} and \ref{QQ plots 2}.

In Figure \ref{Plot_df_vs_logsigmuMLE_othernlist},
$\log\sigma_{\hat\mu}$ roughly follows a power law with respect to $\nu$, and
the scale estimate seems to be consistent for each $n$.
Whereas in Figure \ref{Plot_df_vs_logdfMLE_othernlist},
the shape estimate seems to be consistent for each $n$ only when $0<\nu<0.4$.
However when $0.4\leq\nu\leq1$, there seems to be considerable inconsistencies.
This suggest that the rate of convergence in distribution of $n^{\hat\beta}\mu_n$
is slower for larger $\nu$ compared with smaller $\nu$.
In terms of the trend of the plot,
$\log\nu_{\hat\mu}$ roughly follows an linear trend in the range $\nu>0.2$
for $n=500$,
but curves as $n$ increases.

For additional comparison, we also generated simulation results for $n=20000$ and
plotted the estimated VG parameters of the simulated $n^{\hat\beta}{\hat\mu}_n$ in Figure \ref{Plot_df_vs_estVGmuMLE_othernlist} represented using a black line.
As expected, $\sigma_{\hat\mu}$ is consistent with other sample sizes.
However, $\nu_{\hat\mu}$ curves even more for $\nu>0.4$.
So the slow convergence in distribution might be a possible reason why the estimated optimal rate differ with the proposed optimal rate for $0.4<\nu<1$.
In spite of that, analytically finding the optimal rate of convergence requires further research.

From Figure \ref{QQ plots 1} and \ref{QQ plots 2}, it appears that the VG distribution fits the asymptotic distributions reasonably well for $0.34\leq\nu\leq1$ since the Q-Q plots roughly follow a straight line.
As for $\nu\leq0.32$, the asymptotic distributions appears to have heavier tails and higher density at the centre
than VG distribution as $\nu$ decreases.
More studies are needed to determine which distribution can approximately fit the asymptotic distribution for
the whole range of $\nu$.
Nevertheless, we can construct confidence intervals and
approximate standard errors
for the location parameter of VG distribution
where we use
the estimated VG distribution for $\nu\geq0.34$, and
the simulated $n^{\hat\beta}\hat\mu_n$ samples for $\nu<0.34$.

\section{Conclusion}
\label{Section: Conclusion}

We have proposed an ECM algorithm to accurately estimate parameters from VG distribution while also dealing with the unbounded densities using the LOO likelihood.
The maximum LOO likelihood estimator exhibits consistency and super-efficiency proved by \citet{Podgorski2015}.
We provided simulation results to understand empirically
other asymptotic properties such as the optimal rate of convergence
and asymptotic distribution of the maximum LOO likelihood location estimator, however proving these asymptotic results analytically is still an open question.
Nevertheless, we can construct confidence intervals and approximate standard errors for location parameter of VG distribution using the results in Section \ref{Section: Simulation study of asymptotic distribution}.
Although we only implemented the univariate symmetric case in this paper, the algorithm works well for multivariate skewness case at the expense of additional computation time.

For further research, it is worth considering the asymptotic distribution with skewness and higher dimensions,
or more generally the joint asymptotic distribution and the dependence between the location and other parameters from MSVG distribution.
For more complicated models, finding numerical techniques for estimating the standard error and approximating the asymptotic distribution with the presence of unboundedness to capture strong leptokurtosis is important in real world applications.

The methodology presented in Section
\ref{Section: ECM algorithm for LOO likelihood}
can also be used for estimating parameters from other cusped, unbounded, or even distributions with extreme leptokurtosis such as stable distribution with small stable index, or
leptokurtic financial models for
high frequency data \citep{Kawai2015}.

\begin{appendices}

\addtocounter{section}{-1}
\section{Appendix} \label{Appendix}

\begin{table}[hbp]
	\caption{optimal rate estimates,
        proposed optimal rate, and
        $(\sigma_{\hat\mu},\nu_{\hat\mu})$ values for
        $0.02\leq\nu\leq0.5$ and
        for selected $n=\{500, 3000, 7000, 10000, 20000\}$.}
    \begin{center}
	\begin{tabular}{|c|c|c|c|c|}
		\hline
		$\nu$ & $\hat\beta$ & $\beta$ &
            $\sigma_{\hat\mu}$ & $\nu_{\hat\mu}$ \\ \hline
		0.02 & 24.95 & 25.00 &
            $10^{24}\begin{pmatrix} 1.4 & 1.2 & 0.3 & 72 & 81 \end{pmatrix}$ & $10^{-2}\begin{pmatrix} 1.1 & 1.1 & 1.1 & 1.0 & 1.0\end{pmatrix}$
 \\ \hline
		0.04 & 12.46 & 12.50 &
            $10^{10}\begin{pmatrix} 6.4 & 17 & 6.5 & 3.1 & 4.4 \end{pmatrix}$ & $10^{-2}\begin{pmatrix} 2.4 & 2.3 & 2.4 & 2.5 & 2.4\end{pmatrix}$
 \\ \hline
		0.06 & 8.32 & 8.33 &
            $10^{6}\begin{pmatrix} 5.1 & 8.1 & 8.6 & 5.3 & 19 \end{pmatrix}$ & $10^{-2}\begin{pmatrix} 3.9 & 3.7 & 3.7 & 3.9 & 3.5\end{pmatrix}$
 \\ \hline
		0.08 & 6.23 & 6.25 &
            $10^{4}\begin{pmatrix} 11 & 6.2 & 4.4 & 5.3 & 8.9 \end{pmatrix}$ & $10^{-2}\begin{pmatrix} 5.0 & 5.4 & 5.5 & 5.5 & 5.2\end{pmatrix}$
 \\ \hline
		0.1 & 4.97 & 5.00 &
            $10^{3}\begin{pmatrix} 3.7 & 3.4 & 3.8 & 3.6 & 3.0 \end{pmatrix}$ & $10^{-2}\begin{pmatrix} 7.1 & 7.2 & 7.1 & 7.1 & 7.3 \end{pmatrix}$
 \\ \hline
		0.12 & 4.17 & 4.17 &
            $10^{2}\begin{pmatrix} 9.0 & 11 & 10 & 10 & 9.3 \end{pmatrix}$ & $10^{-2}\begin{pmatrix} 8.8 & 8.5 & 8.6 & 8.7 & 8.8\end{pmatrix}$
 \\ \hline
		0.14 & 3.57 & 3.57 &
            $10^{2}\begin{pmatrix} 3.1 & 3.5 & 3.2 & 2.9 & 3.0 \end{pmatrix}$ & $\begin{pmatrix} 0.10 & 0.10 & 0.10 & 0.10 & 0.10\end{pmatrix}$
 \\ \hline
		0.16 & 3.12 & 3.13 &
            $10^{2}\begin{pmatrix} 1.5 & 1.6 & 1.3 & 1.5 & 1.4 \end{pmatrix}$ & $\begin{pmatrix} 0.12 & 0.12 & 0.12 & 0.12 & 0.12\end{pmatrix}$
 \\ \hline
		0.18 & 2.78 & 2.78 &
            $\begin{pmatrix} 85 & 92 & 94 & 94 & 96 \end{pmatrix}$ & $\begin{pmatrix} 0.13 & 0.13 & 0.13 & 0.13 & 0.13\end{pmatrix}$
 \\ \hline
		0.2 & 2.51 & 2.50 &
            $\begin{pmatrix} 62 & 63 & 66 & 68 & 63 \end{pmatrix}$ & $\begin{pmatrix} 0.15 & 0.15 & 0.14 & 0.14 & 0.15\end{pmatrix}$
 \\ \hline
		0.22 & 2.29 & 2.27 &
            $\begin{pmatrix} 47 & 46 & 47 & 46 & 47 \end{pmatrix}$ & $\begin{pmatrix} 0.16 & 0.16 & 0.16 & 0.16 & 0.16\end{pmatrix}$
 \\ \hline
		0.24 & 2.08 & 2.08 &
            $\begin{pmatrix} 31 & 32 & 32 & 33 & 32 \end{pmatrix}$
 & $\begin{pmatrix} 0.17 & 0.17 & 0.17 & 0.17 & 0.17\end{pmatrix}$
 \\ \hline
		0.26 & 1.92 & 1.92 &
            $\begin{pmatrix} 24 & 25 & 25 & 26 & 25 \end{pmatrix}$
 & $\begin{pmatrix} 0.19 & 0.19 & 0.19 & 0.18 & 0.19\end{pmatrix}$
 \\ \hline
		0.28 & 1.79 & 1.79 &
            $\begin{pmatrix} 22 & 22 & 23 & 24 & 22 \end{pmatrix}$
 & $\begin{pmatrix} 0.21 & 0.20 & 0.21 & 0.20 & 0.21\end{pmatrix}$
 \\ \hline
		0.3 & 1.65 & 1.67 &
            $\begin{pmatrix} 17 & 17 & 16 & 16 & 15 \end{pmatrix}$
 & $\begin{pmatrix} 0.22 & 0.22 & 0.22 & 0.23 & 0.23\end{pmatrix}$
 \\ \hline
		0.32 & 1.54 & 1.56 &
            $\begin{pmatrix} 13 & 13 & 12 & 12 & 12 \end{pmatrix}$
 & $\begin{pmatrix} 0.25 & 0.24 & 0.25 & 0.25 & 0.25\end{pmatrix}$
 \\ \hline
		0.34 & 1.45 & 1.47 &
            $\begin{pmatrix} 11 & 11 & 11 & 11 & 11 \end{pmatrix}$
 & $\begin{pmatrix} 0.27 & 0.27 & 0.27 & 0.28 & 0.28\end{pmatrix}$
 \\ \hline
		0.36 & 1.35 & 1.39 &
            $\begin{pmatrix} 8.9 & 8.6 & 8.6 & 8.3 & 7.6 \end{pmatrix}$
 & $\begin{pmatrix} 0.30 & 0.31 & 0.30 & 0.31 & 0.31\end{pmatrix}$
 \\ \hline
		0.38 & 1.28 & 1.32 &
            $\begin{pmatrix} 7.5 & 7.3 & 7.0 & 6.9 & 6.5 \end{pmatrix}$
 & $\begin{pmatrix} 0.33 & 0.34 & 0.34 & 0.34 & 0.35\end{pmatrix}$
 \\ \hline
		0.4 & 1.21 & 1.25 &
            $\begin{pmatrix} 6.5 & 6.3 & 6.0 & 5.8 & 5.3 \end{pmatrix}$
 & $\begin{pmatrix} 0.36 & 0.38 & 0.37 & 0.39 & 0.40\end{pmatrix}$
 \\ \hline
		0.42 & 1.16 & 1.19 &
            $\begin{pmatrix} 6.1 & 5.9 & 5.6 & 5.7 & 5.1 \end{pmatrix}$
 & $\begin{pmatrix} 0.40 & 0.42 & 0.43 & 0.43 & 0.45\end{pmatrix}$
 \\ \hline
		0.44 & 1.12 & 1.14 &
            $\begin{pmatrix} 5.9 & 5.6 & 5.5 & 5.4 & 4.8 \end{pmatrix}$
 & $\begin{pmatrix} 0.44 & 0.47 & 0.50 & 0.48 & 0.51\end{pmatrix}$
 \\ \hline
		0.46 & 1.07 & 1.09 &
            $\begin{pmatrix} 5.3 & 5.1 & 4.9 & 4.8 & 4.4 \end{pmatrix}$
 & $\begin{pmatrix} 0.47 & 0.53 & 0.54 & 0.56 & 0.57\end{pmatrix}$
 \\ \hline
		0.48 & 1.04 & 1.04 &
            $\begin{pmatrix} 5.1 & 5.0 & 4.8 & 4.7 & 4.2 \end{pmatrix}$
 & $\begin{pmatrix} 0.53 & 0.60 & 0.63 & 0.63 & 0.67\end{pmatrix}$
 \\ \hline
		0.5 & 1.01 & 1.00 & $\begin{pmatrix} 4.9 & 4.8 & 4.7 & 4.6 & 4.2 \end{pmatrix}$ & $\begin{pmatrix} 0.59 & 0.65 & 0.68 & 0.69 & 0.75\end{pmatrix}$
 \\ \hline
	\end{tabular}
    \end{center}	
    \label{Table 1}
\end{table}

\begin{table}[htp]
	\caption{optimal rate estimates,
        proposed optimal rate, and
        $(\sigma_{\hat\mu},\nu_{\hat\mu})$ values for
        $0.52\leq\nu\leq1$ and
        for selected $n=\{500, 3000, 7000, 10000, 20000\}$.}
    \begin{center}
	\begin{tabular}{|c|c|c|c|c|}
		\hline
		$\nu$ & $\hat\beta$ & $\beta$ &
            $\sigma_{\hat\mu}$ & $\nu_{\hat\mu}$ \\ \hline
        0.52 & 0.99 & 0.96 &
            $\begin{pmatrix} 4.7 & 4.8 & 4.7 & 4.5 & 4.1 \end{pmatrix}$
 & $\begin{pmatrix} 0.65 & 0.72 & 0.76 & 0.81 & 0.79\end{pmatrix}$
 \\ \hline
        0.54 & 0.97 & 0.93 &
            $\begin{pmatrix} 4.7 & 4.9 & 4.7 & 4.6 & 4.3 \end{pmatrix}$
 & $\begin{pmatrix} 0.73 & 0.79 & 0.81 & 0.88 & 0.81\end{pmatrix}$
 \\ \hline
        0.56 & 0.94 & 0.89 &
            $\begin{pmatrix} 4.5 & 4.8 & 4.6 & 4.6 & 4.2 \end{pmatrix}$
 & $\begin{pmatrix} 0.77 & 0.84 & 0.89 & 0.85 & 0.85\end{pmatrix}$
 \\ \hline
        0.58 & 0.92 & 0.86 &
            $\begin{pmatrix} 4.5 & 4.9 & 4.8 & 4.7 & 4.4 \end{pmatrix}$
 & $\begin{pmatrix} 0.88 & 0.91 & 0.93 & 0.93 & 0.88\end{pmatrix}$
 \\ \hline
        0.6 & 0.90 & 0.83 &
            $\begin{pmatrix} 4.4 & 4.8 & 4.8 & 4.8 & 4.4 \end{pmatrix}$
 & $\begin{pmatrix} 0.99 & 0.94 & 0.98 & 0.90 & 0.82\end{pmatrix}$
 \\ \hline
        0.62 & 0.88 & 0.81 &
            $\begin{pmatrix} 4.2 & 4.7 & 4.7 & 4.6 & 4.5 \end{pmatrix}$
 & $\begin{pmatrix} 1.0 & 0.98 & 0.98 & 0.94 & 0.83\end{pmatrix}$
 \\ \hline
        0.64 & 0.85 & 0.78 &
            $\begin{pmatrix} 3.8 & 4.3 & 4.3 & 4.4 & 4.1 \end{pmatrix}$
 & $\begin{pmatrix} 1.2 & 1.1 & 1.0 & 0.93 & 0.78\end{pmatrix}$
 \\ \hline
        0.66 & 0.83 & 0.76 &
            $\begin{pmatrix} 3.5 & 4.0 & 4.0 & 4.1 & 3.9 \end{pmatrix}$
 & $\begin{pmatrix} 1.4 & 1.1 & 0.99 & 0.94 & 0.79\end{pmatrix}$
 \\ \hline
        0.68 & 0.79 & 0.74 &
            $\begin{pmatrix} 3.0 & 3.5 & 3.6 & 3.6 & 3.5 \end{pmatrix}$
 & $\begin{pmatrix} 1.6 & 1.2 & 1.0 & 1.0 & 0.81\end{pmatrix}$
 \\ \hline
        0.7 & 0.76 & 0.71 &
            $\begin{pmatrix} 2.6 & 3.1 & 3.2 & 3.2 & 3.2 \end{pmatrix}$
 & $\begin{pmatrix} 2.0 & 1.3 & 1.1 & 1.0 & 0.77\end{pmatrix}$
 \\ \hline
        0.72 & 0.72 & 0.69 &
            $\begin{pmatrix} 2.1 & 2.4 & 2.4 & 2.4 & 2.5 \end{pmatrix}$
 & $\begin{pmatrix} 2.1 & 1.5 & 1.2 & 1.0 & 0.82\end{pmatrix}$
 \\ \hline
        0.74 & 0.69 & 0.68 &
            $\begin{pmatrix} 1.8 & 2.1 & 2.2 & 2.2 & 2.2 \end{pmatrix}$
 & $\begin{pmatrix} 2.5 & 1.9 & 1.2 & 1.1 & 0.85\end{pmatrix}$
 \\ \hline
        0.76 & 0.65 & 0.66 &
            $\begin{pmatrix} 1.4 & 1.6 & 1.7 & 1.7 & 1.7 \end{pmatrix}$
 & $\begin{pmatrix} 2.8 & 2.1 & 1.5 & 1.4 & 0.93\end{pmatrix}$
 \\ \hline
        0.78 & 0.61 & 0.64 &
            $\begin{pmatrix} 1.2 & 1.3 & 1.3 & 1.3 & 1.4 \end{pmatrix}$
 & $\begin{pmatrix} 3.0 & 2.5 & 1.8 & 1.8 & 1.2\end{pmatrix}$
 \\ \hline
        0.8 & 0.59 & 0.63 &
            $\begin{pmatrix} 1.1 & 1.1 & 1.1 & 1.1 & 1.2 \end{pmatrix}$
 & $\begin{pmatrix} 3.6 & 3.3 & 2.5 & 2.5 & 1.6\end{pmatrix}$
 \\ \hline
        0.82 & 0.56 & 0.61 &
            $\begin{pmatrix}0.92 & 0.93 & 0.94 & 0.95 & 0.95\end{pmatrix}$
 & $\begin{pmatrix} 4.2 & 4.6 & 3.5 & 3.7 & 2.9\end{pmatrix}$
 \\ \hline
        0.84 & 0.54 & 0.60 &
            $\begin{pmatrix}0.83 & 0.83 & 0.84 & 0.84 & 0.84\end{pmatrix}$
 & $\begin{pmatrix} 5.5 & 6.9 & 5.6 & 5.4 & 5.1\end{pmatrix}$
 \\ \hline
        0.86 & 0.53 & 0.58 &
            $\begin{pmatrix}0.80 & 0.78 & 0.78 & 0.78 & 0.78\end{pmatrix}$
 & $\begin{pmatrix} 5.7 & 8.4 & 7.4 & 12 & 9.3\end{pmatrix}$
 \\ \hline
        0.88 & 0.51 & 0.57 &
            $\begin{pmatrix}0.74 & 0.72 & 0.72 & 0.72 & 0.72\end{pmatrix}$
 & $\begin{pmatrix} 6.9 & 12 & 16 & 17 & 24\end{pmatrix}$
 \\ \hline
        0.9 & 0.51 & 0.56 &
            $\begin{pmatrix}0.74 & 0.73 & 0.73 & 0.73 & 0.73\end{pmatrix}$
 & $\begin{pmatrix} 8.2 & 13 & 15 & 17 & 19\end{pmatrix}$
 \\ \hline
        0.92 & 0.50 & 0.54 &
            $\begin{pmatrix}0.71 & 0.69 & 0.69 & 0.69 & 0.67\end{pmatrix}$
 & $\begin{pmatrix} 9.3 & 30 & 25 & 30 & 30\end{pmatrix}$
 \\ \hline
        0.94 & 0.51 & 0.53 &
            $\begin{pmatrix}0.73 & 0.72 & 0.72 & 0.71 & 0.72\end{pmatrix}$
 & $\begin{pmatrix} 12 & 30 & 30 & 22 & 30\end{pmatrix}$
 \\ \hline
        0.96 & 0.50 & 0.52 &
            $\begin{pmatrix}0.73 & 0.72 & 0.72 & 0.73 & 0.72\end{pmatrix}$
 & $\begin{pmatrix} 20 & 30 & 30 & 24 & 30\end{pmatrix}$
 \\ \hline
        0.98 & 0.50 & 0.51 &
            $\begin{pmatrix}0.73 & 0.72 & 0.71 & 0.72 & 0.72\end{pmatrix}$
 & $\begin{pmatrix} 24 & 18 & 30 & 30 & 30\end{pmatrix}$
 \\ \hline
        1 & 0.50 & 0.50 &
            $\begin{pmatrix}0.74 & 0.73 & 0.73 & 0.73 & 0.72\end{pmatrix}$
 & $\begin{pmatrix} 13 & 30 & 30 & 28 & 30\end{pmatrix}$
 \\ \hline
	\end{tabular}
    \end{center}
	\label{Table 2}
\end{table}

\begin{figure}[htbp]
    \begin{center}
         \subfigure{\includegraphics[width=0.195\textwidth]
            {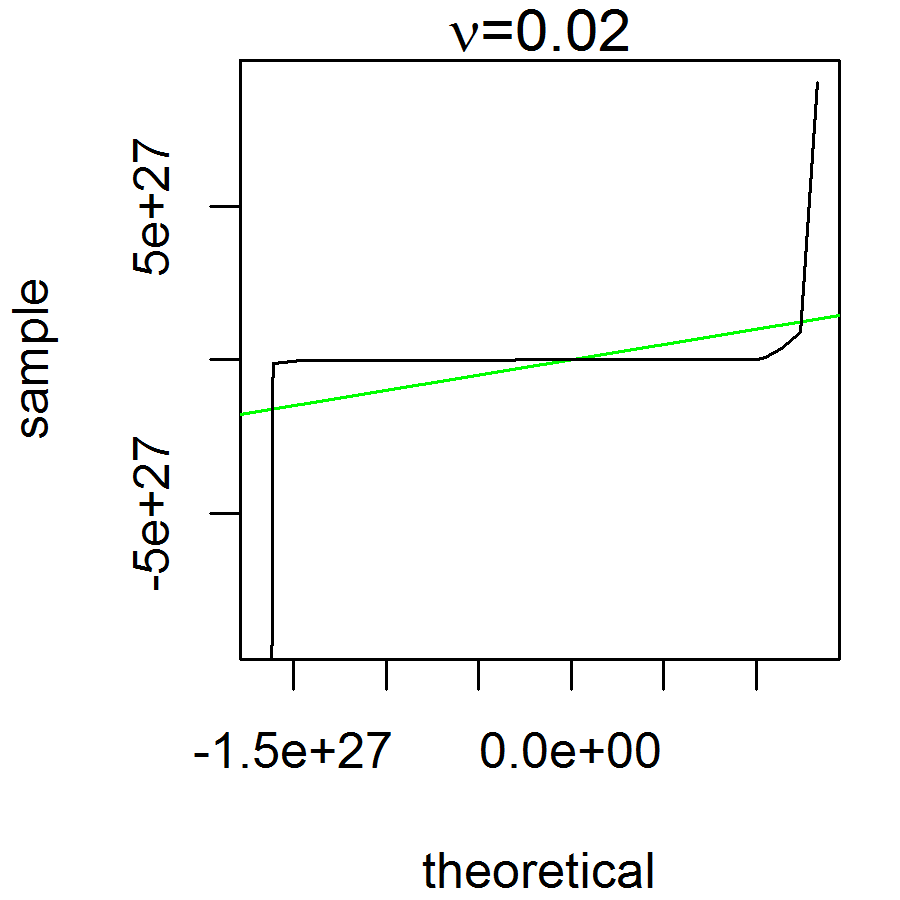}}
         \subfigure{\includegraphics[width=0.195\textwidth]
            {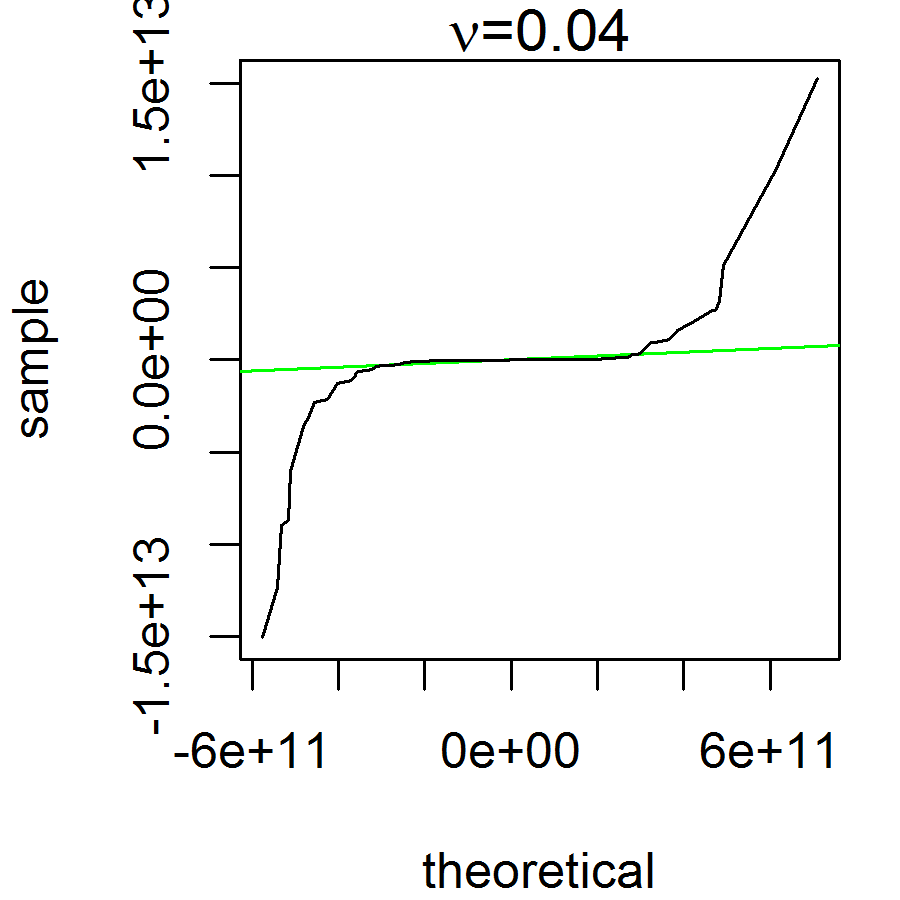}}
         \subfigure{\includegraphics[width=0.195\textwidth]
            {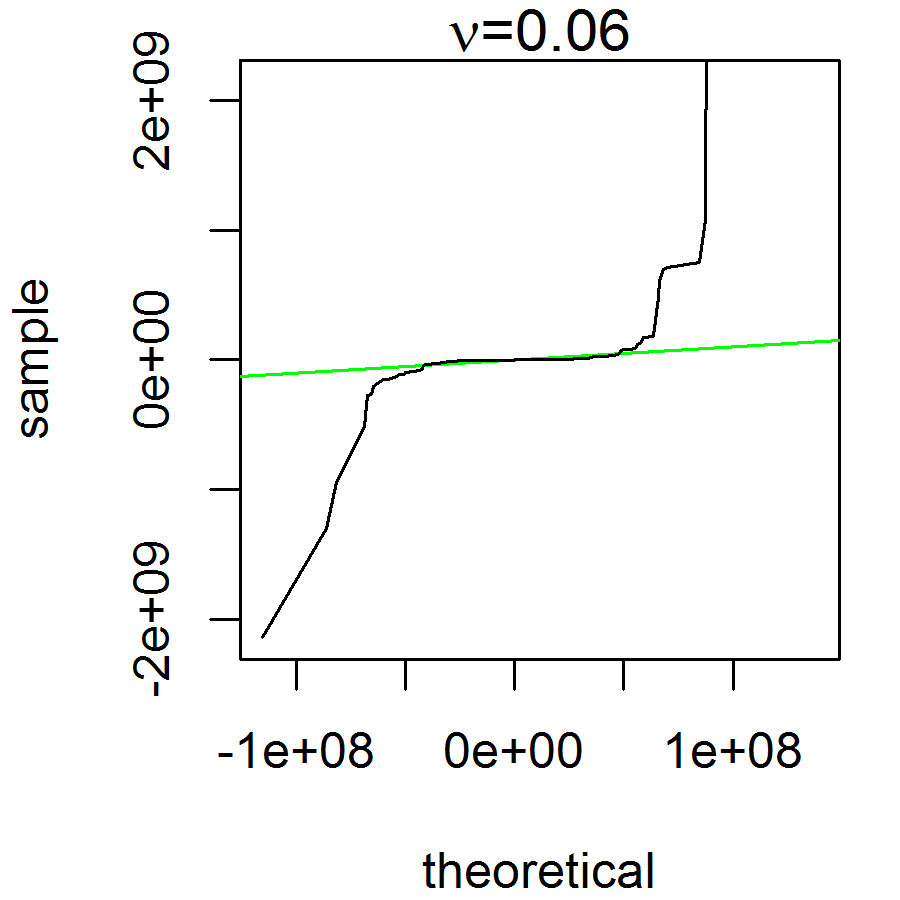}}
         \subfigure{\includegraphics[width=0.195\textwidth]
            {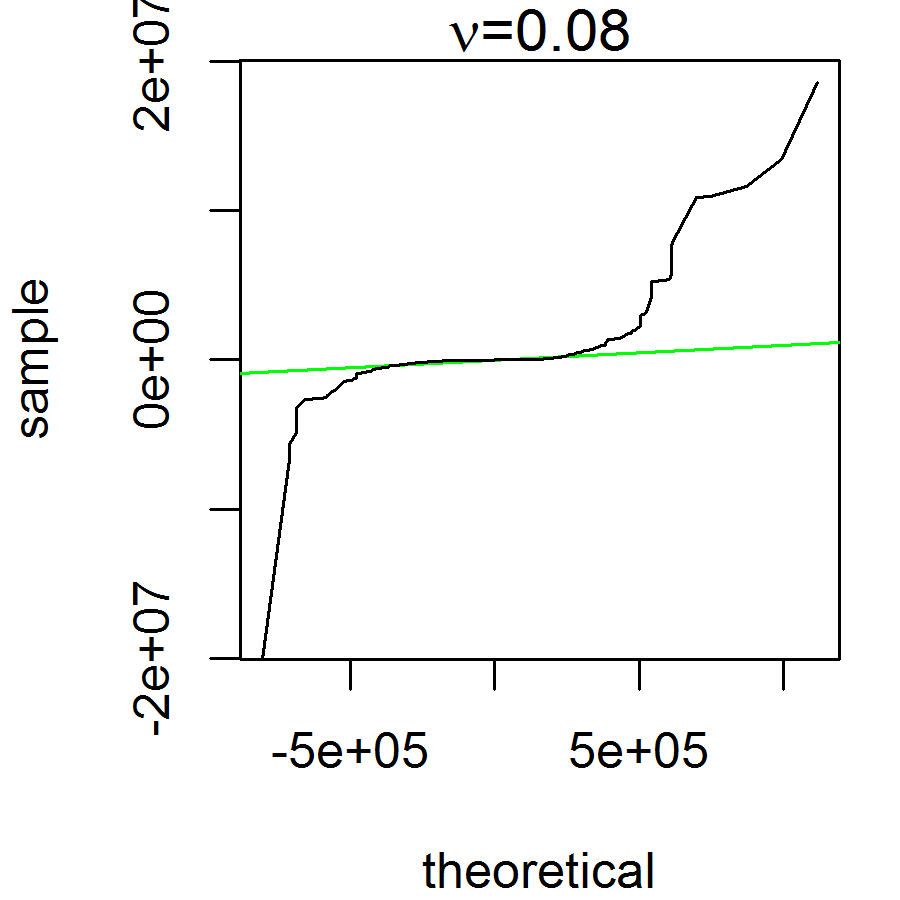}}
        \subfigure{\includegraphics[width=0.195\textwidth]
            {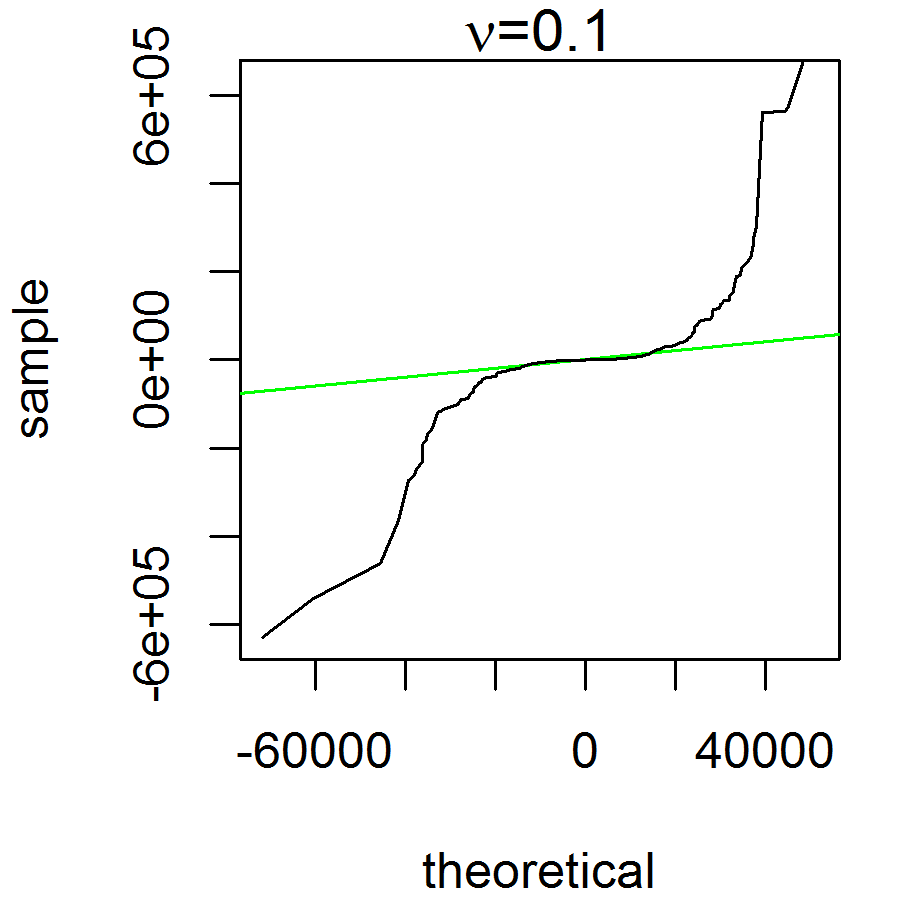}}

        \subfigure{\includegraphics[width=0.195\textwidth]
            {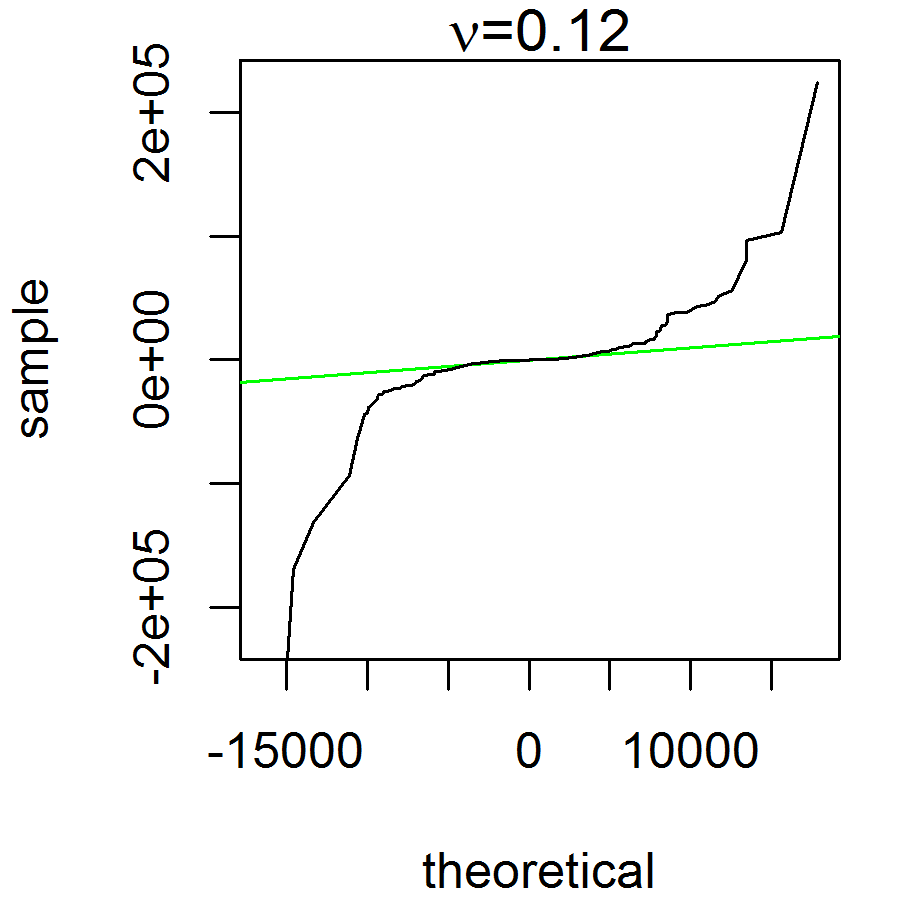}}
         \subfigure{\includegraphics[width=0.195\textwidth]
            {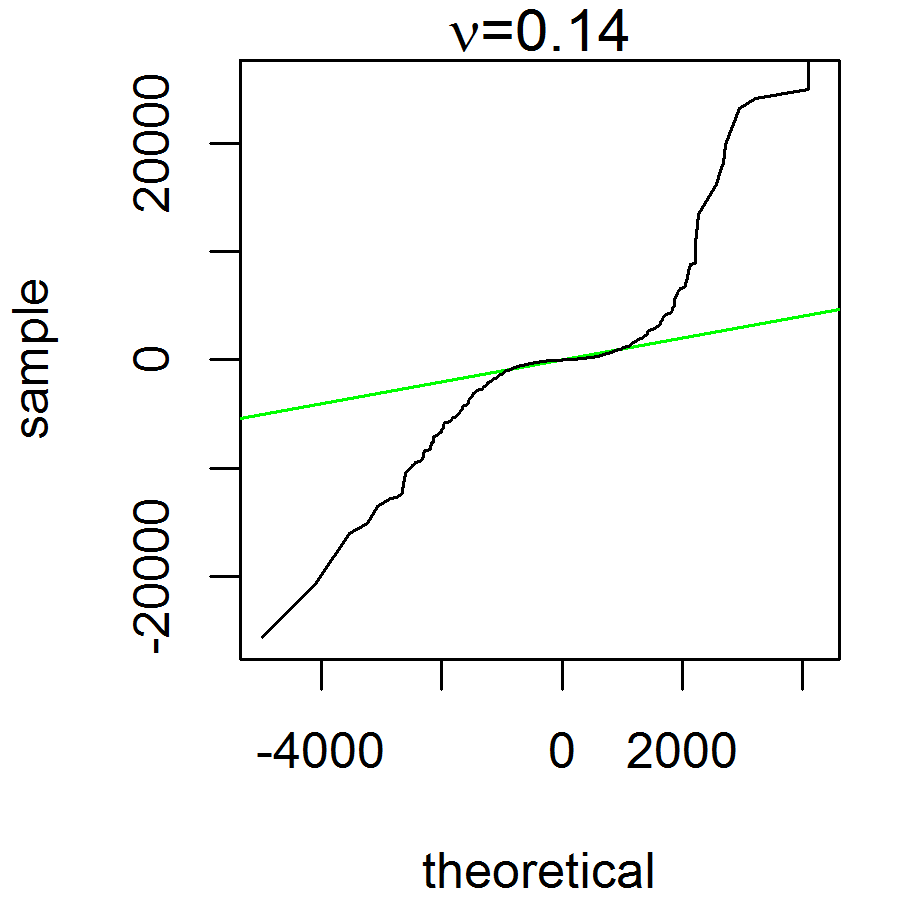}}
         \subfigure{\includegraphics[width=0.195\textwidth]
            {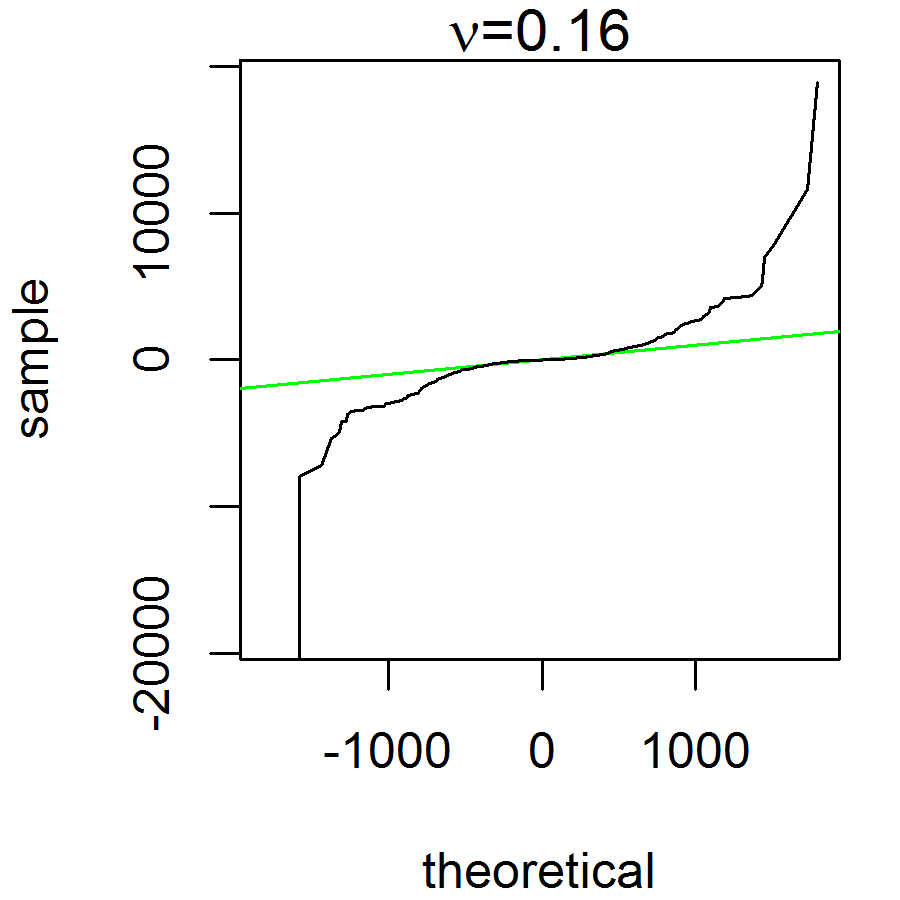}}
         \subfigure{\includegraphics[width=0.195\textwidth]
            {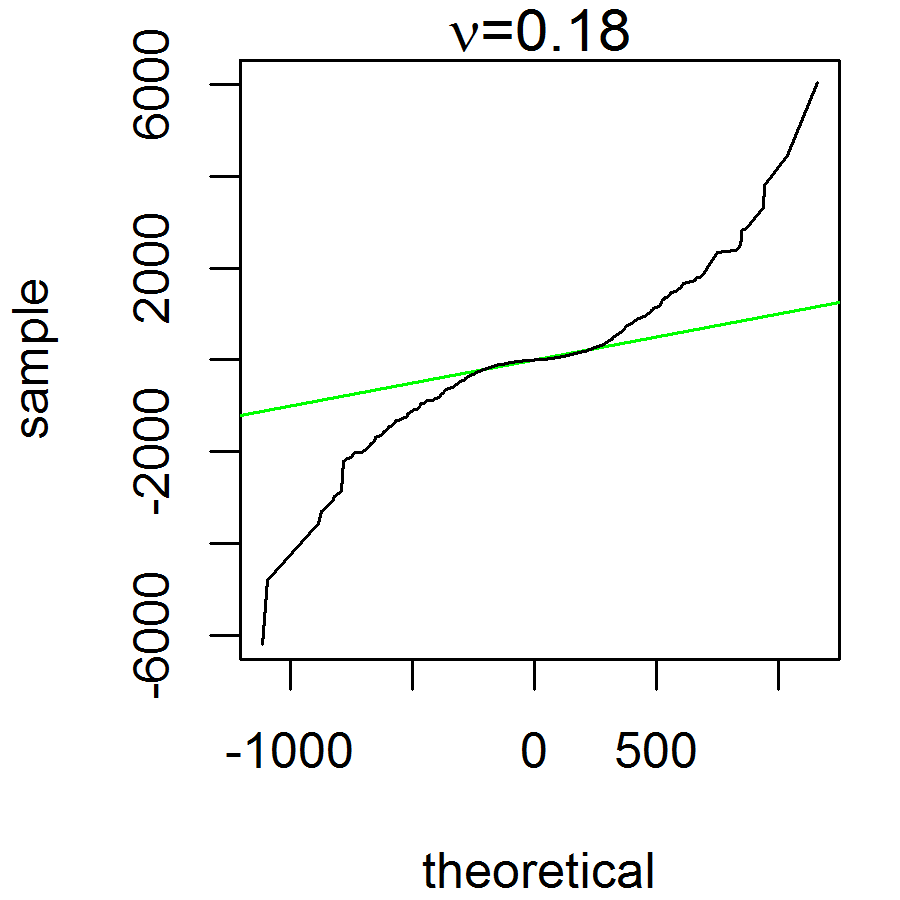}}
        \subfigure{\includegraphics[width=0.195\textwidth]
            {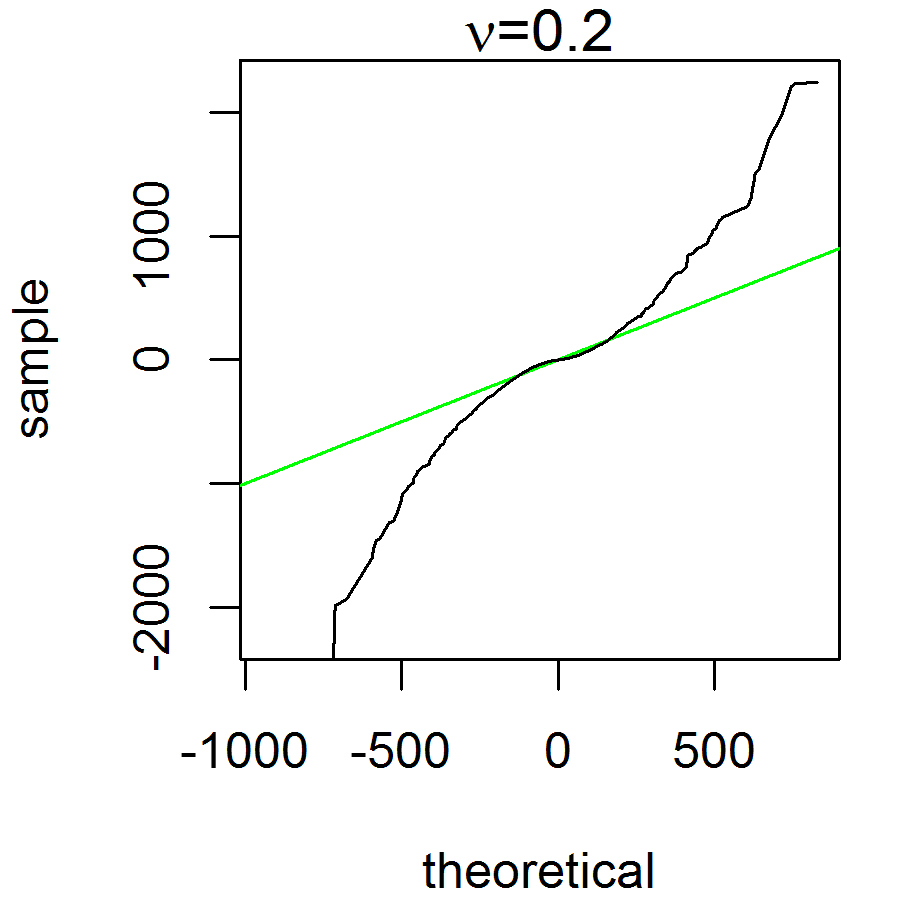}}

        \subfigure{\includegraphics[width=0.195\textwidth]
            {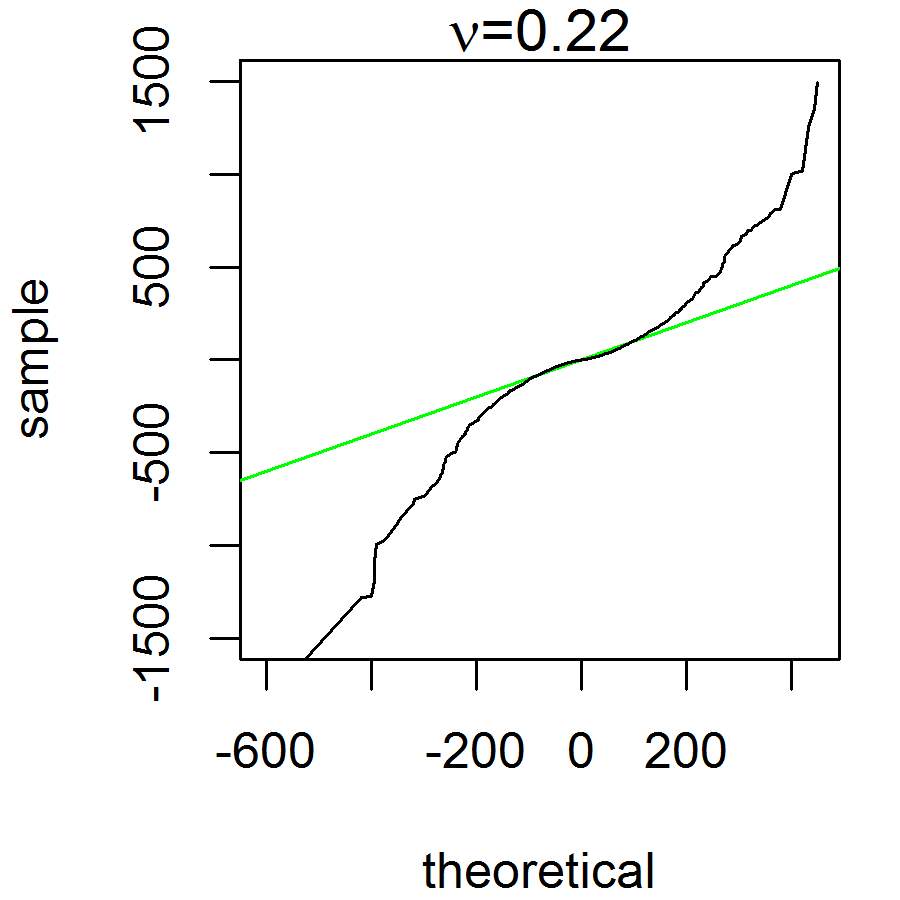}}
         \subfigure{\includegraphics[width=0.195\textwidth]
            {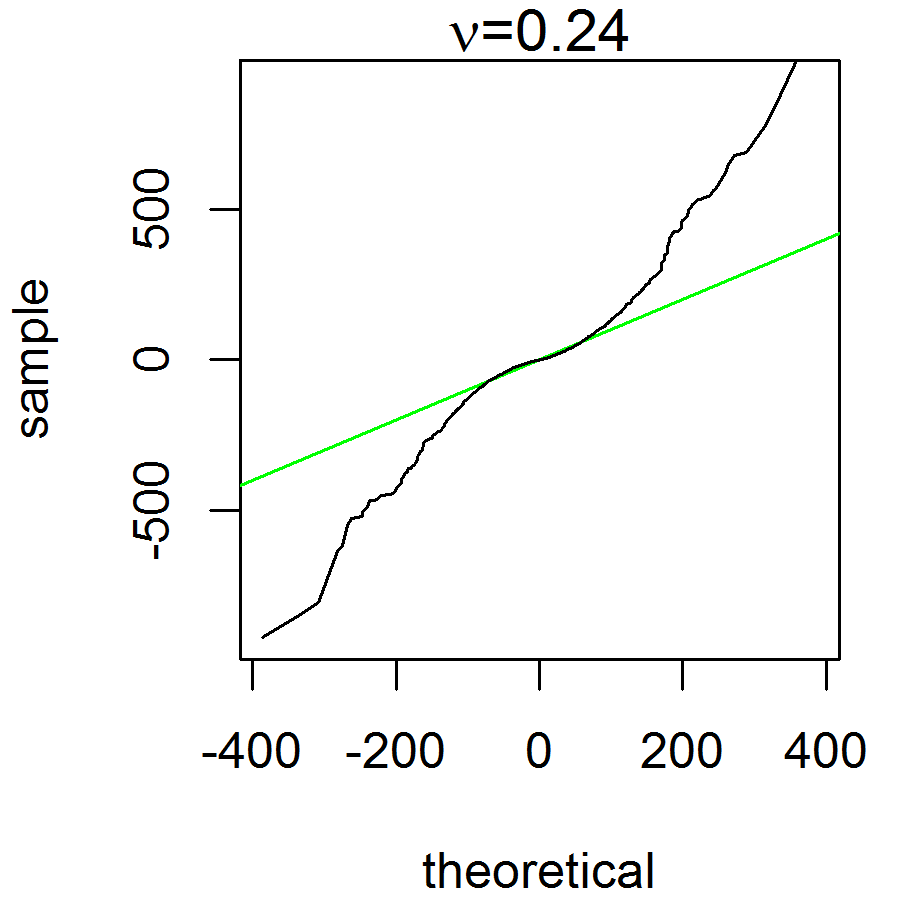}}
         \subfigure{\includegraphics[width=0.195\textwidth]
            {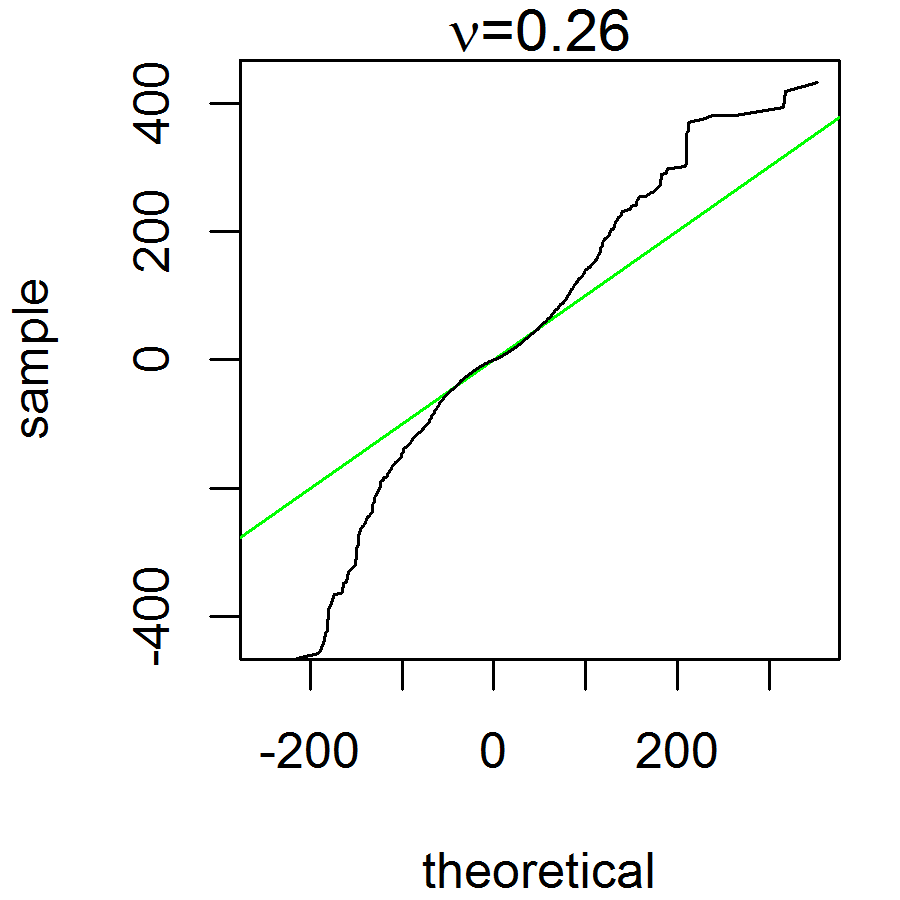}}
         \subfigure{\includegraphics[width=0.195\textwidth]
            {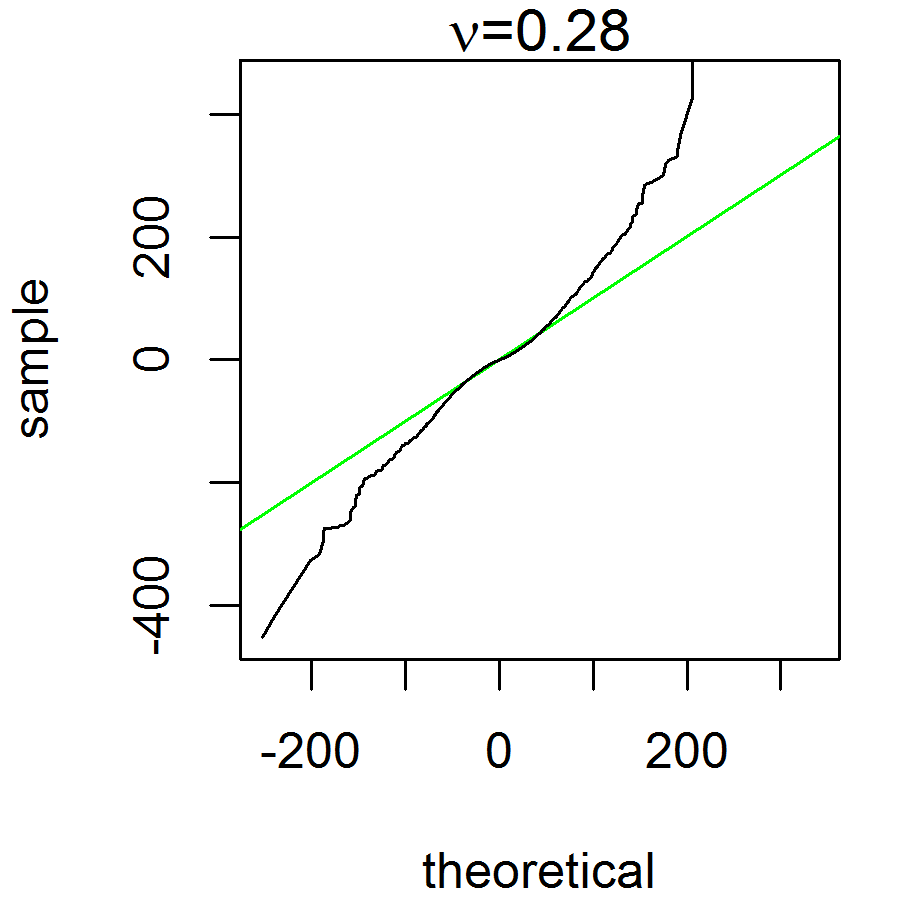}}
        \subfigure{\includegraphics[width=0.195\textwidth]
            {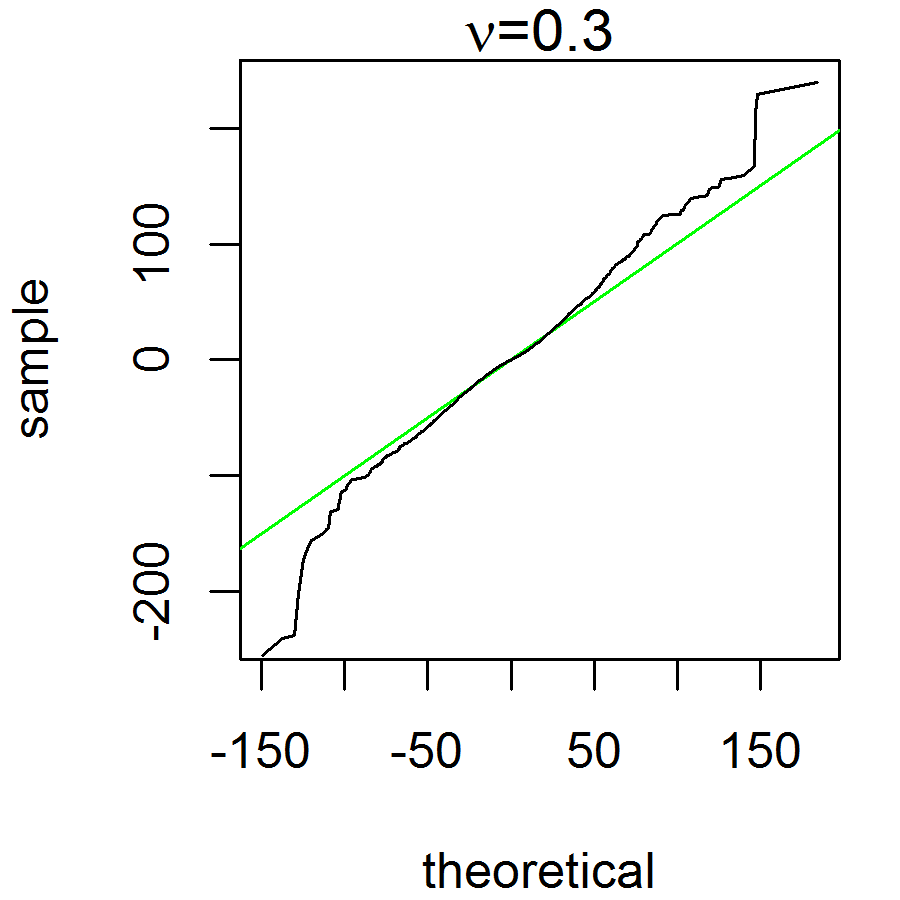}}

        \subfigure{\includegraphics[width=0.195\textwidth]
            {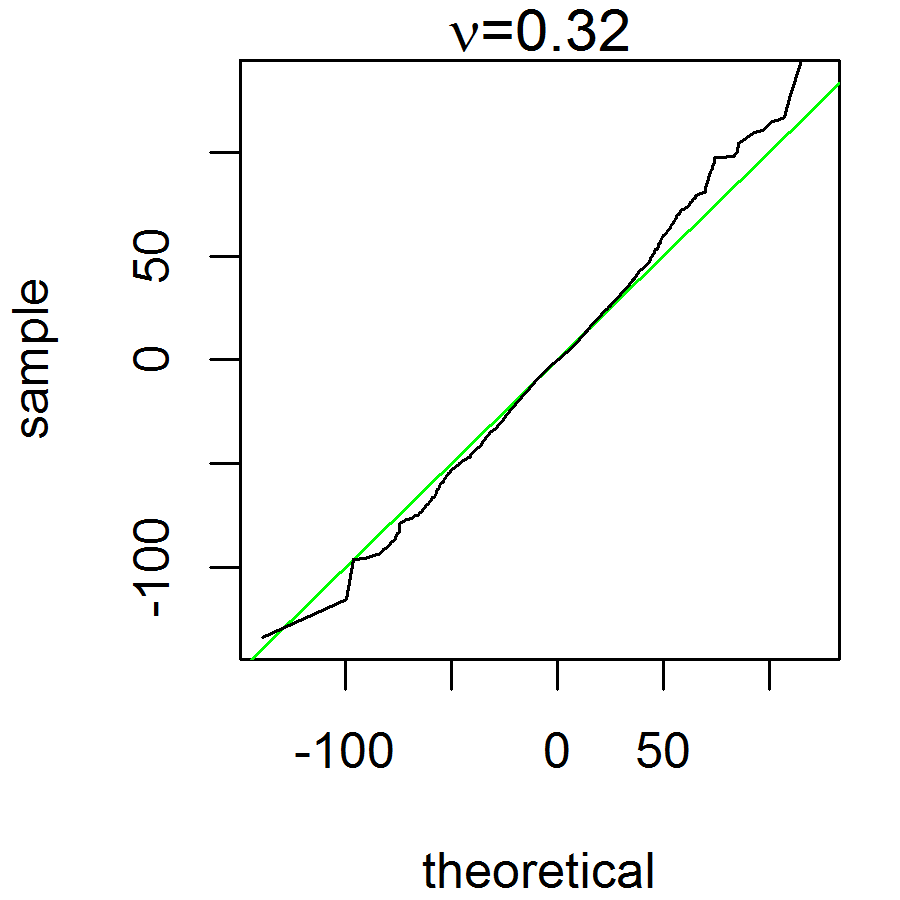}}
         \subfigure{\includegraphics[width=0.195\textwidth]
            {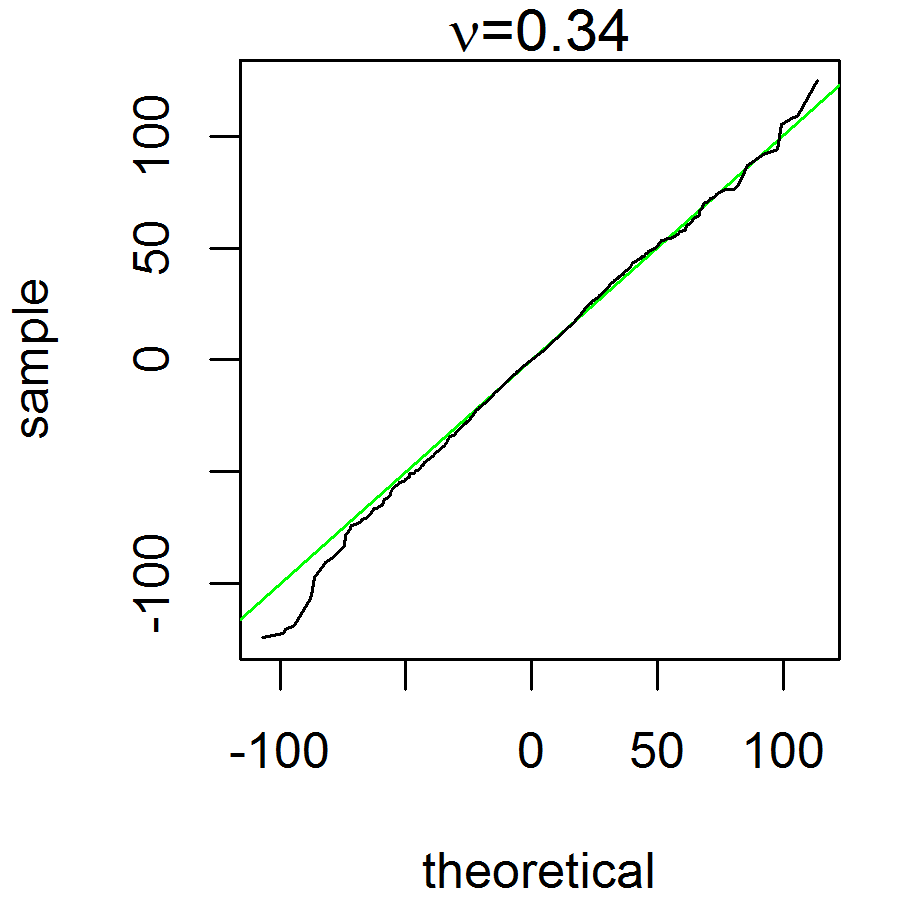}}
         \subfigure{\includegraphics[width=0.195\textwidth]
            {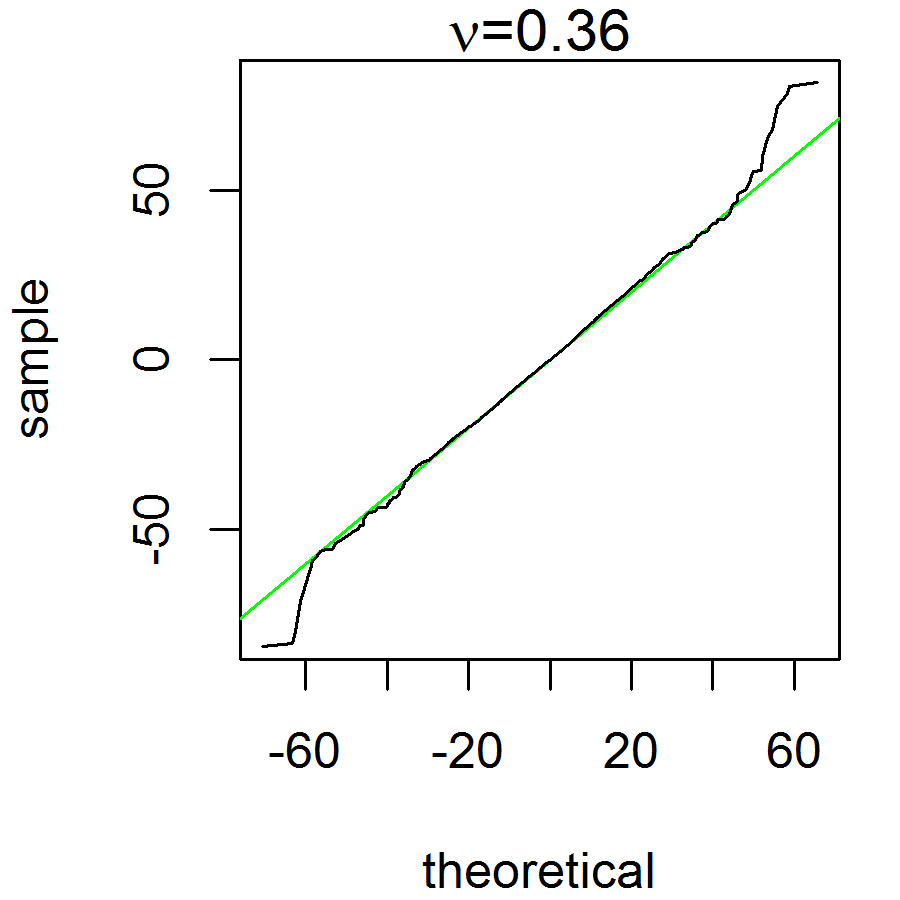}}
         \subfigure{\includegraphics[width=0.195\textwidth]
            {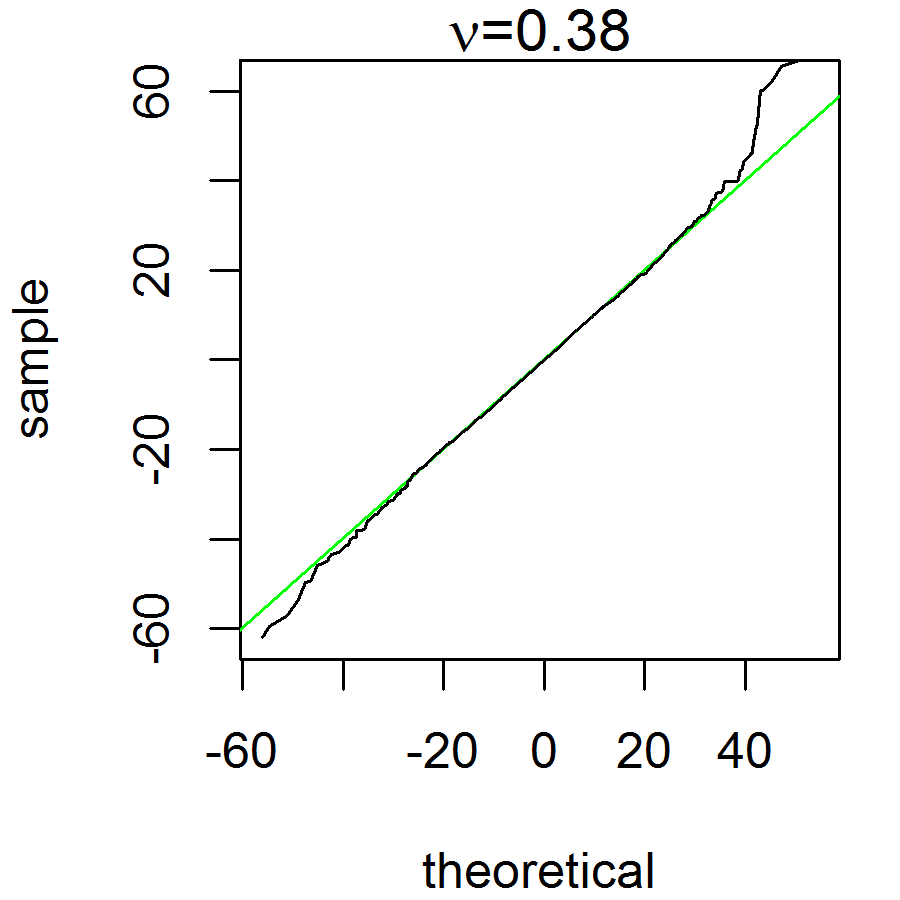}}
        \subfigure{\includegraphics[width=0.195\textwidth]
            {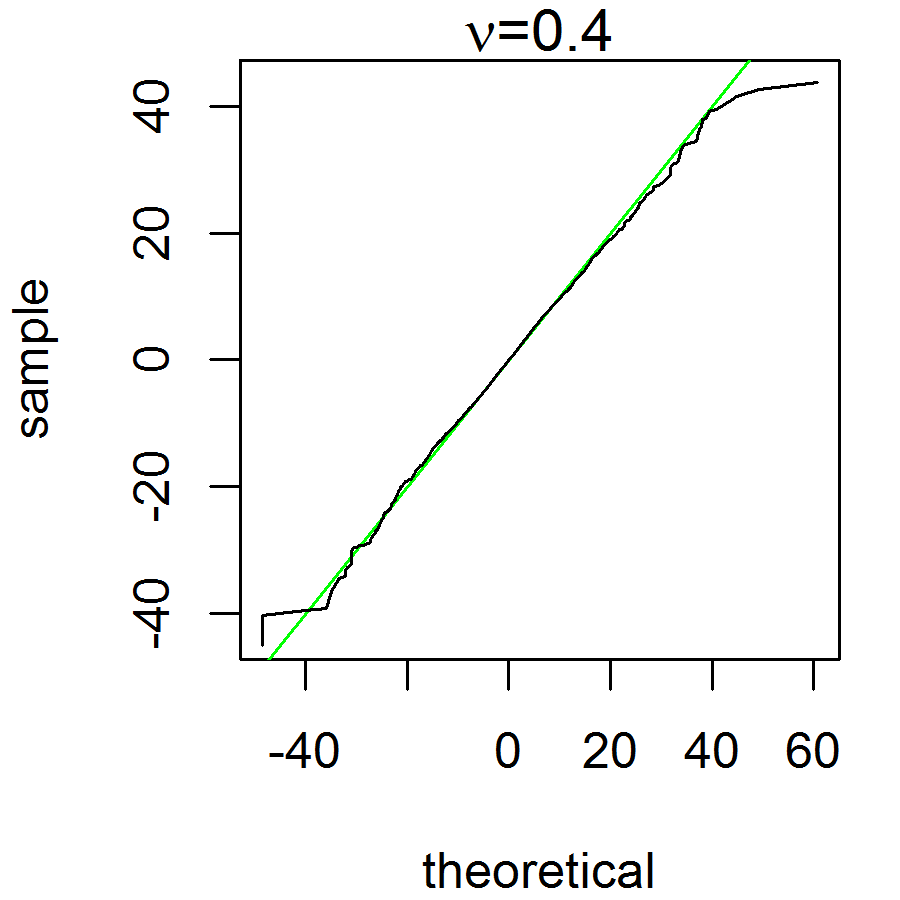}}

        \subfigure{\includegraphics[width=0.195\textwidth]
            {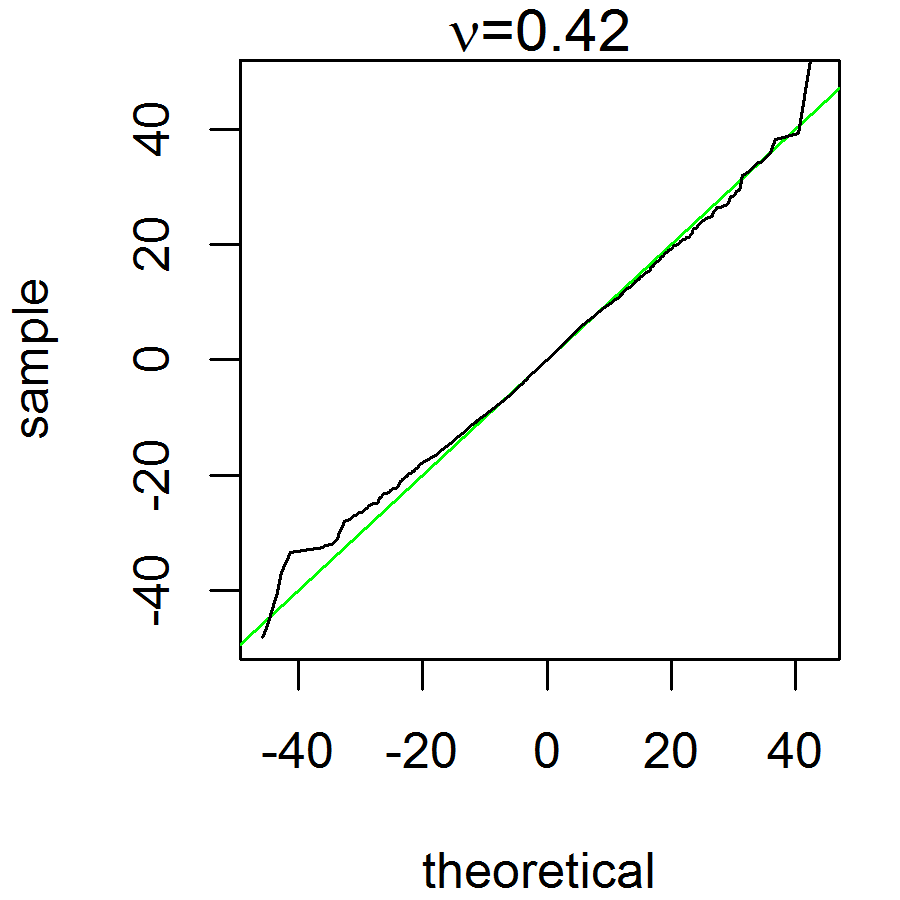}}
         \subfigure{\includegraphics[width=0.195\textwidth]
            {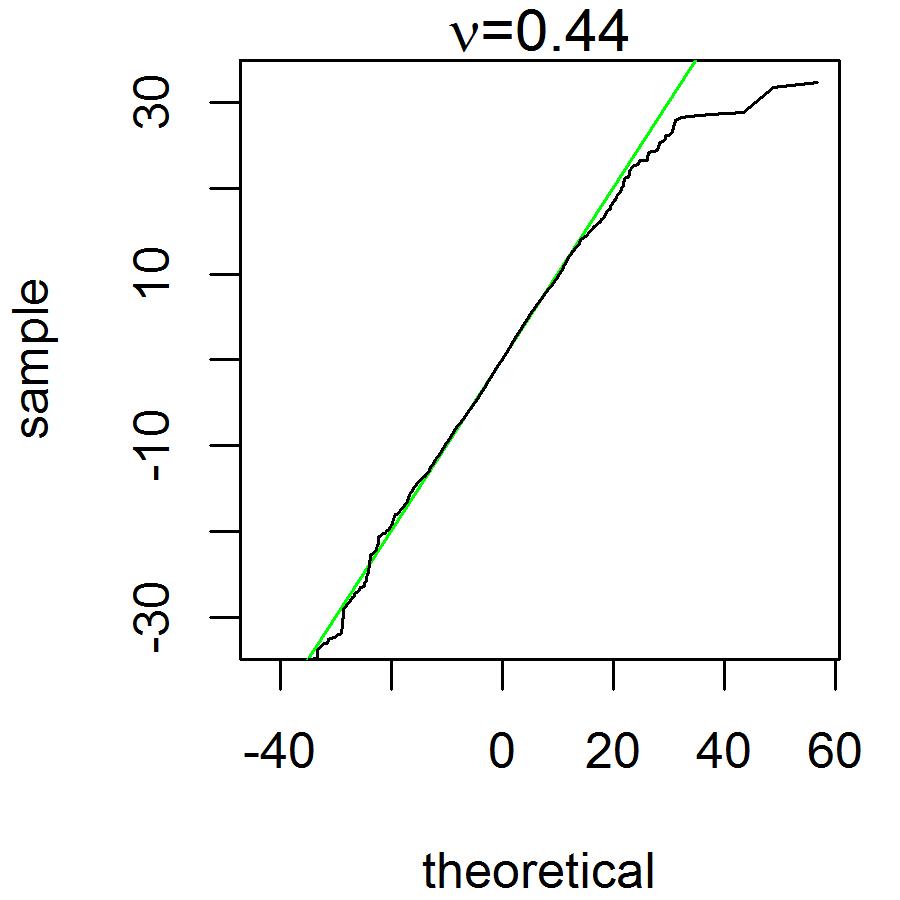}}
         \subfigure{\includegraphics[width=0.195\textwidth]
            {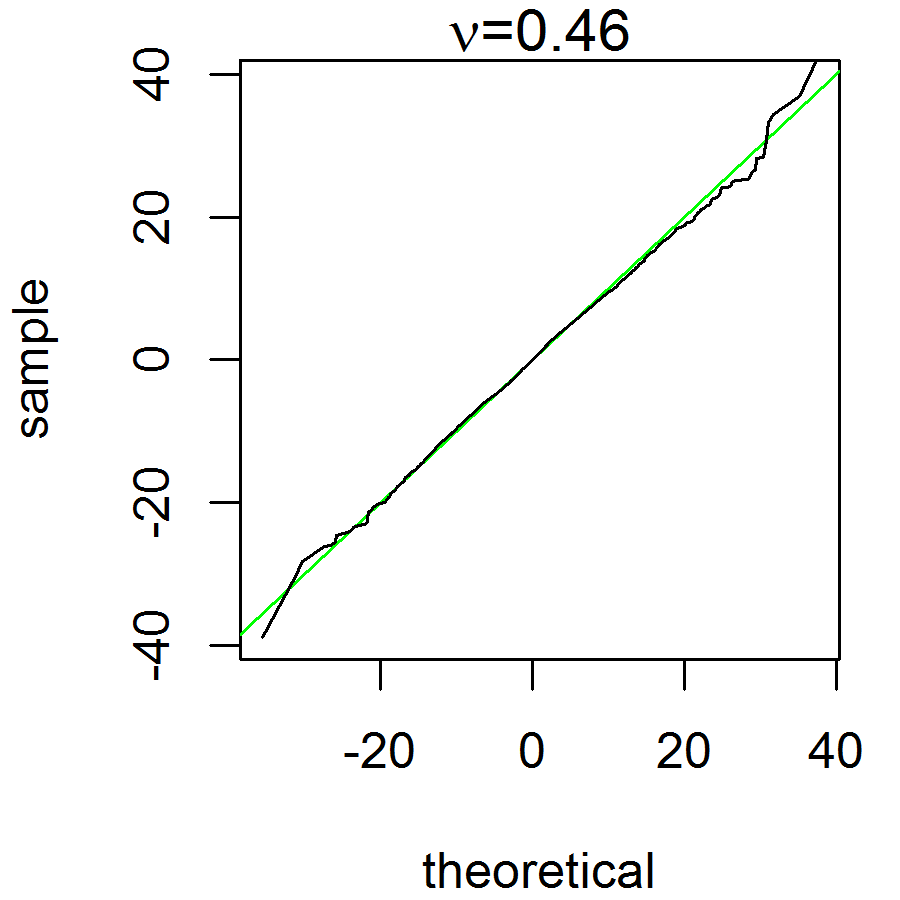}}
         \subfigure{\includegraphics[width=0.195\textwidth]
            {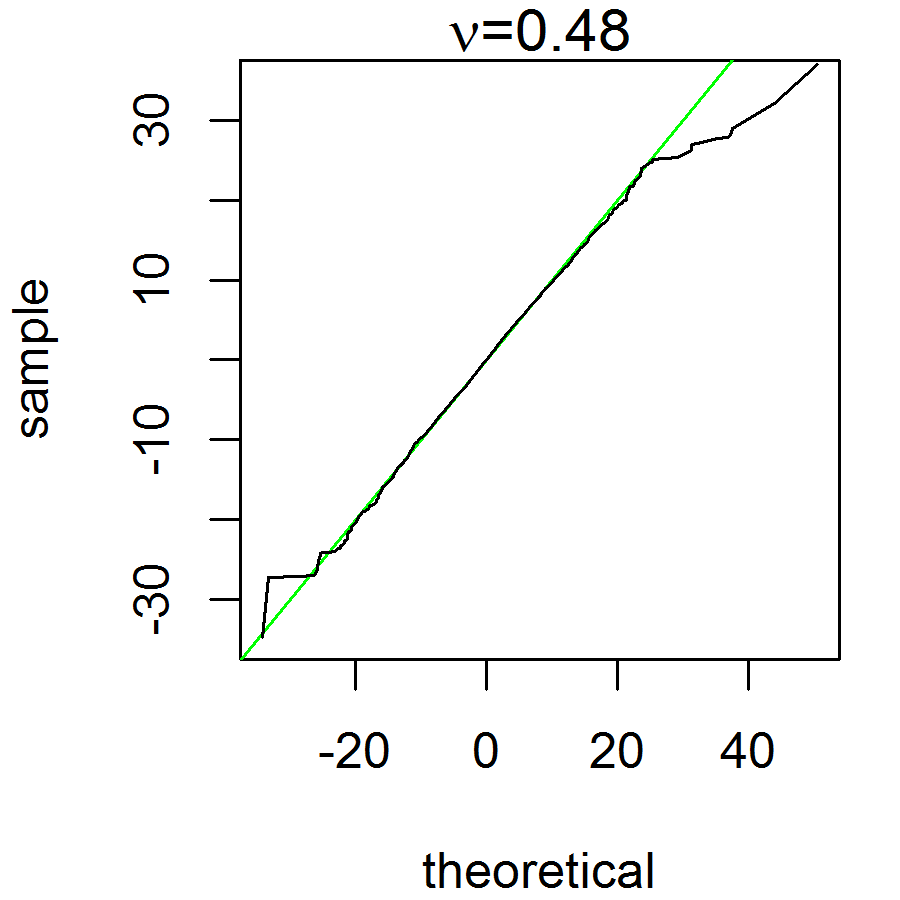}}
        \subfigure{\includegraphics[width=0.195\textwidth]
            {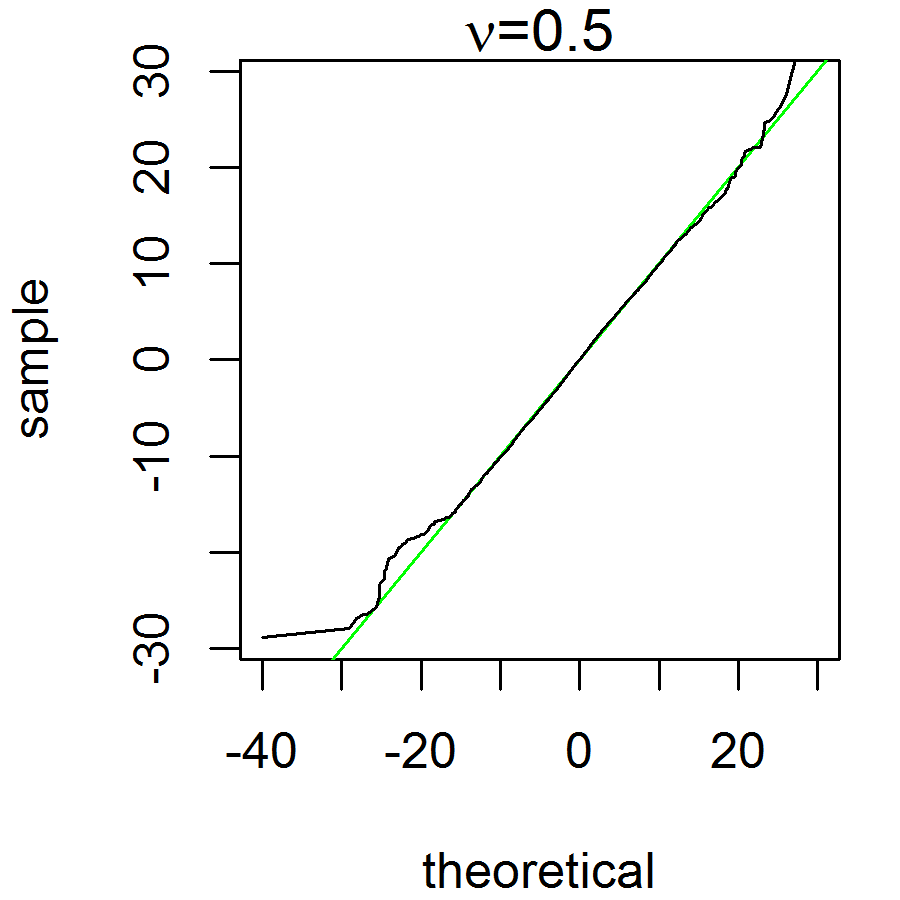}}
    \end{center}
    \caption{Q-Q plots for $0.02\leq\nu\leq0.5$
        where the x-axis represent the theoretical distribution based on ordered monte carlo sample from estimated VG distribution with scale and shape parameters $(\sigma_{\hat\mu}, \nu_{\hat\mu})$,
        and y-axis represents the ordered $n^{\hat\beta}\mu_{n}$ samples for $n=10000$.}
    \label{QQ plots 1}
\end{figure}

\begin{figure}[htbp]
    \begin{center}
        \subfigure{\includegraphics[width=0.195\textwidth]
            {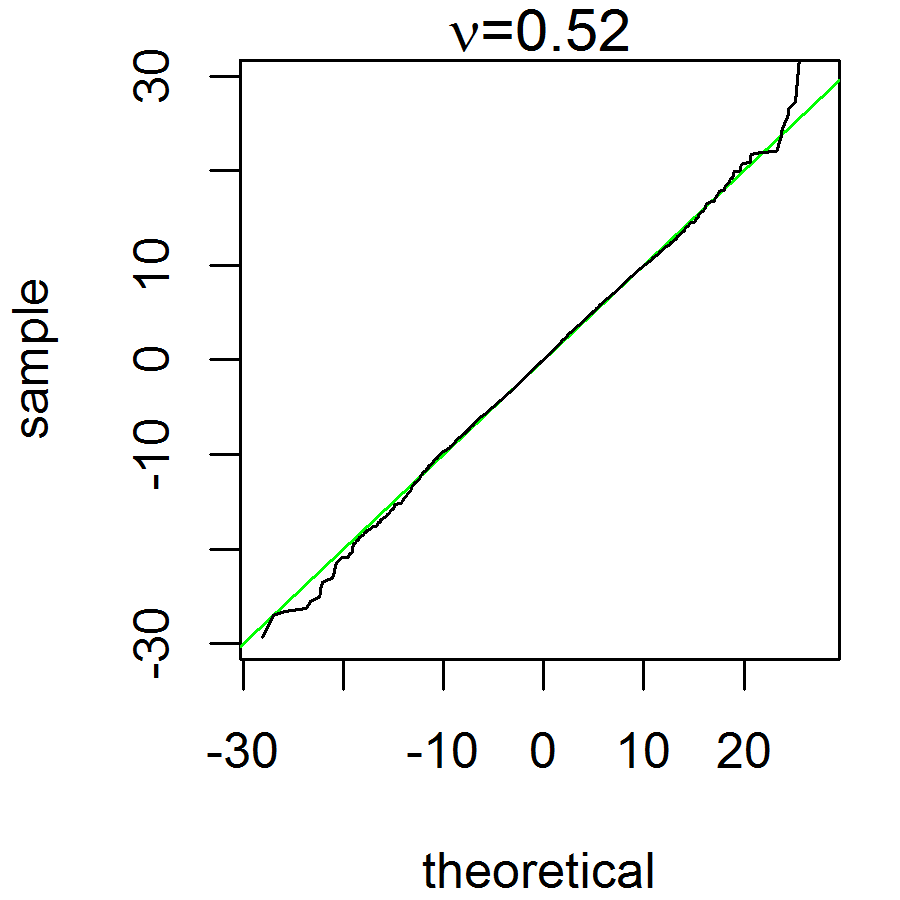}}
         \subfigure{\includegraphics[width=0.195\textwidth]
            {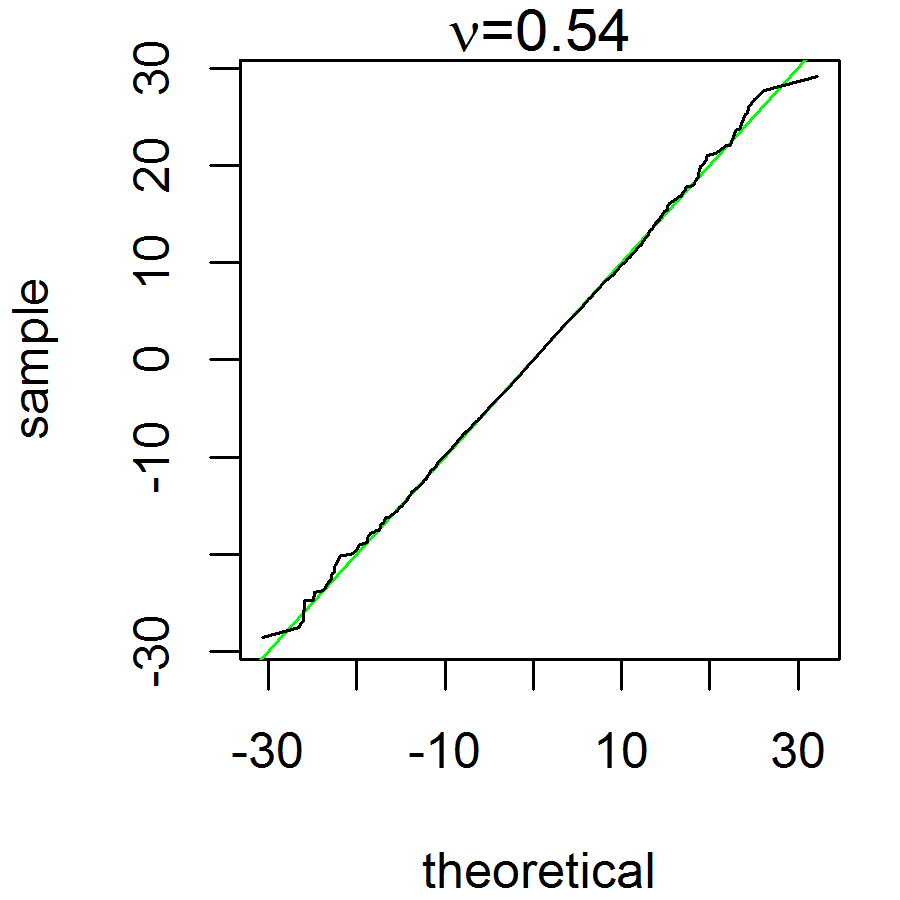}}
         \subfigure{\includegraphics[width=0.195\textwidth]
            {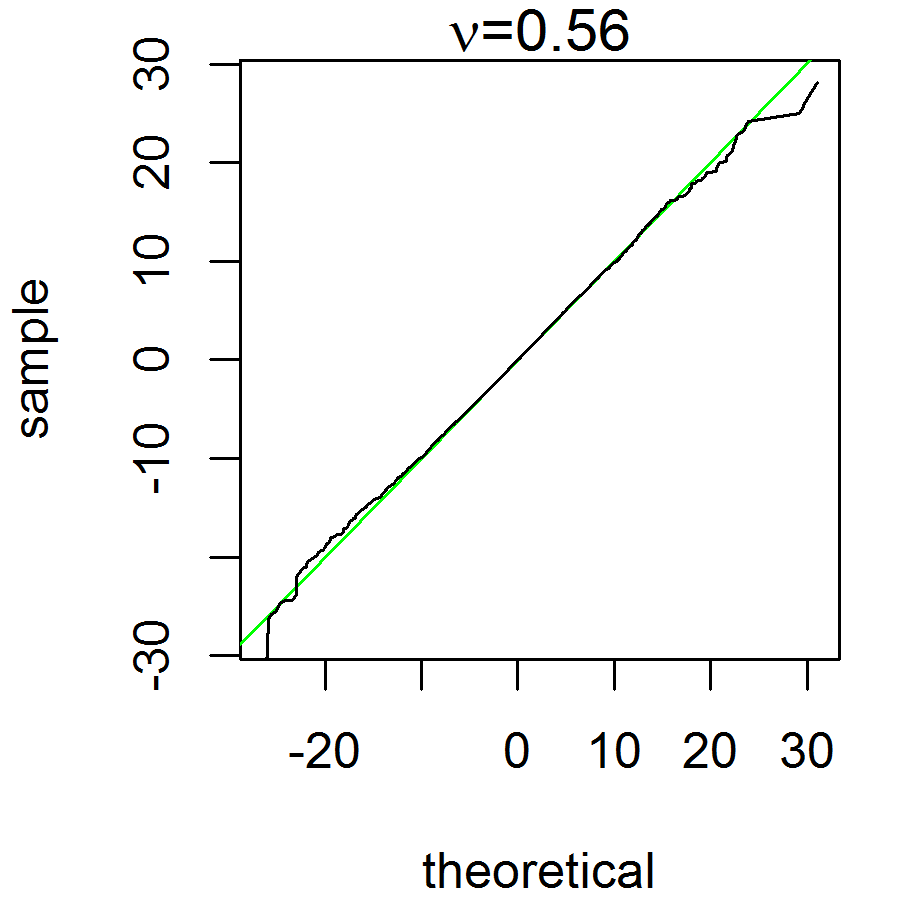}}
         \subfigure{\includegraphics[width=0.195\textwidth]
            {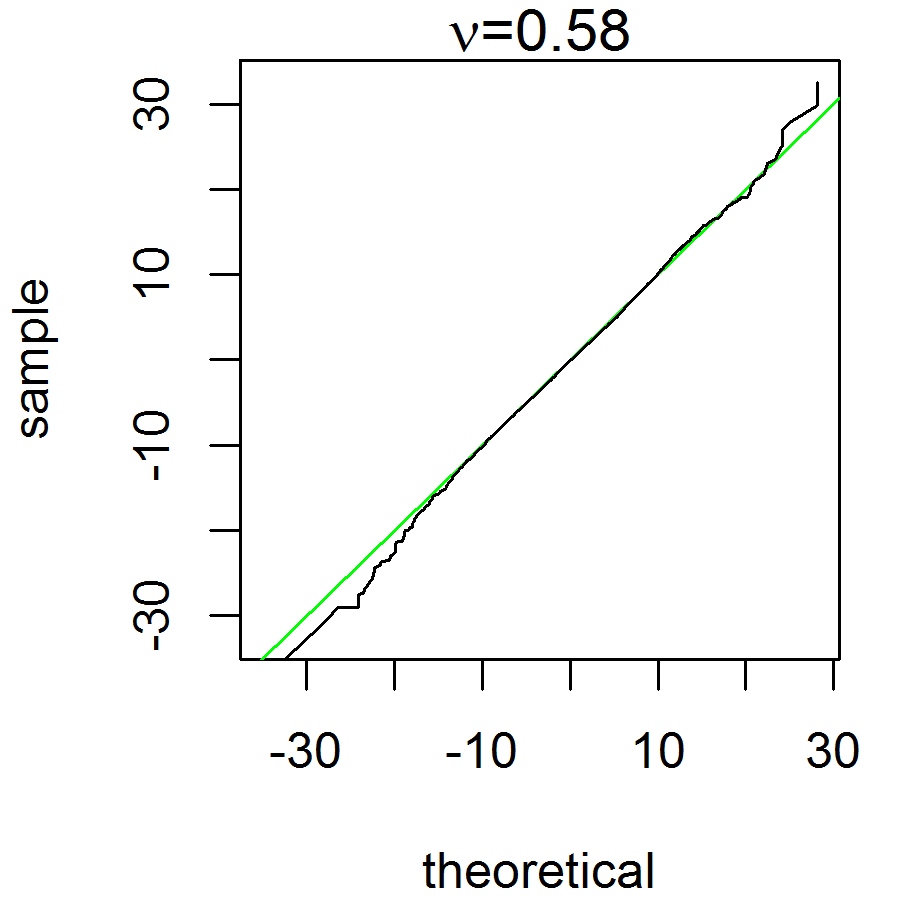}}
         \subfigure{\includegraphics[width=0.195\textwidth]
            {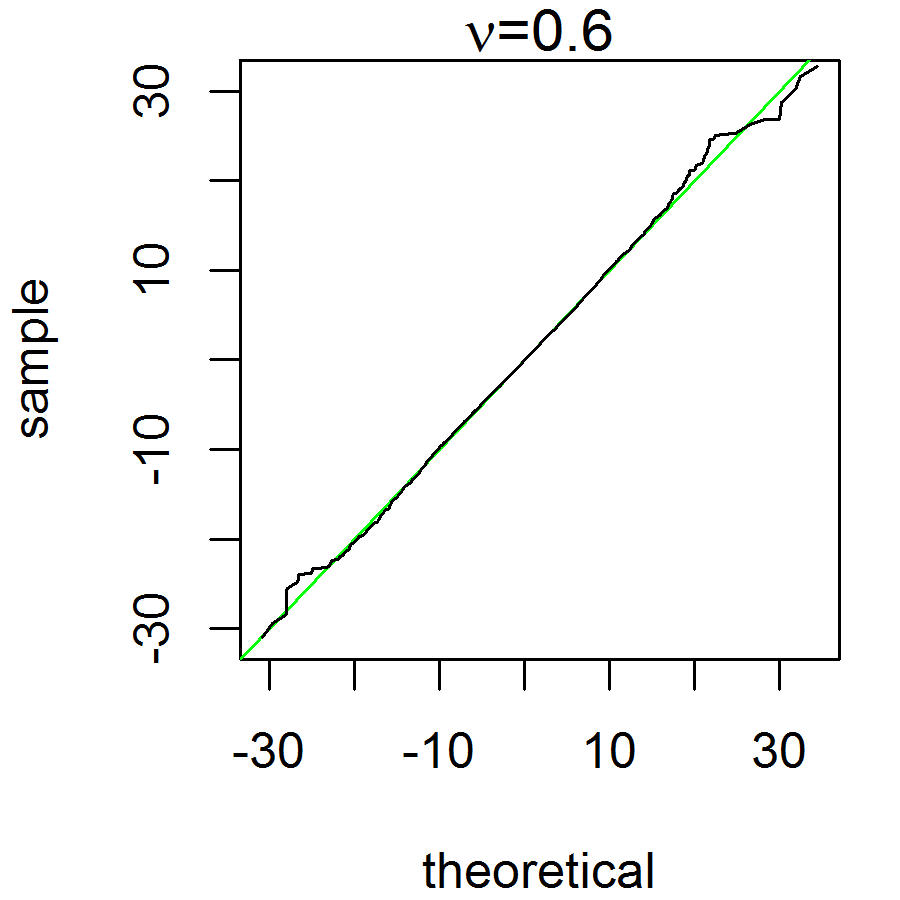}}

        \subfigure{\includegraphics[width=0.195\textwidth]
            {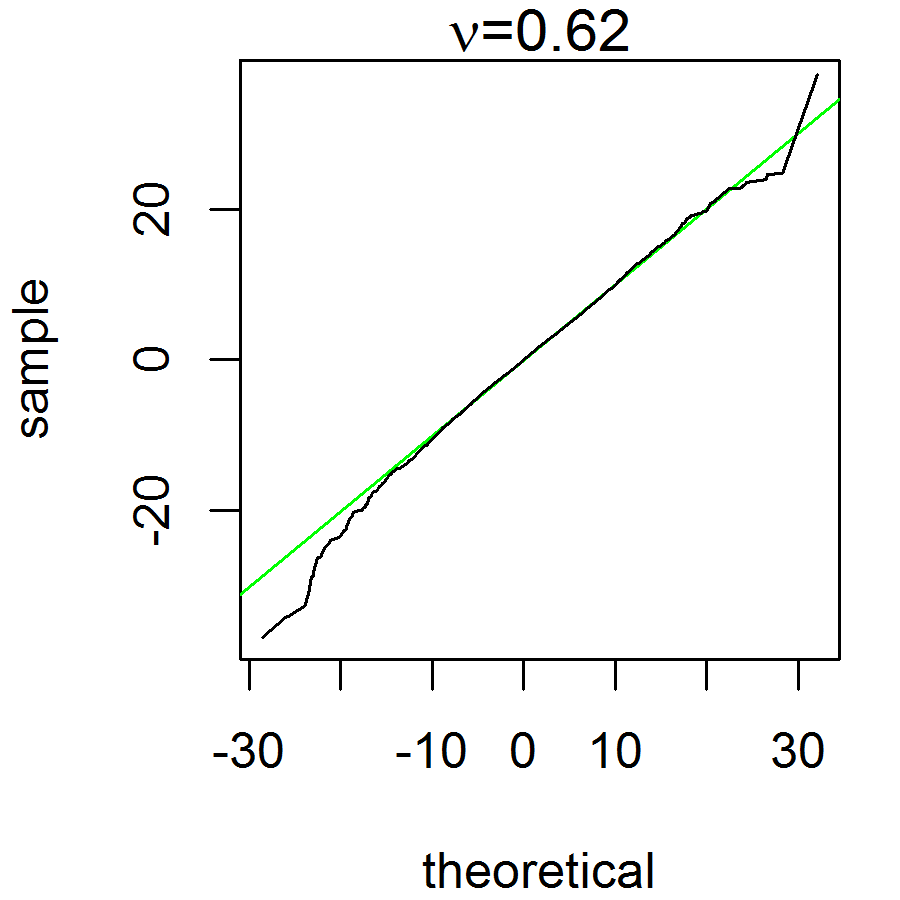}}
         \subfigure{\includegraphics[width=0.195\textwidth]
            {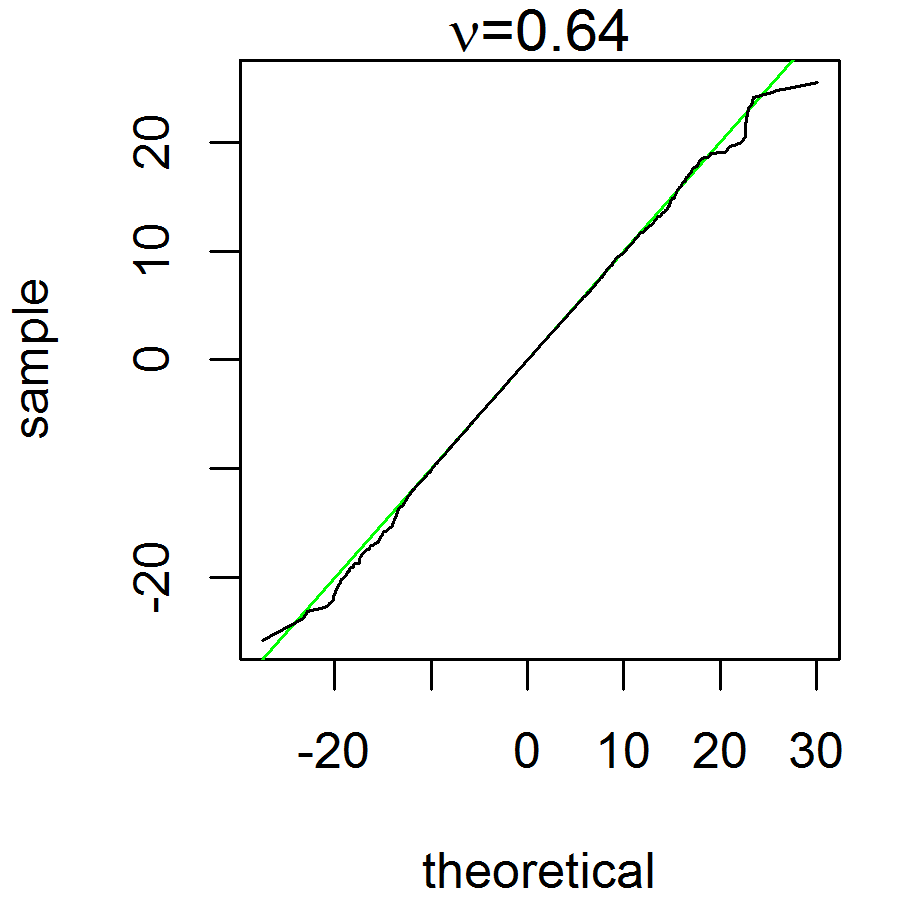}}
         \subfigure{\includegraphics[width=0.195\textwidth]
            {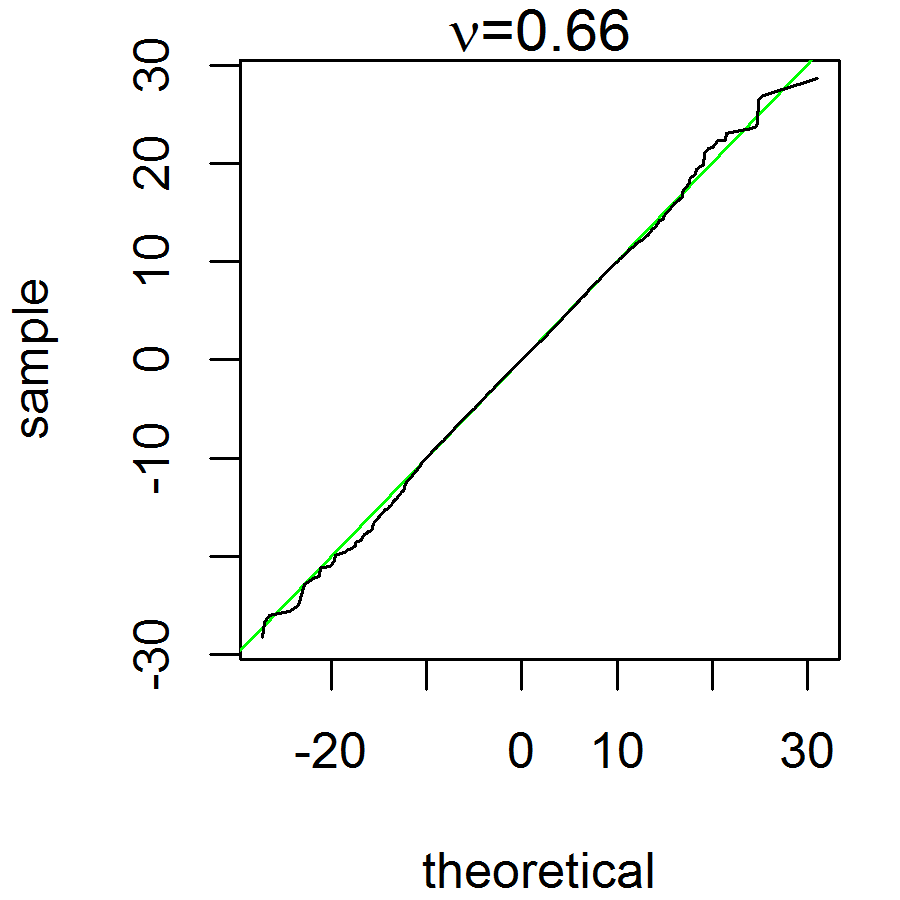}}
         \subfigure{\includegraphics[width=0.195\textwidth]
            {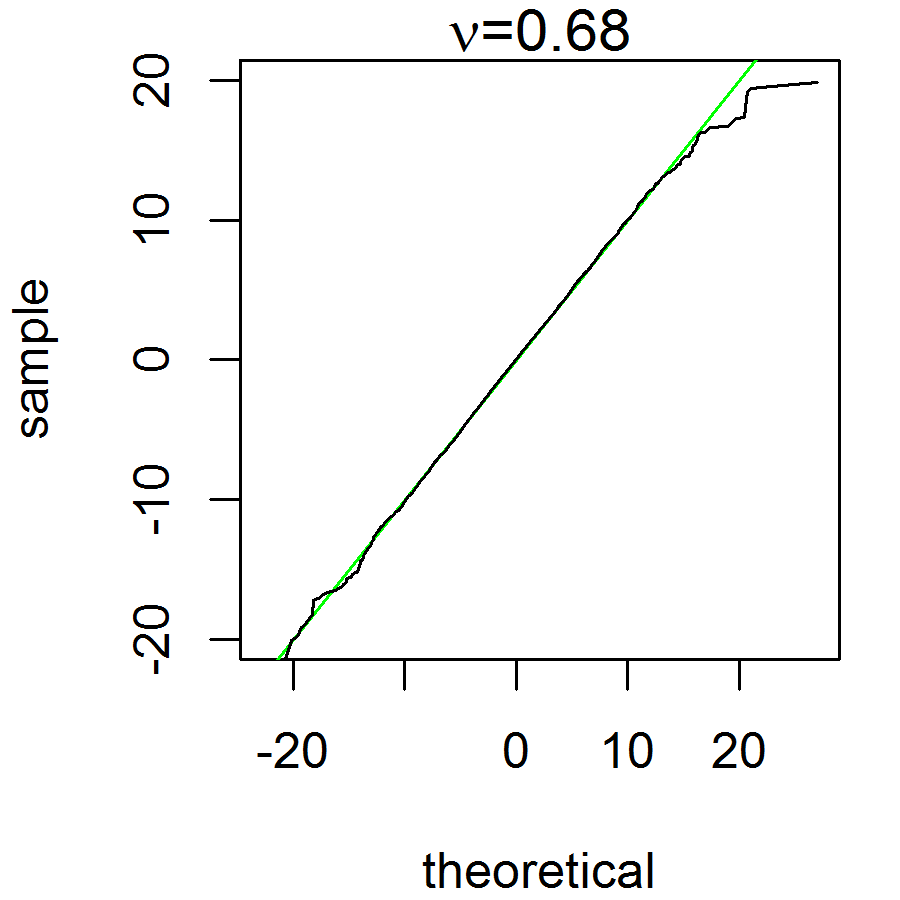}}
        \subfigure{\includegraphics[width=0.195\textwidth]
            {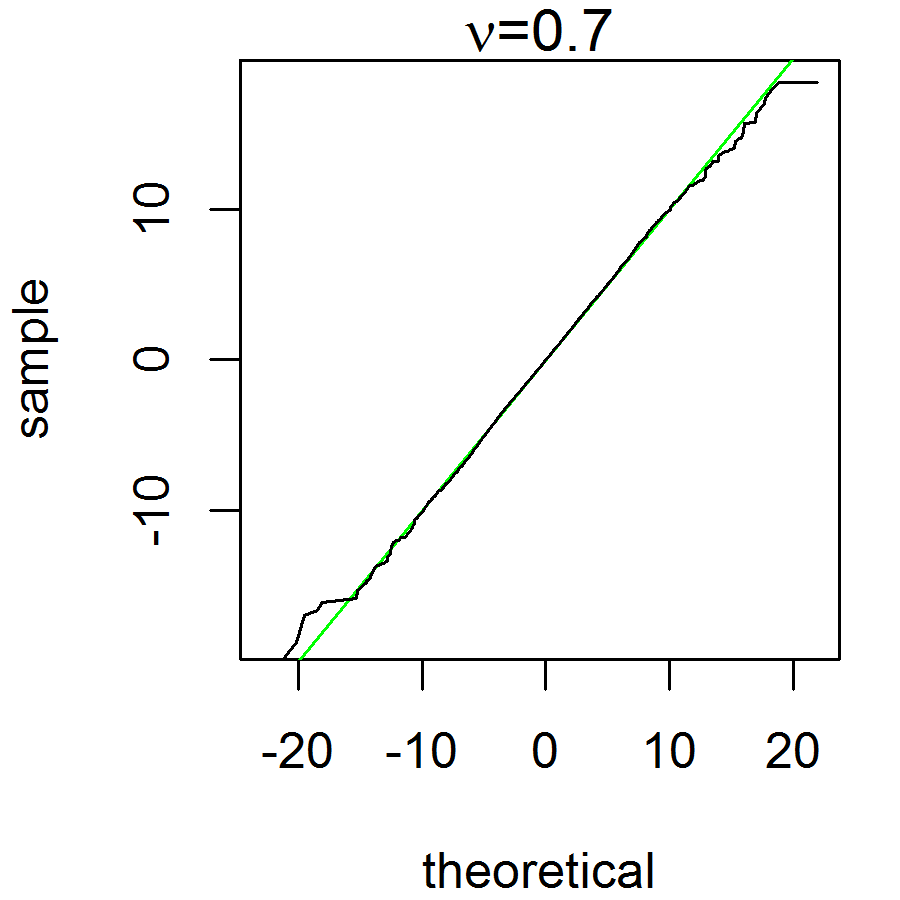}}

        \subfigure{\includegraphics[width=0.195\textwidth]
            {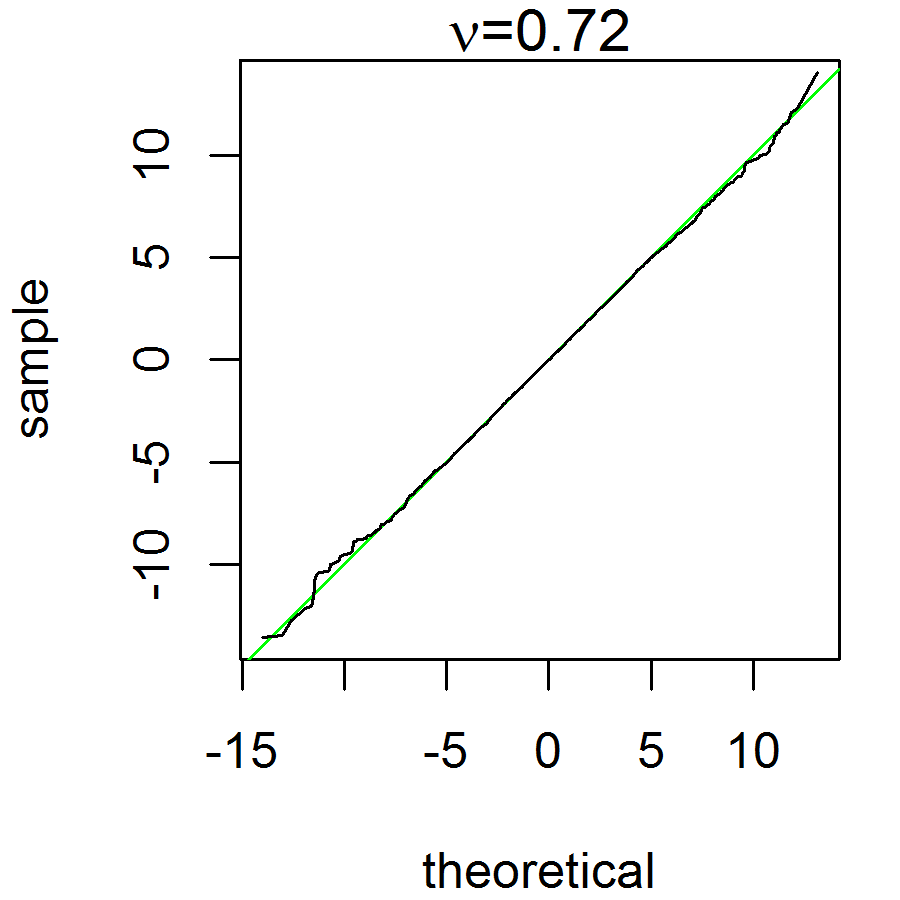}}
         \subfigure{\includegraphics[width=0.195\textwidth]
            {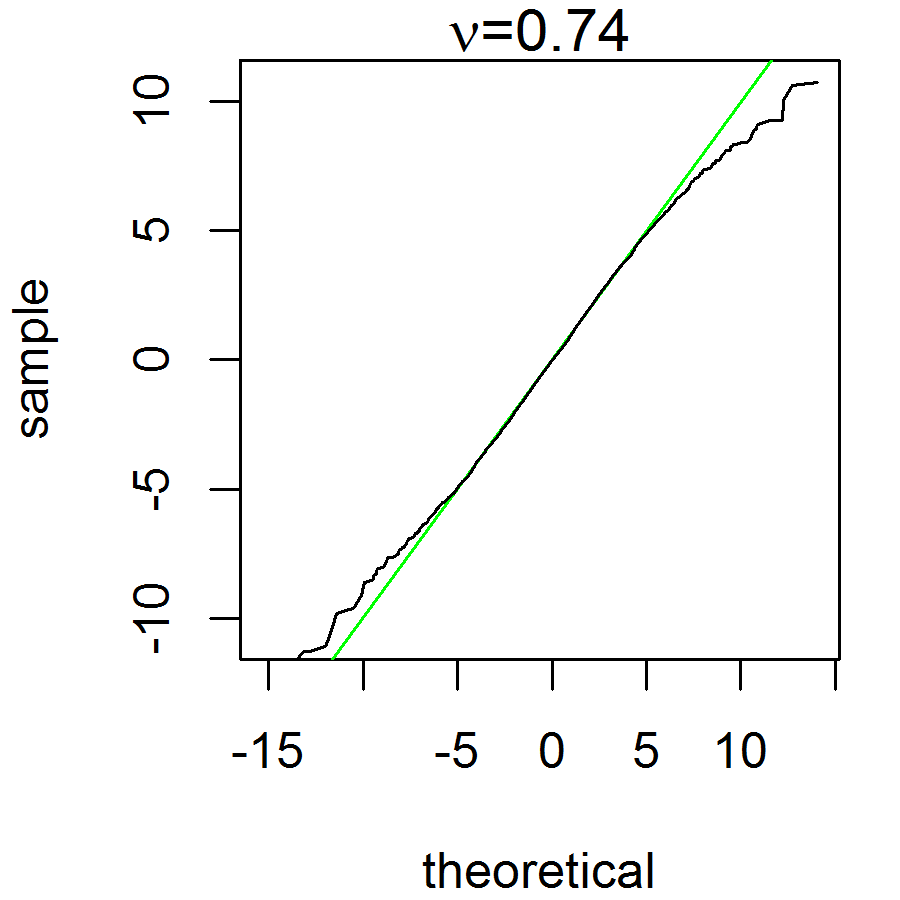}}
         \subfigure{\includegraphics[width=0.195\textwidth]
            {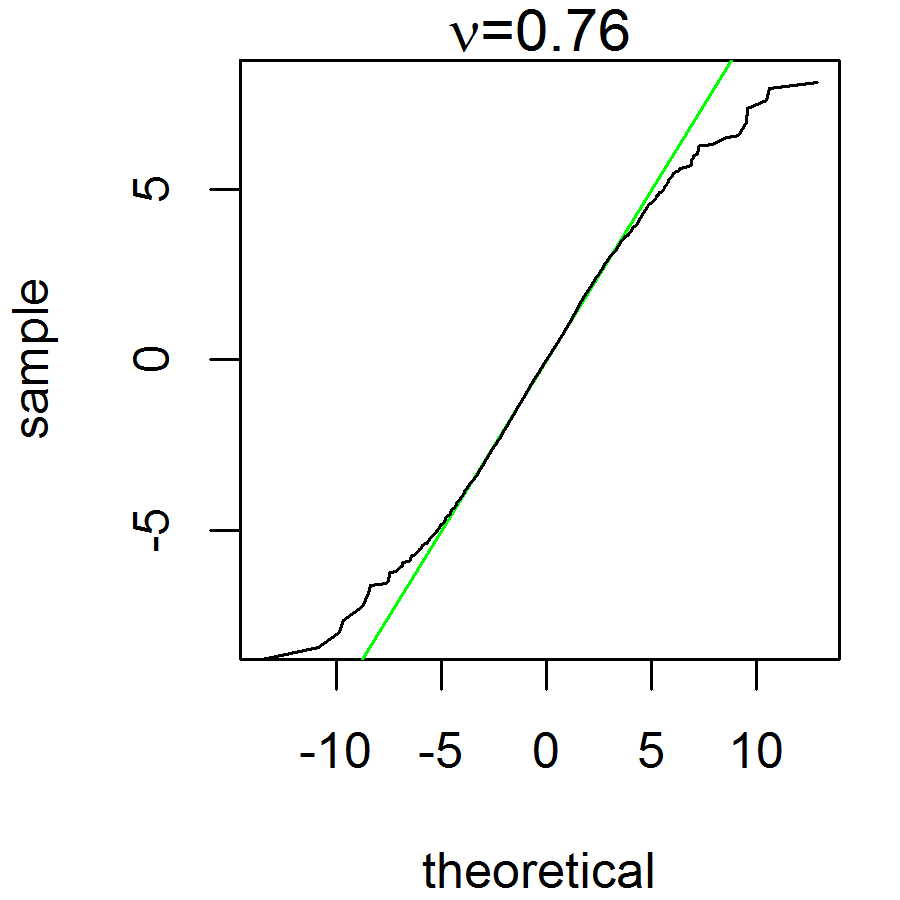}}
         \subfigure{\includegraphics[width=0.195\textwidth]
            {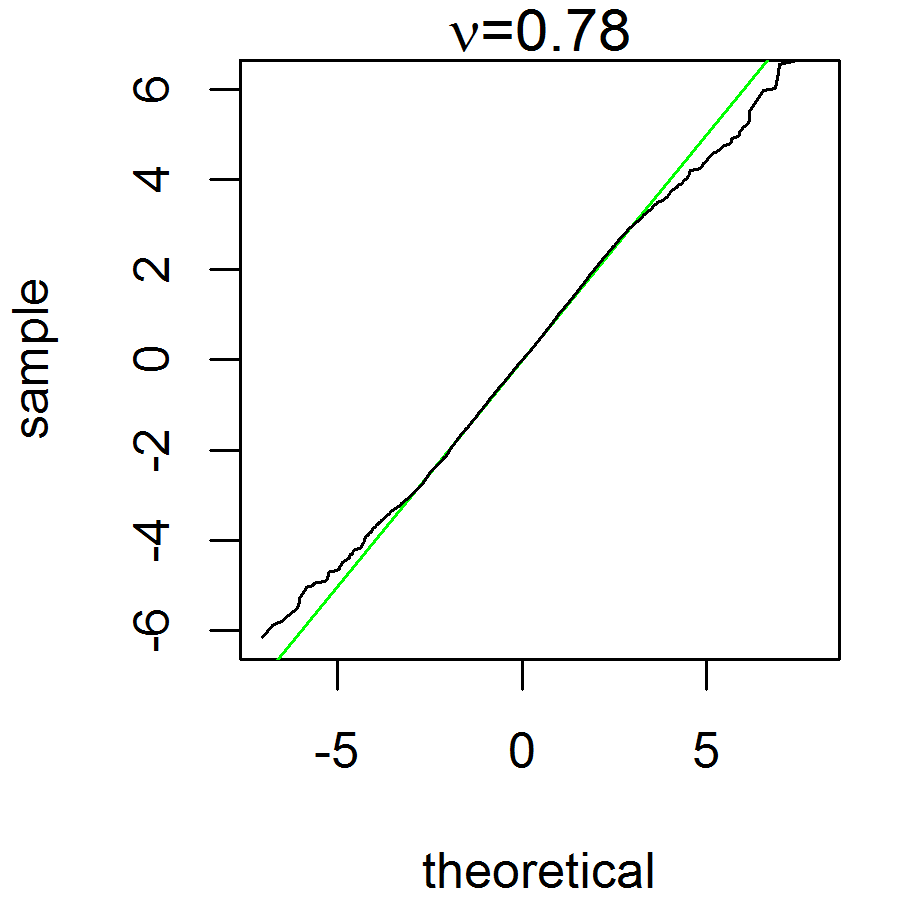}}
        \subfigure{\includegraphics[width=0.195\textwidth]
            {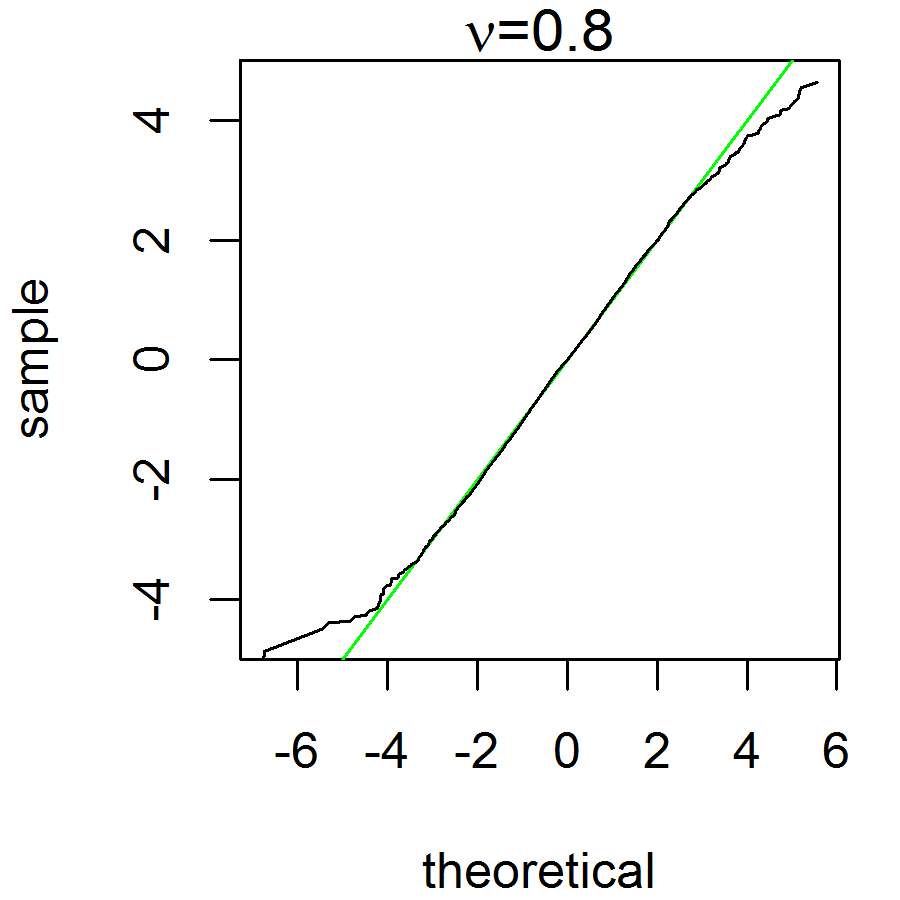}}

        \subfigure{\includegraphics[width=0.195\textwidth]
            {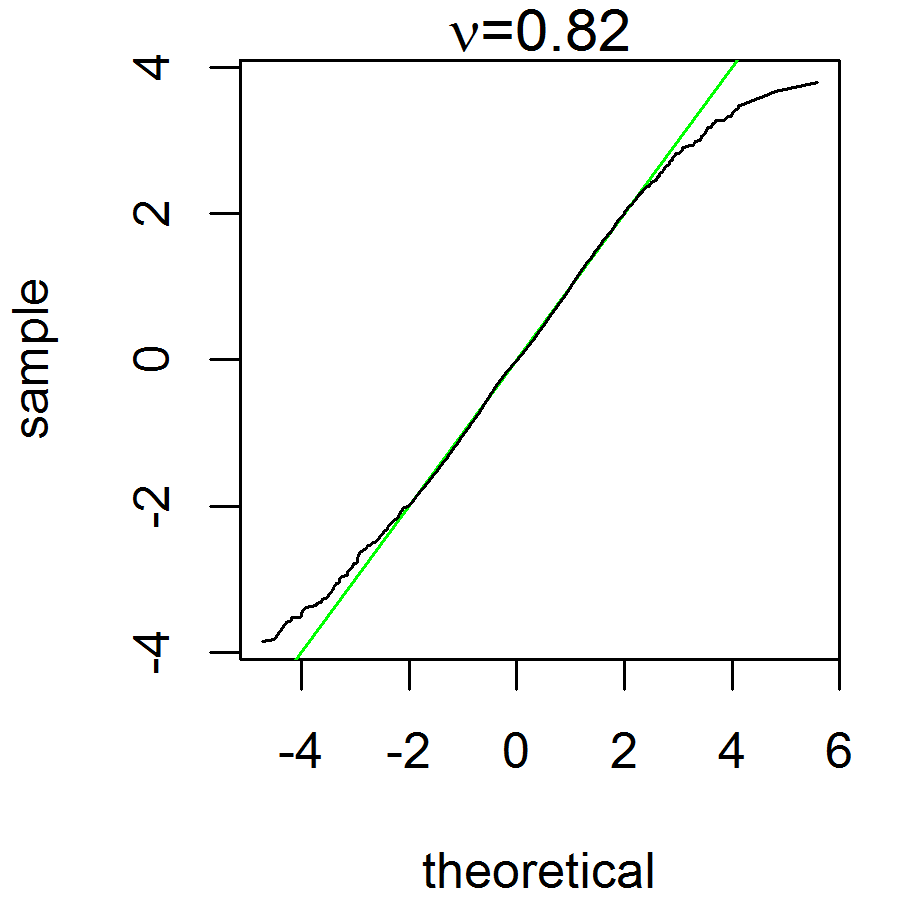}}
         \subfigure{\includegraphics[width=0.195\textwidth]
            {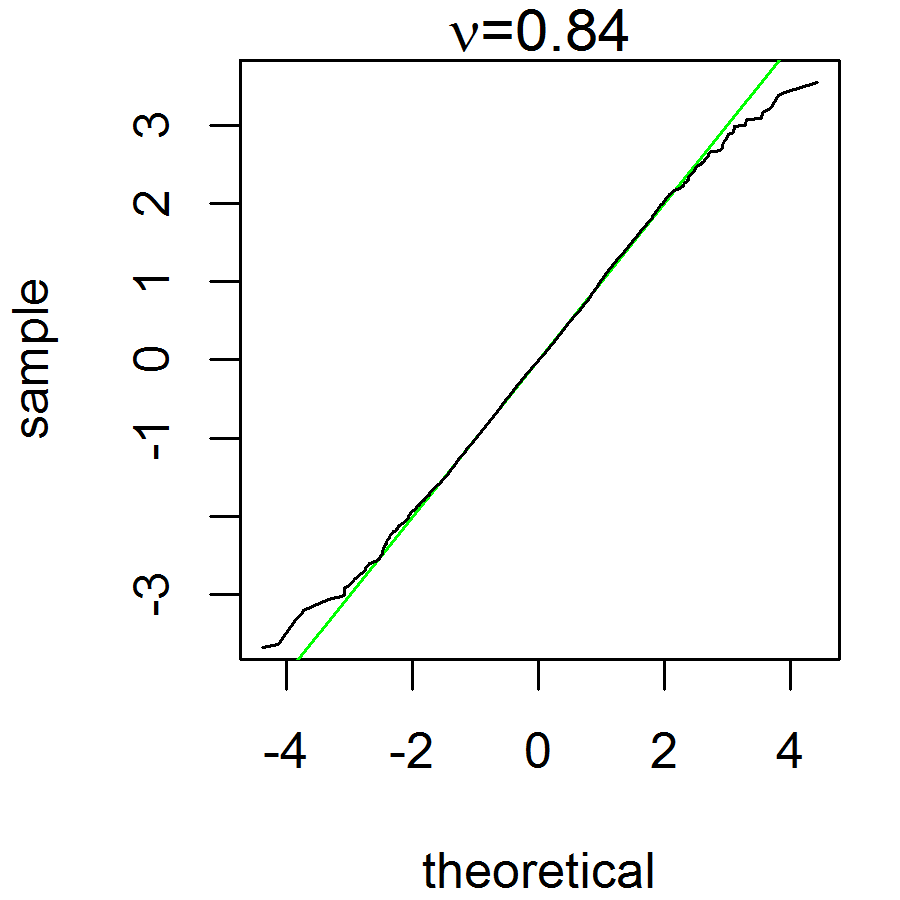}}
         \subfigure{\includegraphics[width=0.195\textwidth]
            {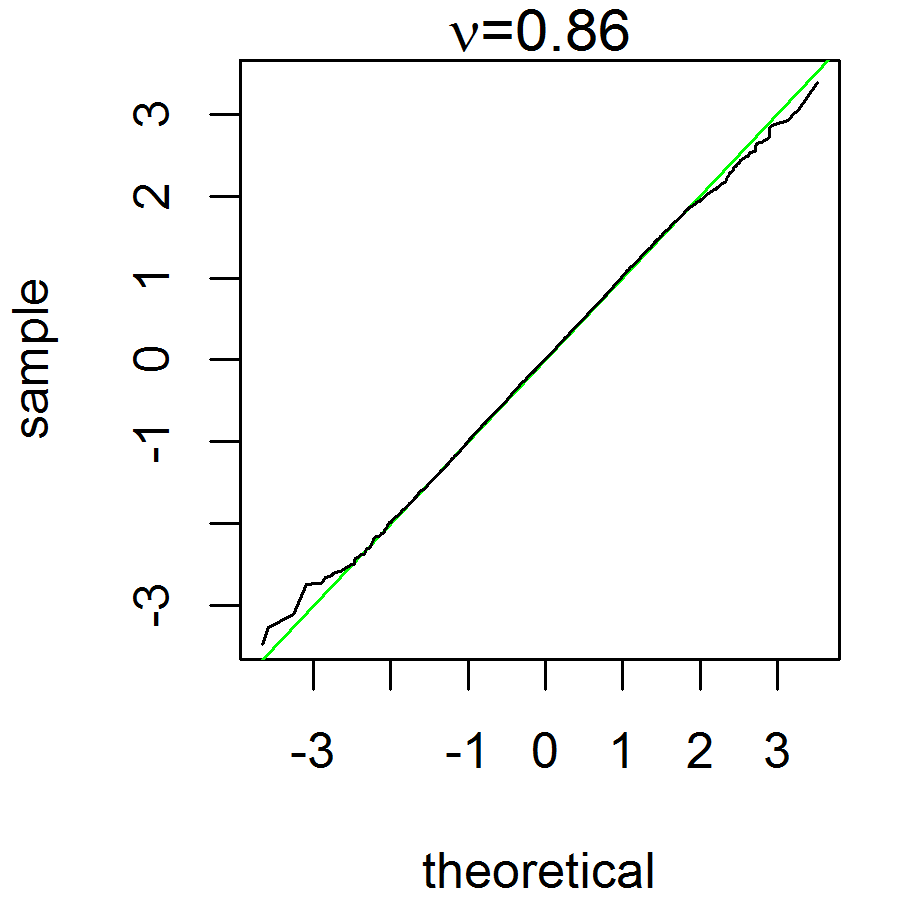}}
         \subfigure{\includegraphics[width=0.195\textwidth]
            {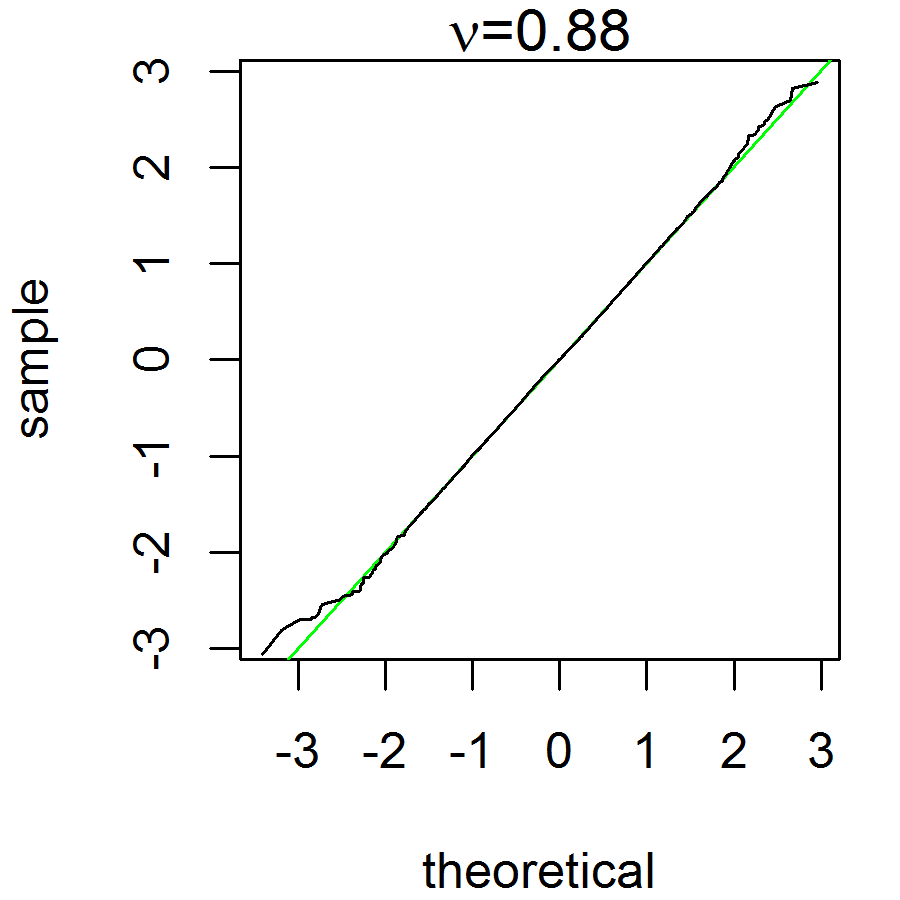}}
        \subfigure{\includegraphics[width=0.195\textwidth]
            {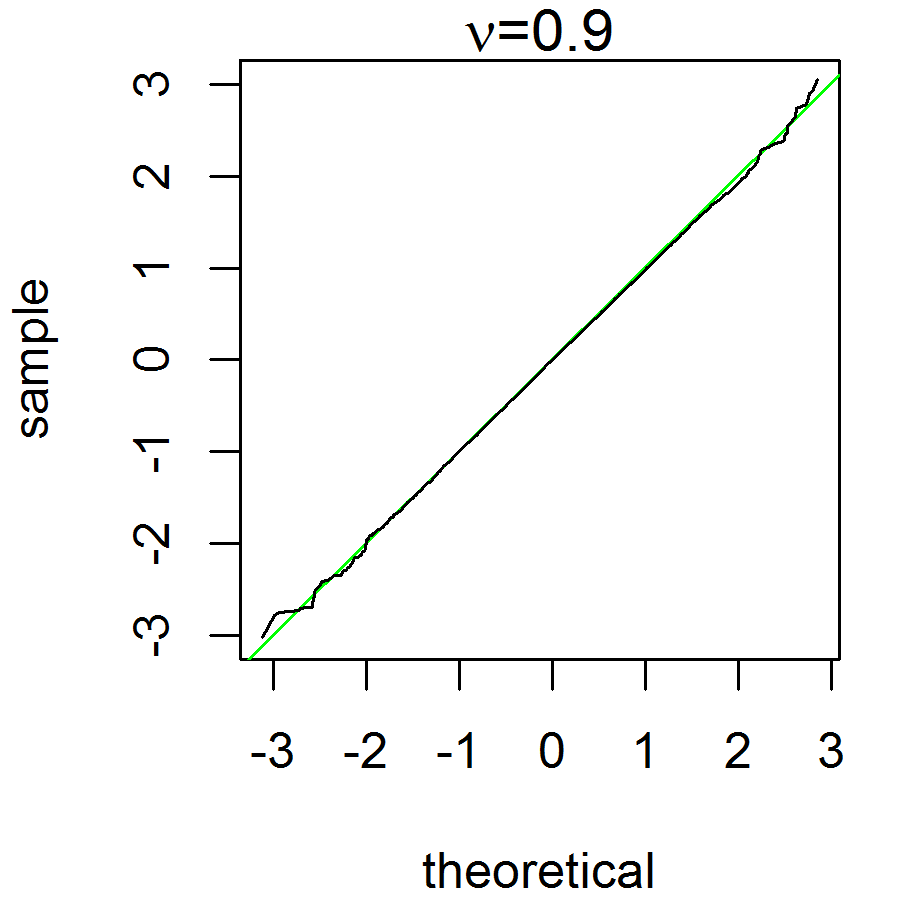}}

        \subfigure{\includegraphics[width=0.195\textwidth]
            {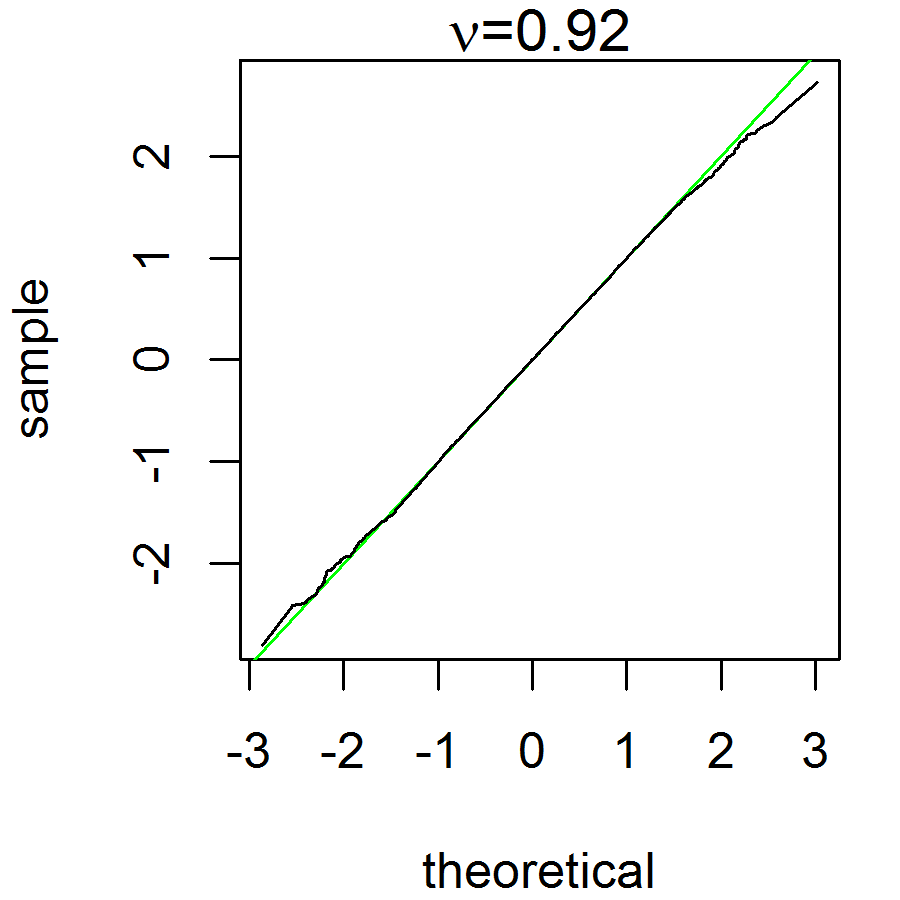}}
         \subfigure{\includegraphics[width=0.195\textwidth]
            {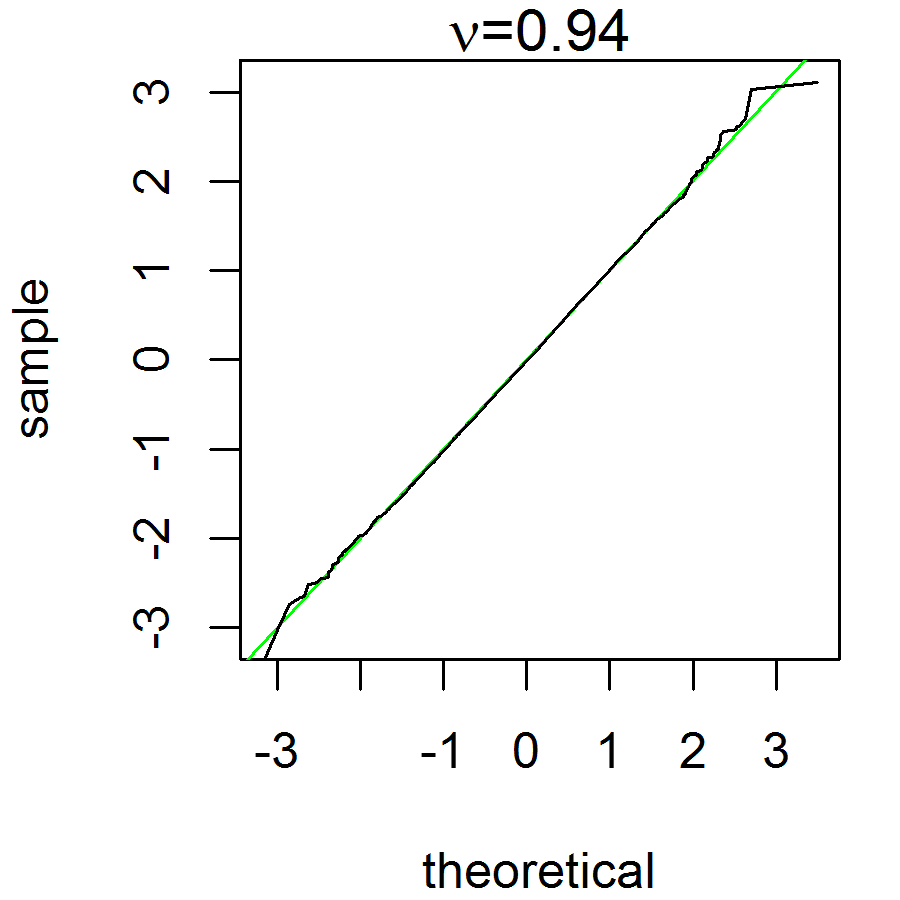}}
         \subfigure{\includegraphics[width=0.195\textwidth]
            {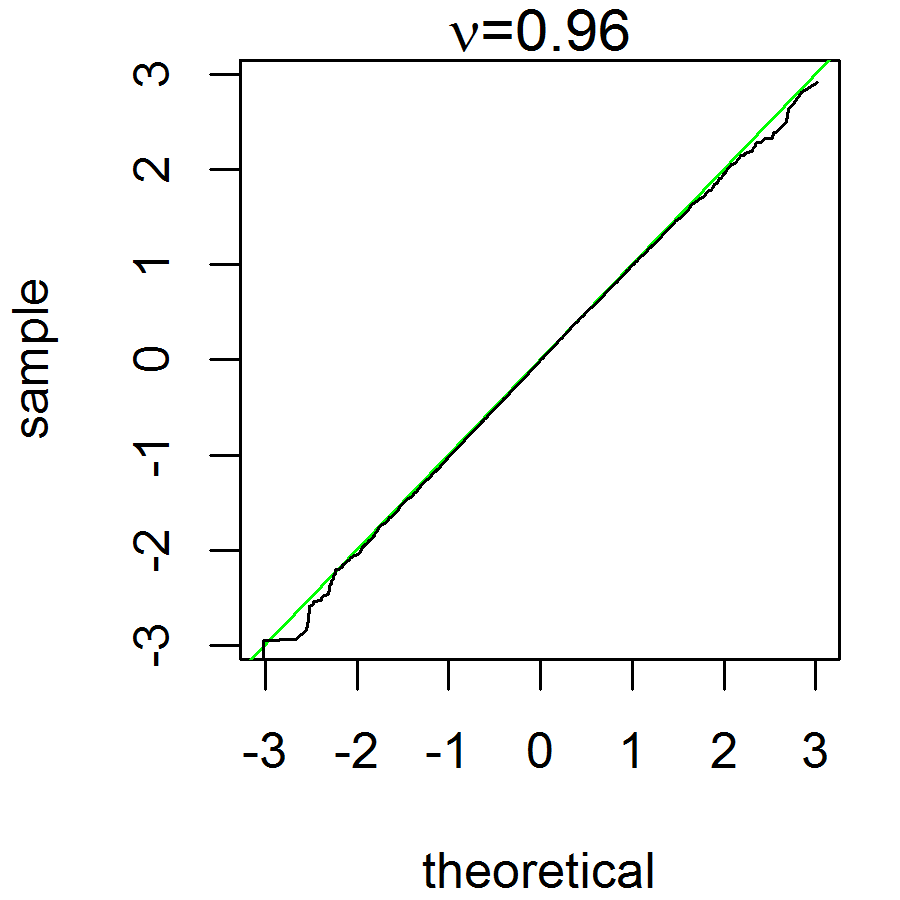}}
         \subfigure{\includegraphics[width=0.195\textwidth]
            {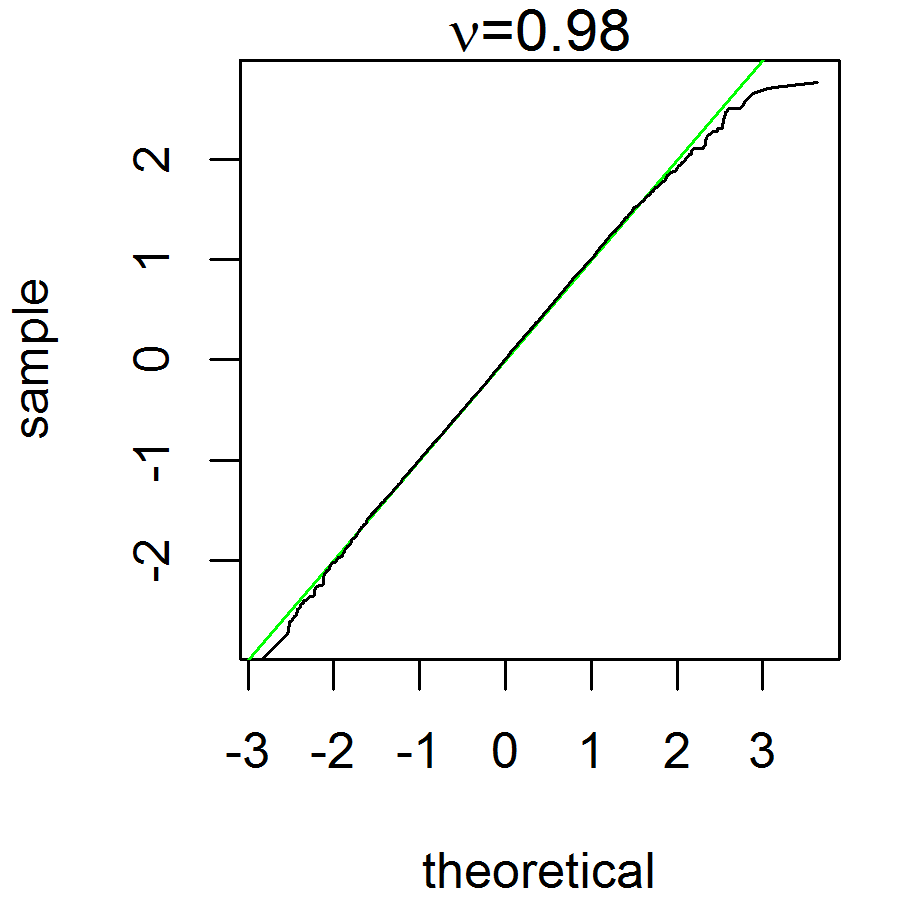}}
         \subfigure{\includegraphics[width=0.195\textwidth]
            {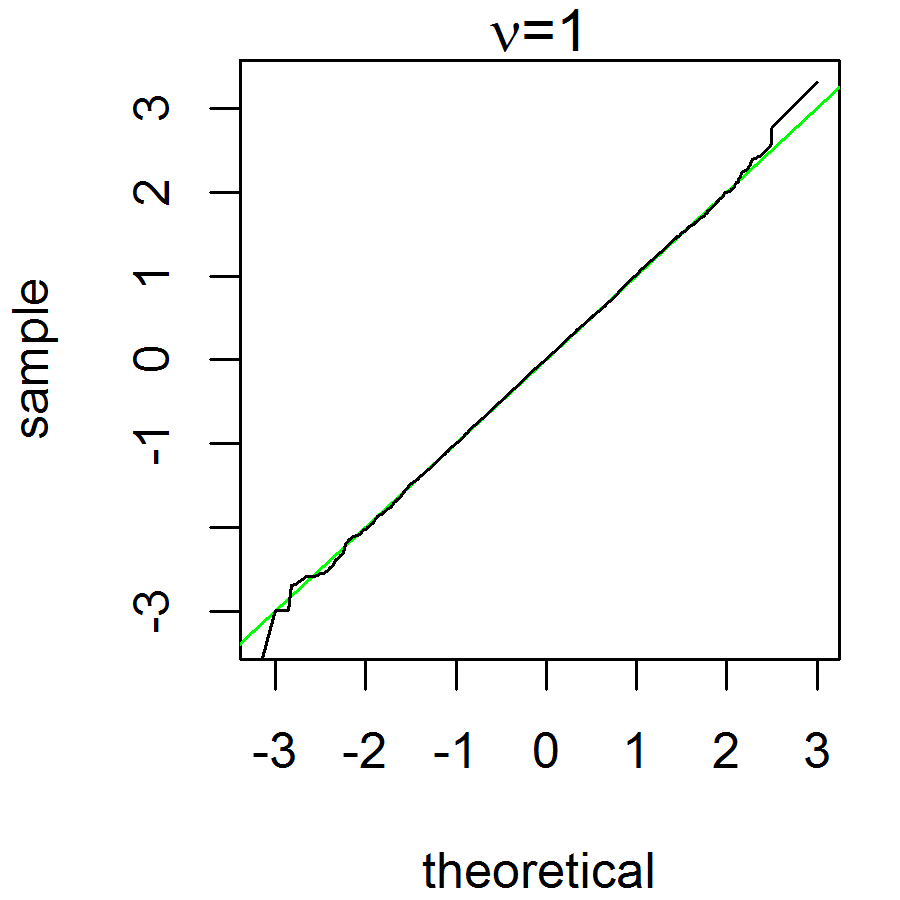}}
    \end{center}
    \caption{Q-Q plots for $0.52\leq\nu\leq1$.}
    \label{QQ plots 2}
\end{figure}

\end{appendices}

\end{document}